\documentclass[a4paper, 12pt]{book}
\usepackage{amssymb}
\usepackage{amsmath}
\usepackage{amsfonts}
\usepackage{amsthm}
\usepackage{latexsym}
\usepackage[mathscr]{eucal}
\usepackage{graphicx}
\usepackage{psfrag}

\theoremstyle{plain}

\newtheorem{proposition}{Proposition}
\newtheorem{lemma}{Lemma}
\newtheorem{theorem}{Theorem}
\newtheorem{corollary}{Corollary}
\newtheorem*{definition}{Definition}
\newtheorem*{bem}{Remark}
\newtheorem*{ann}{Assumption}

\newcounter{mnotecount}[section]

\DeclareMathOperator{\tr}{tr}
\DeclareMathSymbol{\R}{\mathbin}{AMSb}{"52}
\DeclareMathOperator{\sgn}{sgn}

\title{Relativistic Elastodynamics}
\author{Michael Wernig-Pichler \\ Institut f\"ur theoretische Physik der Universit\"at Wien\\ Boltzmanngasse 5, A-1090 Wien, Austria}
\date{}
\begin{document}





\maketitle

\tableofcontents

\chapter*{Acknowledgements}

First of all I want to express my deep gratitude to the supervisor
of this thesis, Prof. Robert Beig. He aroused my interest in the
beautiful field of elasticity. For this and for his continuous
support and inspiration during the last years I am most thankful.
His genuine intuition and deep understanding of physics impressed
and motivated me all the way through the genesis of this thesis.

I am also grateful to the other members of the gravity group. To the
supervisor of my diploma thesis, Prof. P.C. Aichelburg and Prof. H.
Urbantke, whose introductory lectures on relativity opened my eyes
for this marvelous field.

We must not forget the other members: the people from
"Dissertandenzimmer": Michael, Patrick and Roland. Christiane
(formerly known as "little Christiane") and her husband Juan. They
answered my questions and questioned my answers. Florian and
Natascha, who also contributed to the unique atmosphere at the
Vienna gravity group. The former members Walter and Mark who were of
great help during my beginnings in the group. I really enjoyed my
time with all these people at the institute.

It is a pleasure to mention my dear friend and college Anil although
I am sorry that we could not work on relativistic elasticity
together (as was first planned) due to the unsure funding. I truly
hope that the situation of the FWF will improve so that young
scientist like him (or Patrick) are not forced to go abroad for
purely financial reasons.

Leaving the institute my path most naturally leads to the dojo.
Karatedo is an important part of my life and many friendships are
connected with it. Just to mention a few: I am thankful to
Christian, Gunther, Clemens, Frrranzzz, Leo, Markus, Erika, Sonja,
Gernot and Reinhold.

After training there is still enough time to meet more friends: my
flat-mate Manfred (the stoic Gaisacher), J\"orn (the unstoic
German), Bernhard and Sonja (I still got one of their CDs) and
Martin (I smile wile I remember those long evenings we spent
discussing religion, god and girls).

We now hold at the threshold between friends and family. This is the
right point to thank my girl-friend Ceylan. I owe her many important
lessons concerning human relationships. Her belief in me has helped
me a lot during the last months.

At last I mention those who should have come first (I stole this
formulation from Mark) and who will always come first in real life:
from the bottom of my heart I express my deep gratitude to my father
and to my mother as well as to the rest of my family. Thank you,
Udo, Brigitte, Johannes, Leonhard, Eva, Anna-Oma, Milli-Oma, \dots

\vspace{1cm} This work was supported by the by the FWF (Project
number P16745).

\chapter{Introduction}

To grasp the importance of understanding elastic materials one just
has to take a look around. Many of the things one will see then are
"elastic". These include all kinds of fluids (fluid dynamics is a
special case of elasticity), as well as solid materials. From very
large to very small scales elasticity theory applies. Phenomena like
earth-quakes can be understood as the propagation of waves in an
elastic medium (see \cite{aki1}). The very same theory applies to
complete different field like human medicine: tumor growth (see
\cite{sle1}) or the elastic properties of arteria (see \cite{hum1})
can be considered.

These fields are covered by classical non-relativistic elasticity
theory. There the material is identified with a certain region in
space and its motion is modeled by time-dependent mappings between
this region and the actual position of the material. Usually one
assumes the existence of a flat (Riemannian) metric on space.

See \cite{gur1} or \cite{her2} for an introductory textbook into
non-relativistic theory. A more mathematical approach is given in
\cite{mh1}, \cite{ogd1} or \cite{cia2}. For a (very) short but
nevertheless enlightening compendium including information on global
properties we refer to \cite{sid2}.

Attempts to formulate a relativistic theory of elasticity can be
traced back as early as 1911 for special relativity (\cite{her1})
and 1916 for general relativity (\cite{nor1}). The promptness of
these results (being published only a short time after the advent of
special (general) relativity) indicates the importance of elasticity
to the scientific community back then.

Since these early attempts there have been a number of approaches to
obtain a relativistic theory of elastic materials. We ought to
mention the work of Carter and Quintana \cite{cq1} together with its
review by Ehlers \cite{ehl1}. We also refer to the works of Maugin
\cite{maugin}, Kijowski and Magli \cite{km1} and Tahvildar-Zadeh
\cite{tz1}.

In this thesis we follow the formulation of relativistic theory as
given by Beig and Schmidt \cite{bs1} or Christodoulou \cite{chr1}.
This means that we treat relativistic elasto-dynamics as a
Lagrangian field theory. We derive the corresponding equations of
motion and prove local well-posedness for a variety of different
setups.

The following are the basic ingredients:
\begin{itemize}
\item a space-time manifold equipped with a smooth Lorentzian
metric
\item a three-dimensional "material manifold" equipped with a smooth
volume form represents the material
\item a stored energy function, which plays the role of an "equation
of state"
\end{itemize}
The motion of the elastic medium is described by mappings from
space-time onto the material manifold (space-time description) or by
time-dependent mappings from the material manifold onto space-like
hyper-surfaces of space-time (material description). The equations
of motion are derived from an action principle. In this sense we
restrict ourselves (in terms of non-relativistic theory) to
hyper-elastic materials. In difference to standard non-relativistic
theory we do not assume the material manifold to be equipped with a
flat reference metric, since the volume form turns out to be
sufficient for all kinematical purposes. An overview on the
formulation of relativistic elasticity as used in this thesis can be
found in \cite{beig2}. For a review emphasizing the hyperbolic
structure of the equations of motion we recommend \cite{beig1}.

While in this thesis we consider elasto-dynamics (i.e.
time-dependent motions), we note that there exist related work
within the same framework by Beig and Schmidt on stationary
scenarios. They investigate the boundary value problem for
time-independent formations of elastic matter (like a solid star in
rigid rotation). For the non-relativistic case we refer to
\cite{bs3}, while the relativistic problem is treated in \cite{bs2}.
An application of the formalism to static elastic shells can be
found in \cite{losert}.

This thesis is organized as follows:

In chapter \ref{elvis1} we review basic definitions from the theory
of hyperbolic quasi-linear second order systems of partial
differential equations and collect local well-posedness results for
both bounded and unbounded domains.

Basic kinematics for relativistic elastic materials in developed in
chapter \ref{chaptergeokin}. For both space-time ("Euler")
description and material ("Lagrange") description we identify the
basic unknowns and derive a covariant formulation of relativistic
elasticity in terms of these unknowns. For both descriptions the
Lagrangian is derived in terms of material quantities. The
energy-momentum tensor is given and a discussion of the reference
state (a "realization" of the (abstract) material in the actual
world) is included.

Chapter \ref{elvisdynamics} deals with the dynamic properties of
elastic materials. For both descriptions we derive the equations of
motion and prove local well-posedness results under various
assumptions. We include a study of conserved quantities in the
presence of space-time symmetries and compute an explicit formula
for the solution of the initial value problem for the linearized
equations of motion. Then the self-interaction due to gravity is
included and a well-posedness result for this setup is proved.
Finally we consider the initial-boundary value problem and obtain
well-posedness for natural boundary conditions.

A summary of the results of this thesis is given in chapter
\ref{elvisconc1}. We finish by stating various open problems and
possible ways to generalize the results obtained in the course of
the present work.

The appendix finally collects remarks on energy conditions satisfied
by the energy-momentum tensor, non-relativistic limit for the
elastic equations of motion and 3+1 decomposition for the Chistoffel
symbols. It also provides us with an easy 1+1 dimensional example to
get used to the basic definitions as introduced in chapter
\ref{chaptergeokin}.

\chapter{Hyperbolic Second Order Systems}\label{elvis1}

As soon as one tries to describe a phenomenon in objective terms,
mathematics comes necessarily into play. If the phenomenon is
dynamical it is very likely that one will not be able to proceed
without the tool of partial differential equations. If these
equations governing this dynamical process arise from an action
principle (which usually is the case in mathematical physics) these
will naturally form a quasi-linear (second order) system of partial
differential equations. We will find various examples for such
systems (arising from action principles) in the course of this
thesis. It is thus favorable to have a well-oiled machinery, when it
comes to questions concerning well-posedness. To provide the
necessary tools to prove such results for the systems arising in the
context of elasticity in chapter \ref{elvisdynamics} is the aim of
this chapter:

We will essentially state two results. The first is concerned with
the pure initial value problem and is due to Kato et al \cite{hkm1}.
The second deals with the initial boundary value problem and is due
to Koch \cite{koch1}. We will not use these results directly but
rather give corollaries, which will turn out to be more appropriate
for later application.

\section{Basics}\label{quasisec}

We start with a some general definitions concerning quasi-linear
second order systems of partial differential equations: the latter
are of the form
\begin{equation}\label{qso1}
A^{\mu\nu}_{AB}\partial_{\mu}\partial_{\nu}f^{B}+B_{A}=0
\end{equation}
where the unknown $f^{A}$ appears linearly in the highest order
derivatives, i.e. $\partial_{\mu}\partial_{\nu}f^{B}$. The
coefficients $A^{\mu\nu}_{AB}$ and $B_{A}$ are allowed to depend on
lower order derivatives of the unknown and may also depend on
$x^{\mu}$, the coordinates on the base space. One finds that most
information about the system can be read off from the highest order
terms. We thus define
\begin{definition}
The principal part of a system of PDE is the part involving the highest order derivatives.
\end{definition}
For the case of equation (\ref{qso1}) the principal part is given by
\begin{equation}\label{pp1}
A^{\mu\nu}_{AB}\partial_{\mu}\partial_{\nu}f^{B}
\end{equation}
\begin{definition}
The principal symbol corresponding to the principal part (\ref{pp1})
(and thus to the system (\ref{qso1})) is given by
\begin{equation}
A_{AB}(k):=A^{\mu\nu}_{AB}k_{\mu}k_{\nu}
\end{equation}
Finally the characteristic polynomial is given by the determinant of the principal symbol
\begin{equation}
A(k):=\det A_{AB}(k)
\end{equation}
\end{definition}
Let upper case Latin indices range from one to $N$. Then the
characteristic polynomial is homogenous of order $2N$ in the
co-vector $k$. Note that both the principal symbol and the
characteristic polynomial depend on the unknown $f$ and its first
order derivatives. In this sense the unknown determines the
character of the equation. This fact has to be kept in mind for the
following definitions, where the dependence on the unknown will be
suppressed.


In the following we restrict ourselves to a certain type of systems:
\begin{definition}
The system (\ref{qso1}) is called weakly hyperbolic w.r.t. a
co-vector $\tilde k$, if $A(\tilde k )\ne 0$ and the polynomial
$A(k+s\tilde k)$ admits only real zeros $(s_{1},\dots , s_{2N})$ for
any choice of $k$, i.e. if the characteristic polynomial is a
hyperbolic polynomial. Such a $\tilde k$ is then called
sub-characteristic.
\end{definition}
This conditions turns out to be sufficient to obtain well-posedness
for linear systems (see \cite{john111}), but it is too weak to
establish well-posedness for the quasi-linear equations
(\ref{qso1}). There are various stronger concepts of hyperbolicity,
which are not equivalent. Without the attempt to give an overview
(which can f.e. be found in \cite{beig1} or \cite{chr1}) we state
the notion used in this thesis.
\begin{definition}
We call the system (\ref{qso1}) regular hyperbolic if it is regular
hyperbolic w.r.t. a pair $(X^{\mu},\omega_{\nu})$, i.e. if there
exist a pair $(X^{\mu},\omega_{\nu})\in TM \times T^{*}M $
satisfying $X^{\mu}\omega_{\mu}>0$ and constants
$(\epsilon_{1},\epsilon_{2})>(0,0)$ such that for arbitrary
$\psi^{A}$ we have
\begin{equation}
-A^{\mu\nu}_{AB}\omega_{\mu}\omega_{\nu}\psi^{A}\psi^{B}\ge
\epsilon_{1} |\psi|^2
\end{equation}
while
\begin{equation}
A^{\mu\nu}_{AB}k_{\mu}k_{\nu}\psi^{A}\psi^{B}\ge \epsilon_{2}|\psi
|^2|k|^2
\end{equation}
for all $k_{\mu}\in T^{*}M$ satisfying $k_{\mu}X^{\mu}=0$.
\end{definition}
\begin{bem}
This is an open condition (it involves real inequalities). Suppose
everything is sufficiently smooth and the system is regular
hyperbolic for the unknown $f$. Then the system is also regular
hyperbolic in some neighborhood of this $f$ (in a suitable
topology).
\end{bem}
The second condition in the definition above is of some importance on its own. For later need we define:
\begin{definition}
An objects $C^{ij}_{AB}$ is called strongly elliptic if the following inequality holds
\begin{equation}\label{LH1}
C^{ij}_{AB}\mu_{i}\mu_{j}\lambda^{A}\lambda^{B}\ge \epsilon |\mu|^2 |\lambda|^2
\end{equation}
for a positive real number $\epsilon$. This condition is also called Legendre-Hadamard condition or rank one positivity.
\end{definition}
The two notions of hyperbolicity given so far are connected via the following lemma:
\begin{lemma}
Regular hyperbolicity implies weak hyperbolicity. Every w.r.t. a
pair $(X,\omega)$ regular hyperbolic operator is also weakly
hyperbolic w.r.t the co-vector $\omega$.
\end{lemma}
A proof of this statement can be found f.e. in \cite{beig1}. In
short the argument works like this: Hyperbolicity is defined as a
condition on the zeros of a determinant. This is equivalent to an
eigenvalue problem, which due to the symmetry of the involved
objects can be shown to admit only real eigenvalues.

There exists another even weaker notion of hyperbolicity:
\begin{definition}
A co-vector $\tilde k$ is called non-characteristic, if $A(\tilde k)
\ne 0$. In turn a co-vector $k$, for which $A(k)=0$ is called
characteristic.
\end{definition}
Clearly a sub-characteristic co-vector is non-characteristic but not
the contrary has not necessarily to be true. This can be seen
directly from the definitions.
\begin{bem}
Note that hyperbolicity in the forms given here is a point-wise
(i.e. local in particular) condition.
\end{bem}
To get used to the notation we discuss the scalar wave equation on Minkowski space as an easy example:
\begin{equation}
\eta^{\mu\nu}\partial_{\mu}\partial_{\nu}\phi =\mbox{l.o.}
\end{equation}
The principal part is simply given by the d'Alembert operator acting
on the unknown: $\eta^{\mu\nu}\partial_{\mu}\partial_{\nu}\phi$.
Thus the principal symbol is $\eta^{\mu\nu}k_{\mu}k_{\nu}$. Because
we are only dealing with the scalar case in this example, the
characteristic polynomial coincides with the principal symbol. We
find that light-like co-vectors are characteristic and that all
other co-vectors are non-characteristic. Choosing $\tilde k = dt$ we
get that the zeros of
\begin{equation}
-(k_{0} +s)^2 + \vec{k}^2
\end{equation}
are all real. Since we can always find coordinates such that $\tilde
k =dt$ at a certain point for time-like $\tilde k$, we find that the
wave equation is weakly hyperbolic w.r.t. every time-like co-vector.
(Thus every time-like co-vector is sub-characteristic.) It is easy
to see that it is not weakly hyperbolic w.r.t. any space-like or
light-like vector. Finally we note that the wave equation is regular
hyperbolic w.r.t. any pair $(X^\mu, \tilde k_\nu)$ where $X^\mu$ and
$\tilde k_\nu$ are both time-like.

We have mentioned, that the notion of weak hyperbolicity is not
strong enough to guarantee well-posedness of the Cauchy problem. In
the next sections we will see that regular hyperbolicity is a
sufficient assumption for that matter.

\section{Local existence with no boundaries present}\label{loceubd11111}

The ultimate goal of this section is to give a local existence
result based purely on the concept of regular hyperbolicity. To this
purpose we first state a local existence result which is a
simplification of a theorem due to Kato et al. \cite{hkm1} and then
modify it. The final corollary will in some sense be weaker than the
original theorem, but will be more natural from a space-time point
of view, since no preferred foliation is included. For this purpose
we make use of an domain of dependence result due to Christodoulou
\cite{chr1}. The original theorem by Kato et al. reads as follows:

Given a quasi-linear second order system
\begin{equation}\label{qso2}
A^{\mu\nu}_{AB}\partial_{\mu}\partial_{\nu}f^{B}+B_{A}=0
\end{equation}
for an unknown $fÂ$ which is a mapping from $[0,T]\times \R^3 $ to
$\R^3 $.  Greek indices run from $0$ to $3$, where
$x^{0}=t\in[0,T]$. The coefficients $A^{\mu\nu}_{AB}$ and $B_{A}$
depend smoothly on the unknown, its first order derivatives as well
as on time $t$ and on the point $x^{i} \in\R ^3$, i.e.
\begin{equation}
 A^{\mu\nu}_{AB} \in C_{b}^{\infty}([0,T]\times \R^3 \times\Omega ) \ \mbox{and} \ B_{A}\in C_{b}^{\infty}([0,T]\times \R^3 \times\Omega )
\end{equation}
Here $\Omega\subset \R^3\times\R^{12} $ is an open set. We assume that $(f,\partial f)$ take values in $\Omega$.

In addition to these differentiability properties we require the principal part to show the following symmetry:
\begin{equation}\label{symmqn1}
A^{\mu\nu}_{AB}=A^{\nu\mu}_{BA}
\end{equation}
We also assume that the principal part is regular hyperbolic w.r.t.
$(\partial_{t},dt)$ for some reference state in $\Omega$. This means
that there exists a real $\epsilon>0$ such that
\begin{eqnarray}
-A^{00}_{AB} \psi^{A}\psi^{B} \ge \epsilon |\psi|^2   \\
A^{ij}_{AB}\xi_i \xi_j  \psi^{A}\psi^{B} \ge \epsilon |\xi |^2 |\psi|^2
\end{eqnarray}
for all $(t,x,g,\partial g)\in [0,T]\times \R^3 \times\Omega$ as
well as for all $\xi\in T^{*}\R^3$ and all $\psi \in T\R ^3$.

Under these assumptions the Cauchy problem with initial conditions
\begin{eqnarray}
f(t=0)=f_{0}\in H^{s+1}(\R^3,\R^3)\\
\partial_{t}f(t=0) =f_{1}\in H^{s}(\R^3,\R^3)
\end{eqnarray}
is considered, where we assume that $(f_{0}, f_{1},\partial_{i}f_{0})\in\Omega $. Then the following theorem holds:
\begin{theorem}
Assume $s \ge 3$ and let the preceding assumptions hold. Then the
Cauchy problem for (\ref{qso2}) is well-posed in the following
sense: Given $(f_{0},f_{1})\in H^{s+1}(\mathbb{R}^3
,\mathbb{R}^3)\times H^{s}(\mathbb{R}^3 ,\mathbb{R}^3 )$ such that
$(f_{0}, f_{1},\partial_{i}f_{0})\in\Omega $, there is a
neighborhood $V$ of $(f_{0},f_{1})$ in $H^{s+1}\times H^s $ and a
positive number $T^{*} \le T$ such that, for any initial conditions
in $V$, (\ref{qso1}) has a unique solution $f(t,\cdot)$ for $t\in
[0,T^{*}]$ with $(f,\partial  f) \in \Omega $. Moreover $f\in
C^{r}([0,T^{*}],H^{s+1-r})$ for $0\le r\le s$ and the map
$(f_{0},f_{1})\mapsto (f, \partial_{t}f )$ is continuous in the
topology of $H^{s+1}\times H^s $ uniformly in $t\in [0,T^{*}]$.
\end{theorem}
A proof of a more general version can be found in \cite{hkm1}.

We will alter this result in two directions. On the one hand we
would like to generalize it to arbitrary regular hyperbolic systems.
This on the other hand brings up a new problem, namely the need for
a version of the theorem, which is local in space. This need arises
since regular hyperbolicity in general does not lead to a situation
covered directly by the theorem. This will be seen in the next
lemma. Then we will show in a final lemma how to deal with this
problem by giving a way how to continue initial data on an open set
$U\subset\R^3$ to initial data on all of $\R^3$ in a proper way so
that the theorem can still be applied. But first we show how the
requirement of regular hyperbolicity locally leads to a problem of
the above type.
\begin{lemma}
Assume that the system (\ref{qso2}) obeys the symmetry
(\ref{symmqn1}) and is regular hyperbolic w.r.t. a pair $(X^\mu,
\omega_\nu)$. If the coefficients are regular (smooth for example)
there exists a coordinate system, every point of space-time has a
neighborhood where we can find coordinates such that the
requirements of the theorem are met.
\end{lemma}
In the proof we will not deal with the differentiability
requirements but will only care about the geometry. First note that
the symmetry (\ref{symmqn1}) is unchanged under coordinate
transformations. The goal is that of finding a foliation, such that
$\omega$ is the co-normal and $X$ is tangent to the time-flow, i.e.
a coordinate system in which $(X,\omega)\propto(\partial_{t},dt)$.
Having this picture in mind, we first assume that $\omega$ is
closed, i.e. that $d\omega=0$. This implies that $\omega=d\alpha$
for a scalar potential $\alpha$. By choosing this $\alpha$ as new
time coordinate, we obtain $\omega=dt$. For the other coordinates we
have to solve
\begin{equation}
X^\nu \partial_\nu y^i=0
\end{equation}
This means the new coordinate has to be constant along the flow of
$X$. Let $\phi^\nu_s (x^\mu)$ denote this flow. Then (since $X$ is
time-like) the equation $\phi_s (t,x)=(0,z)$ uniquely determines the
parameter $s(t)$. Now we define
\begin{equation}
y^i(t,x):=\phi^i_{-s(t)}(t,x)
\end{equation}
Defined this way, $y$ is constant along the integral curves of $X$
and thus obeys the above equation. Hence $(\alpha, y^i)$ constitute
a coordinate system satisfying the claims of the lemma.

If $\omega$ is not closed we have to proceed along a slightly
different route. Fix an arbirtary point on the manifold we can find
a hyper-surface with co-normal $\tilde \omega$ such that at the
chosen point $\tilde \omega$ coincides with $\omega$. Since we
assumed the system to be regular hyperbolic w.r.t. $(X,\omega)$, by
continuity there exists a neighborhood of the chosen point, where
the system is regular hyperbolic w.r.t. the pair $(X, \tilde
\omega)$. Since $\tilde \omega$ is hyper-surface-orthogonal by
definition, we can use the first part of the proof to conclude the
result of the lemma.

\begin{bem}
For a regular hyperbolic system we can find data, such that the
requirements of the theorem are locally met. To apply the theorem,
we have to extend the initial data to all of $\R^3$. This is the
gaol of the following lemma.
\end{bem}
\begin{lemma}
Let $\tilde U$ be a non-empty open subset of $\R^3$ and let
$(g_0,g_1)\ \in H^{s+1}(\tilde U,\mathbb{R}^3)\times H^{s}(\tilde
U,\mathbb{R}^3 )$ be small initial data on $\tilde U$, i.e. a pair
of functions satisfying the requirements for the theorem restricted
to the open set $\tilde U$. Let $U$ be a non-empty open subset of
$\tilde U$. We also assume the existence of a reference solution,
which induces initial data in the sense of the theorem on all of
$\R^3$. Then there exists a pair of functions $(f_0,f_1) \in
H^{s+1}(\mathbb{R}^3 ,\mathbb{R}^3)\times H^{s}(\mathbb{R}^3
,\mathbb{R}^3 )$, which is an initial data set in the precise sense
of the theorem with the property that $(g_0,g_1)=(f_0,f_1)$ in $U$.
The development of the initial data in the domain of dependence of
$U$ is independent of the continuation and depends solely on the
choice for $(g_0,g_1)$.
\end{lemma}
\begin{bem}
 This lemma allows us to use the theorem also for the case when data are only given in a subregion of
 $\R^3$. For the notion of "domain of dependence" we refer to
 \cite{chr1}. For the given context we might sloppily translate the last part
 of the lemma as: "there exists a unique local solution".
\end{bem}
We denote the initial data on $\R^3$ coming from the reference
solution by $(d_0, d_1)$. Then on $\tilde U$ we define
\begin{equation}
(\delta g_0 ,\delta g_1)=(g_0,g_1)-(d_0, d_1)
\end{equation}
One could think of multiplying this with the characteristic function
for $\tilde U$ and adding this then to $(d_0, d_1)$. These data
would satisfy the algebraic conditions, but would not meet the
differentiability requirements in general. We thus have to smoothen
these data. Let $\chi$ be a smooth nonnegative function on all of
$\R^3 $ such that
\begin{equation}
\chi=\left\{\begin{array}{cc} 1 \ \mbox{on}\ U  \\    0  \ \mbox{on}
\R^{m+1}\setminus \tilde U     \end{array} \right.
\end{equation}
Then $\chi (\delta g_0 ,\delta g_1)$ is defined on all of $\R^3$ and has the required differentiability. Thus
\begin{equation}
(f_0,f_1):=(d_0, d_1)+\chi (\delta g_0 ,\delta g_1)
\end{equation}
has the required differentiability too. We have to check that the
hyperbolicity conditions are not violated. By assumption they
obviously hold in $U$ and $\R^3\setminus \tilde U$. They also hold
in the region $\tilde U\setminus U$, since the contribution of $\chi
(\delta g_0 ,\delta g_1)$ can be made arbitrarily small by choosing
$(\delta g_0 ,\delta g_1)$ small enough (in the proper norms, f.e.
the supremum norm). This works by the smoothness of $\chi$. In the
region $U$ $(f_0,f_1)$ and $(g_0,g_1)$ clearly coincide. This proves
the first part of the lemma. The second part of the lemma
(concerning domain of dependence) is actually proven in theorem 5.10
of \cite{chr1} which we can directly apply to the given setup.

\begin{corollary}
Given a regular hyperbolic system of the form (\ref{qso2}). Then the Cauchy problem is locally well-posed in the sense, that for local (in space) initial data there exists a unique local (in space and time) solution for the initial value problem, which depends continuously on the data.
\end{corollary}
The involved function spaces are the same as in the theorem.
\begin{bem}
Note that strong ellipticity of the spatial part (i.e. the
Legendre-Hadamard condition) was sufficient to guarantee
well-posedness. In the presence of boundaries the need for more
restrictive conditions will arise.
\end{bem}

\section{Local existence in the presence of
boundaries}\label{kochsect1}

In analogy to the previous section we are going to state a local existence result on the initial-boundary value problem for quasi-linear second order systems. The main theorem is due to Koch \cite{koch1}. A similar result by Chrzestzczyk can be found in \cite{chrz1}, although the Koch version is much closer to the Lagrangian approach used in this work as we will see. Although the theorem in its genuine form allows for mixed boundary conditions we will only state a version involving just one kind of boundary conditions namely such of Neumann-type. As before we will first state the main theorem and then discuss some of its properties and point out the main differences to the unbounded case.  \\
Let $\Omega$ be a bounded domain in $\R^3$ with smooth boundary
$\partial\Omega$. Let $\Omega_T:=[0,T)\times \Omega$ and
$\partial\Omega_T:=[0,T)\times\partial\Omega$ for $0<T\le \infty$.
Let $n_{\mu}$ denote the outer co-normal to $\Omega_T$. Greek
indices run from zero to three, while Latin indices range from one
to three. The independent variables are $(x^\mu)= (t,x^k)$.

Suppose we are given a system of the form
\begin{equation}\label{qso3}
\frac{d}{dx^{\mu}}\mathscr{F}_A^\mu=B_A
\end{equation}
in $\Omega_T$ subjected to the boundary condition
\begin{equation}\label{qsobc3}
\mathscr{F}_A^\mu n_\mu|_{\partial\Omega_T}=0
\end{equation}
at $\partial\Omega_T$. Here both $\mathscr{F}$ and $B$ are functions
of $x^i$, the unknown $u^A$ and its first order derivatives
$\partial_\mu u^A$. Thus the given system is quasi-linear. We seek
solution for the following initial conditions at $\{t=0\}\times
\Omega$:
\begin{equation}\label{qsoic3}
u|_{t=0}=u_0 \ \ \ \ \ \partial_t u|_{t=0}=u_1
\end{equation}
We assume that $(u_0 ,u_1)\in H^{s+1}\times H^s $ for $s\ge 3$. The
principal part is obtained from $\mathscr{F}$ by variation w.r.t.
the gradient of the unknown:
\begin{equation}
A^{\mu\nu}_{AB}:=\frac{\partial\mathscr{F}^\mu_A}{\partial (\partial_\nu u^B)}
\end{equation}
Let $U$ be an open neighborhood of
$\mbox{graph}(u_0,u_1,\partial_iu_0)$ in $(\Omega\cup \partial
\Omega )\times\R^3\times \R^{12}$. We assume that $\mathscr{F}$ and
$B$ are smooth functions on $U$. We additionally assume the
following symmetry of the principal part:
\begin{equation}
A^{\mu\nu}_{AB}=A^{\nu\mu}_{BA}
\end{equation}
Note that this relation is automatic, if the system is derived from
an action principle. Now comes the crucial input, namely the
hyperbolicity requirements. From the previous section we expect
point-wise positivity conditions on certain portions of the
principal part. This view will only partly persist. For the
time-components we require that for any $v\in U$
\begin{equation}
-A^{00}_{AB}(v)\phi^A\phi^B\ge \epsilon |\phi|^2
\end{equation}
for some positive constant $\epsilon$. This is the same condition as
we had before in the unbounded case. The difference comes with the
spatial part: Instead of a point-wise condition we now have the
following coerciveness condition: for any $v_0\in C^1(\bar\Omega)$,
$v_1\in C(\bar\Omega)$ with $\mbox{graph}(v_0,v_1,\partial_i v_0)\in
U$ there exist constants $\epsilon >0$ and $\kappa \ge 0$, such that
for all smooth $\phi$
\begin{equation}\label{G1}
\int_\Omega A_{AB}^{ij}(x,v_0,v_1,\partial_i v_0)\partial_i\phi^A\partial_j\phi^B\ge \epsilon ||\phi||^2_{H^1}-\kappa||\phi||^2_{L^2}
\end{equation}
This inequality is usually called Garding inequality (see f.e.
\cite{wrl1}).
\begin{bem}
We will see later in this section, that the Garding inequality
implies the Legendre-Hadamard condition (\ref{LH1}). The converse
statement however is in general not true.
\end{bem}
If $\kappa=0$, then this coercivity condition tells us that the
$H^1$ norm can be estimated from above by some bilinear form. If
$A^{ij}_{AB}$ is smooth, then it is bounded too and the left had
side in (\ref{G1}) can itself be estimates from above by the $H^1$
norm. Thus the bilinear form generates an equivalent norm. Along
this way, one can try to prove existence results. In the general
case however $\kappa\ne0$ and this does not work immediately.

If one is only interested in Dirichlet boundary conditions, the
unknown can be assumed to vanish at the boundary. In that case the
coercivity condition is equivalent to the Legendre-Hadamard
condition. This will be shown in a lemma following the main theorem.

Since boundaries are present, a smallness assumption on the initial
data will not be sufficient anymore. We will have to guarantee that
the initial data go along with the boundary conditions. This
requirement is reflected in the so called compatibility conditions,
which represent restrictions on the initial data. A discussion of
the actual meaning of these conditions will be given following the
theorem.

\begin{theorem}
Suppose the compatibility conditions are satisfied up to order $s$.
Then under the above assumptions there exists a unique $0<T_*\le T$
and a unique classical solution $u\in C^2(\Omega_{T_*}\cup
\partial\Omega_{T_*})$ of the system (\ref{qso3})-(\ref{qsoic3})
with $\partial^{\sigma} u (t)\in L^2(\Omega)$ if $0\le\sigma\le s+1$
($\partial$ denotes all derivatives). In addition the solution
depends continuously on the initial data.
\end{theorem}
\begin{bem}
Note that the assumptions of the theorem involve mainly open
conditions. Thus if we can show them to hold for some initial data,
we immediately obtain a neighborhood of the original data as well as
admissible data. Consequently the existence of a reference solution
implies the existence of solutions, which are small perturbations.
Then the continuity part of the theorem tells us, that these
solutions remain close to the reference state at least for short
times.
\end{bem}
As was mentioned in the introduction, a proof of (an extended
version of) this theorem by Koch can be found \cite{koch1}.

We still need to discuss the compatibility conditions. Roughly
speaking they guarantee, that the initial data satisfy the boundary
conditions up to a certain order in time. Higher time derivatives of
the initial data are expressed via the field equations. F.e. we say
that the compatibility conditions are satisfied to zeroth order, if
the boundary conditions hold, where $(u,\partial_t u , \partial_i
u)$ is substituted by $(u_0,u_1,\partial_i u_0)$. It is quite
lengthy to actually write down these conditions for general order,
so at this stage we only give the recipe how they are obtained.

We introduce (suppressing the $x^{i}$-dependence)
\begin{equation}\label{KB1}
n_\mu\partial_{t}^{m} \mathscr{F}^{\mu}_{A} =: n_{\mu} \mathscr{F}^{(m)\mu}_{A}(u^{A},u^{A},_{i},\partial_{t} u^{A}, \partial_{t}u^{A},_{i},\dots, \partial_{t}^{m} u^{A},_{i},\partial_{t}^{m+1}u^{A})
\end{equation}
To express the time derivatives in terms of the initial data we use
the differential equations (\ref{qso3}) and their time derivatives.
$
\partial_{t}^{m} u^{A}$ for example is recovered as a function of
the lower time derivatives by solving
\begin{equation}\label{4cc3}
\partial_{t}^{m-2} \left( \frac{d}{dx^{\mu}}\mathscr{F}_A^\mu-B_A  \right) =0
\end{equation}
for the highest order in time. This is possible, since the
coefficient is given by $A^{00}_{AB}$ which is non-degenerate by the
assumptions of the theorem. The solution is called $u^{(m)A}$. By
induction we can eliminate all higher time derivatives and are left
with functions, which are completely determined by the initial data
and their spatial derivatives. Abusing notation we denote these
again by $u^{A(m)}$. Then the compatibility condition reads
\begin{definition}
We say, the compatibility condition is satisfied up to order $s$, provided
\begin{equation}\nonumber
n_\mu\mathscr{F}^{(k)\mu}_{A}(u_0^A,u_1^{A},u_0^{A},_{i}, u_1^{A},_{B},\dots ,u^{(k)A},_{B},u^{(k+1)A})|_{\partial\Omega}=0
\end{equation}
is satisfied for all $0\le k\le s$.
\end{definition}
\begin{bem}
This intuitive definition only works if the ingredients are sufficiently smooth.
\end{bem}
In general one will have to deal with non-continuous functions, i.e.
elements of certain Sobolev spaces. In this setup the restriction of
a function to the boundary is not defined in a strict sense. This
problem is overcome by introducing the trace operator (see f.e.
\cite{evans1} or \cite{wrl1}), which assigns boundary values in a
proper way. One can show (see \cite{evans1} p. 259) that the trace
of a function $f\in H^1$ does vanish on the boundary if and only if
$ f\in H^1_0$. Using this we give an alternative definition for the
compatibility condition:
\begin{definition}
We say, the compatibility condition is satisfied up to order $s$, provided
\begin{equation}\nonumber
n_\mu\mathscr{F}^{(k)\mu}_{A}(u_0^A,u_1^{A},u_0^{A},_{i}, u_1^{A},_{B},\dots ,u^{(k)A},_{B},u^{(k+1)A}) \in H^{s-k}(\Omega)\times H^{1}_0(\Omega)
\end{equation}
for all $0\le k\le s$.
\end{definition}
In general it is not clear whether the compatibility conditions can
be satisfied. In the actual case we will show, that under certain
assumptions a large class of initial data can be given satisfying
the compatibility condition. Assume we are given a time-independent
reference solution for the system (\ref{qso3})-(\ref{qsobc3}). We
call this solution $f$. It induces initial data of the form
$(f,\partial_t f)|_{t=0}=(f_0,0)$. Further assume that the principal
part satisfies the Legendre-Hadamard condition w.r.t. the co-normal
$n_\mu$:
\begin{equation}
A(x,f,\partial f)^{\mu\nu}_{AB}n_\mu n_\nu\phi^A\phi^B> C |\phi|^2
\end{equation}
for arbitrary $\phi$ and a positive constant $C$. Then we have the following lemma
\begin{lemma}\label{fuerteeforever}
Assume that $A^{0i}_{AB}(x,f,\partial f)=0$. (There are no mixed
components in the reference deformation.) Then under the two
assumptions made above, we can derive initial conditions close to
those for the reference deformation, which obey the compatibility
conditions.
\end{lemma}
First of all note that the zero-component of the co-normal vanishes
since the co-normal is annihilated by $\partial_t$ by virtue of the
setup:
\begin{equation}
n_0=0
\end{equation}
Let us first prove the lemma for the zeroth order. Let
$\partial_n:=n^\mu
\partial _\mu$ denote a vector field transversal to the boundary
$\partial \Omega$. (If a metric $g^{\mu\nu}$ is given we can choose
f.e. $n^\mu =g^{\mu\nu}n_\nu$) Then the zeroth order condition reads
$n_\mu \mathscr{F}^{(0)\mu}_A|_{\partial\Omega}=0$. Clearly
\begin{equation}
\frac{\partial n_\mu \mathscr{F}^{(0)\mu}_A }{\partial (\partial_n u^B_0)}= n_\mu\frac{\partial \mathscr{F}^{\mu}_A }{\partial (\partial_\nu u^B)}n_\nu
\end{equation}
The right hand side is positive definite near the reference solution
by the Legendre-Hadamard condition. Thus the finite dimensional
inverse function theorem (see f.e. \cite{mh1}) allows us to solve
$n_\mu \mathscr{F}^{(0)\mu}_A|_{\partial\Omega}=0$ for $\partial_n
u_0$. This means that the zeroth order compatibility condition is
satisfied by a suitable choice of a transversal derivative of the
initial data at the boundary.

For the higher orders we first note that
\begin{equation}
n_\mu\mathscr{F}^{(k)\mu}_{A}=n_\mu A^{\mu\nu}_{AB}\partial_t^ku^B_\nu + \mbox{l.o.t.}
\end{equation}
Again the Legendre-Hadamard condition allows us to use the implicit
function theorem to solve this for $\partial_{t}^k\partial_n u$.
This can be rewritten as $\partial_n u^{(k)}$. For $k=1$ this is
nothing but the normal derivative of the initial datum $\partial_n
u_1$. For higher orders $k\ge 2$ we have to show that $\partial_n
u^{(k)}$ can be solved for higher normal derivatives of the initial
data. For this we use the differentiated field equation. For $k\ge
2$ we can write (\ref{4cc3}) as
\begin{equation}
A^{\mu\nu}_{AB}\partial_\mu\partial_\nu \partial^{k-2}_t u^{B}=\mbox{l.o.t.}
\end{equation}
Using that $n_0=0$, this can be rewritten as
\begin{equation}
-A^{00}_{AB}\partial_n u^{(k)B}=n_\mu n_\nu A^{\mu\nu}_{AB}\partial_{n}^3u^{(k-2)B} +\mbox{l.o.t.}
\end{equation}
where the terms denoted by $\mbox{l.o.t.}$ may be of the same order
but involve $A^{0i}_{AB}$ and thus cause no trouble as we will see.
Note that both $A^{00}_{AB}$ and $ n_\mu n_\nu A^{\mu\nu}_{AB}$ are
non-degenerated in some neighborhood of the reference solution by
the hyperbolicity conditions. Thus we can iteratively use the above
equation to determine $u^{(k)B}$ in terms of normal derivatives. We
have that
\begin{equation}
\partial_n u^{(k)B}=M^B_{(k)A}\partial_n^{k+1} u^A +\mbox{l.o.t.}
\end{equation}
for even $k$ where $M^B_{(k)A}$ denotes a non-degenerated matrix
formed from $A^{00}_{AB}$ and $n_{\mu}n_\nu A^{\mu\nu}_{AB}$
(non-degenerated at least close to the reference solution). Finally
for odd $k$ we obtain
\begin{equation}
\partial_n u^{(k)B}=M^B_{(k-1)A}\partial_n^k u^{(1)A} +\mbox{l.o.t.}
\end{equation}
Now we can use the implicit function theorem on both these
expressions because: For the reference solution this equation is
identically satisfied by assumption. The $\mbox{l.o.t.}$-terms are
either lower order in the normal derivative or do vanish due to the
$A^{0i}_{AB}=0$ assumption. On the other hand $M^B_{(k)A}$ involves
only the initial data and first order normal derivatives. Thus all
the requirements are met to apply the implicit function theorem.

We have found that the compatibility condition can in fact be
interpreted as a condition on the odd-numbered normal derivatives of
the initial data at the boundary. Thus by a suitable choice of these
derivatives, the compatibility conditions can be satisfied. Besides
these restrictions the data can be given freely, as long as they
stay in a certain neighborhood of the reference solution. In
particular this means that the field, its first order time
derivative and all of their even normal derivatives are free data.
\begin{bem}
Note that by the above constructions we can choose a large class of
initial data close to those generated by the reference solution,
which satisfy the compatibility condition up to arbitrary order.
\end{bem}

In the next lemma we establish the connection between the
Legendre-Hadamard condition (\ref{LH1}), which was sufficient for
local well-posedness in the unbounded case and the Garding
inequality (\ref{G1}):
\begin{lemma}\label{prince}
Let $\Omega$ be a bounded domain in $\R ^3$. Then the Garding
inequality (\ref{G1}) for $u\in H^1_{0} (\Omega )$ and the
Legendre-Hadamard condition (\ref{LH1}) are equivalent.
\end{lemma}
First we show, how the Legendre-Hadamard condition implies the
Garding inequality: For this we will follow Giaqunita \cite{gia1}.
The argument consists of three parts: first we assume that
$A^{ij}_{AB}(x)=\tilde A^{ij}_{AB}$ is constant, then we derive the
inequality for a finite region and in the last step we cover
$\Omega$ by such regions. The basic tool used for the first step is
Fourier transformation. Because $H^{1}_{0}(\Omega)$ is the closure
of $C^{\infty}_{0} (\Omega)$ it suffices to prove the result for
$u\in C^{\infty}_{0} (\Omega)$.  By the Plancherel theorem (see f.e.
\cite{evans1}) the $L^2$ norm of a square-integrable function is
unchanged under Fourier-transform, thus
\begin{equation}
||\partial_{i} u^{A}||_{L^2}^2 =||k_{i}\hat u^{A} ||_{L^2}^2
\end{equation}
where an upper hat denotes the Fourier transform. Thus
\begin{equation}
\tilde A(u,u)= 4 \pi ^2 \tilde A^{ij}_{AB}\int_{\Omega}
k_{i}k_{j}\hat u^{A}\hat u^{B} \mbox{d} k \ge c \int_{\Omega} |k| ^2
|\hat u|^2 \mbox{d} k=c \int_{\Omega} |\partial u|^2 \mbox{d}x
\end{equation}
Next suppose that $u\in C^{\infty}_{0} (\Omega)$ such that the
support is a subset of $U_{\epsilon}(x_{0})$. On
$U_{\epsilon}(x_{0})$ we can add and subtract a term involving
$A^{ij}_{AB}(x_{0})$. Using the inequality above this leads to
\begin{equation}
\int_{\Omega}A^{ij}_{AB}(x)\partial_{i}u^{A}\partial_{j}u^{B}\mbox{d}x\ge c \int_{\Omega}|\partial u|^2 \mbox{d}x-\omega\int_{\Omega}|\partial u|^2\mbox{d}x
\end{equation}
The first term comes from the Garding inequality for the constant
operator $A^{ij}_{AB}(x_{0})$, while the second is obtained by
uniform continuity (on $U_{\epsilon}(x_{0})$, the difference of
$A^{ij}_{AB}(x)$ from $A^{ij}_{AB}(x_{0})$ must not exceed
$\omega$). By choosing $\epsilon$ small enough we can obtain
$c-\omega >0$.

Since $\Omega$ is bounded subset of $\R ^3$ it is compact and can
hence be covered by a finite number of balls
$B_{\epsilon}(x_{j})\subset\Omega$ for $j=1,\dots, N$. Then we
choose $\phi_{j}\in C^{\infty}_{0}(B_{\epsilon}(x_{j}))$ such that
$\sum\limits_{j=1}^{N} \phi_{j}^2=1$. Then $A(u,u)$ can be written
as
\begin{equation}
\int_{\Omega} A^{ij}_{AB}\left( \sum\limits_{l=1}^{N} \phi_{l}^2\right) \partial_{i}u^{A}\partial_{j}u^{B}= \sum\limits_{l=1}^{N}\int_{\Omega}A^{ij}_{AB}\partial_{i}( \phi_{l}u^{A})\partial_{j}(\phi_{l}u^{B})+\dots
\end{equation}
The dots stand for terms which are at most linear in derivatives of
$u$. Those terms can be handled easily by the triangle inequality
using the fact that both $A^{ij}_{AB}$ and $\phi_{l}$ are bounded
functions. Thus we obtain bounds from below by the $L^2$ norms of
$u$ and $\partial u$ respectively. we obtain
\begin{equation}
A(u,u)\ge c\sum\limits_{l=1}^{N}\int_{\Omega}|\partial (\phi_{l}u )|^2\mbox{d}x-d||u||_{L^2}||\partial u||_{L^2}-e||u||_{L^2}
\end{equation}
As indicated above the constants (especially $d$ and $e$) depend on
$\phi$ and the operator $A^{ij}_{AB}$. We now may pull $\phi_{l}$
out of the bracket in the first term which can lead to a change in
$e$ and $d$. We obtain
\begin{equation}
A(u,u)\ge  c\int_{\Omega}|\partial u|^2-d||u||_{L^2}||\partial u||_{L^2}-e||u||_{L^2}
\end{equation}
Note that $e$ is non-negative in this procedure.We now can apply
Young's inequality (see f.e. \cite{evans1}) to obtain the desired
result.

The other direction is much easier to prove. We only have to find an
appropriate function to insert into the Garding inequality. A good
choice is
\begin{equation}
\phi^{A}=\lambda^{A}\sin (\alpha \mu_{i}x^{i})\chi_{U}
\end{equation}
where $\chi_{U}$ denotes an arbitrary smooth function with support
in an open subset $U \subset \Omega$. Inserting this into the
inequality leads to
\begin{equation}
 \alpha ^2 \int_{U} A^{ij}_{AB} \lambda^{A}\lambda^{B}  \mu_{i}  \mu_{j}\chi_{U}^2 \cos^2 (\alpha \mu_{l}x^{l}) \ge c \alpha ^2 \int_{U}  | \mu |^2 |\lambda |^2 \chi_{U}^2\cos^2 (\alpha \mu_{l}x^{l}) + \mbox{ l.o.}
\end{equation}
The terms denoted by l.o. are those of lower order in $\alpha$. Note
that the coefficients of the $\alpha$ are all bounded, since we
integrate smooth functions over bounded regions. We get rid of these
lower order terms by a high frequency limit. Since all terms are
bounded, there exists an $\alpha_{0}$ such that for all
$\alpha>\alpha_{0}$ the above inequality holds without the lower
order terms. We have that
\begin{equation}
 \int_{U} \cos^2 (\alpha \mu_{l}x^{l})\chi_{U}^2 \left(  A^{ij}_{AB} \lambda^{A}\lambda^{B}  \mu_{i}  \mu_{j}-c  | \mu |^2 |\lambda |^2 \right) \ge 0
\end{equation}
Recall that $U$ was taken to be an arbitrary open subset of
$\Omega$. This implies together with the smoothness of the integrand
that the above estimate holds point-wise. This in turn allows us to
drop the cosine and the function $\chi_{U}$. The result is the
desired Legendre-Hadamard condition.

Although this section is devoted to boundary value problems, we add
the following remark: the Lemma can be used to prove the analogous
result for $\Omega =\R ^3$:
\begin{corollary}
The lemma still holds with the bounded domain $\Omega$ being
replaced by $\R^3$.
\end{corollary}
To show that the Legendre-Hadamard condition follows from the
Garding inequality in this more general context we can use the very
same argument as above without any change at all. The only
difficulty arises when we want to prove the other direction. The
techniques used in the proof of the lemma do not carry over, since
there we made use of the compactness of $\Omega$, which now is not
given anymore. We have to work along a different route.

Let therefore $\phi\in C^{\infty}_{0}$. Then there exists a compact
subset $\Omega$ such that the support of $\phi$ is contained within
$\Omega$. Hence by lemma \ref{prince} the Garding inequality holds
on $\Omega$. But since $\phi$ has compact support all integrals can
be taken over all of $\R ^3$. The Garding inequality on all of
$\R^3$ is obtained. The analogous procedure can be carried out for
all $\phi\in C^{\infty}_{0}$. The desired result follows.

\begin{bem}
This lemma tells us that the Legendre-Hadamard condition is still sufficient to treat the Dirichlet problem.
\end{bem}

We conclude this section by stating two other positivity notions
beside the Legendre-Hadamard condition (\ref{LH1}) and discussing
their relevance in the given context:
\begin{definition}
An objects $C^{ij}_{AB}$ is called rank-two-positive if the following inequality holds
\begin{equation}\label{r2p1}
C^{ij}_{AB}\mu^A_i \mu^B_j \ge \epsilon |\mu|^2
\end{equation}
for a positive real number $\epsilon$. Here $|\mu|$ denotes an
appropriate norm, f.e. $|\mu|^2:=\mu^A_i\mu^A_i$. We finally call an
object $C_{ABCD}$ strong point-wise stable, if
\begin{equation}\label{sps1}
C_{ABCD}\mu^{AB}\mu^{CD} \ge\epsilon |L\mu |^2
\end{equation}
for a positive real number $\epsilon$ where $L\mu^{AB}=\frac{1}{2}(
\mu^{AB}+\mu^{BA})$ denotes the symmetric part of $\mu$.
\end{definition}
Note that by raising or lowering the indices we can write the
rank-two positivity condition as well as the Legendre-Hadamard
condition in the form of the strong-point-wise-stability condition.
Then we have the following picture: The Legendre-Hadamard condition
is the weakest of these three, followed by strong point-wise
stability. The strongest condition is rank-two-positivity.

Note that this rank-two condition is strong enough to guarantee the
validity of the Garding inequality (\ref{G1}). To show this we
simply choose $\mu^A_i =\partial_i \phi^A$ and integrate
(\ref{r2p1}) over a region $\Omega$. This immediately gives the
Garding inequality with $\epsilon=\kappa>0$.

We will however show in a later chapter that in fact
point-wise-stability suffices to obtain the Garding inequality and
hence satisfies the requirements needed for the theorem.

\chapter{Geometry and Kinematics}\label{chaptergeokin}
There have been several accounts to deal with relativistic
elasticity, the best known maybe that of Carter and Quintana
\cite{cq1}. We however will proceed along the route of Beig and
Schmidt first published (for the space-time setup) in \cite{bs1}.

In this chapter we will introduce the basic concepts of relativistic
elasticity due to the latter authors. The basic unknown will be the
configuration of the elastic material, which is described as a
mapping between manifolds. We will work in two different settings.
On the one hand we will use the space-time-description (also known
as Euler picture) where we assign to each point of space-time a
particle. On the other hand we will work with the spatial
description (or Lagrange picture), where for every instant of time
the spatial position of the particle is determined. We will write
down the Lagrangian for both cases. Later in chapter \ref{eav231}
the field equations will be derived from these Lagrangians.

As we will see it is more natural to introduce the main objects of
the theory in the space-time picture. This is due to the fact that
the material description requires a foliation of space-time,
therefore lacking the obvious covariance properties of the
space-time description.

There are two manifolds naturally involved in the formulation of
this theory. One of them is space-time $\mathcal{M}$, which is a
four-dimensional manifold equipped with a Lorentzian metric
$g_{\mu\nu}$. We assume that $\mathcal{M}$ is time-orientable. Let
$x^{\mu}$ ($\mu=0,\dots ,3 $) denote the coordinates on
$\mathcal{M}$.

The other one is called material manifold or simply body. We assign
it the letter $\mathcal{B}$. It is a three-dimensional manifold
equipped with a volume-form $V_{ABC}$. The coordinates on
$\mathcal{B}$ are $X^{A}$ ($A=1,2,3$).

Other manifolds, which appear in this work are constructed out of
these two.

The material manifold can be viewed as an abstract collection of the
particles making up the continuous material. (It is the idea of the
material, as Platon would say.) Usually people assume $\mathcal{B}$
to be a Riemannian manifold, but we find that the existence of a
volume form (which is of course a weaker assumption) is sufficient.
In standard non-relativistic elasticity the metric on $\mathcal{B}$
is used to define the natural (reference) deformation. In our
approach we go the other way round: the natural deformation
(possibly) defines the metric on $\mathcal{B}$. This is more
satisfactory, since one would not expect the natural deformation to
have the same body metric for different space-time metrics, which
would be a consequence of the standard approach.

\section{Space-time description}

In this section we treat elasticity in the space-time description.
The unknowns are a set of scalar fields on space-time taking values
on the body. We will define the basic quantities and the
energy-momentum tensor as well as the Lagrangian are derived.

\subsection{Basic setup}\label{mausi}
The basic unknown in this setup is $f^{A}$, a mapping from
space-time onto the body:
\begin{eqnarray}
f\colon \mathcal{M} &\to& \mathcal{B}  \\
x^{\mu}&\mapsto& f^{A}(x)
\end{eqnarray}

\begin{ann}We place the following restrictions on $f^{A}$:
\begin{itemize}
\item $\partial_{\mu} f^{A}$ as a mapping from $T\mathcal{M} \to T\mathcal{B}$ has maximal rank.
\item the null space of $\partial_{\mu} f^{A}$ is time-like w.r.t. the space-time metric.
\end{itemize}
\end{ann}
\begin{definition}
We call $f^{A}$ configuration. $\partial_{\mu}f^{A}$ will be called
configuration gradient.
\end{definition}
Given a configuration $f^{A}$. Then the conditions on the
configuration gradient uniquely determine a time-like direction in
the tangent space $T\mathcal{M}$. Thus there exists a unique
time-like future-pointing vector-field $u^{\mu}\in T\mathcal{M}$
with the properties
\begin{eqnarray}
u^{\mu}\partial_{\mu}f^{A}=0 \\
g_{\mu\nu}u^{\mu}u^{\nu}=-1
\end{eqnarray}
\begin{definition}
The $u^{\mu}$ as defined above is called four-velocity.
\end{definition}
\begin{bem}
It is important to note that in this context $u^{\mu}$ is not a
given vector-field, but rather a functional of the configuration.
\end{bem}
Next we look at the pullback of $V_{ABC}$, the volume form on
$\mathcal{B}$, along the configuration, i.e. we consider
\begin{equation}
v_{\mu\nu\lambda}:=f^{A},_{\mu}f^{B},_{\nu}f^{C},_{\lambda}V_{ABC}
\end{equation}
which by definition is orthogonal to $u^{\mu}$. Let
$\epsilon_{\mu\nu\lambda\rho}$ be the volume-form coming from the
space-time metric. Then
\begin{equation}
\epsilon^{\mu\nu\lambda\rho}v_{\nu\lambda\rho}
\end{equation}
is time-like and non-zero. Without loss of generality we assume that
it is future-pointing. One easily finds that is orthogonal to
$\partial_{\mu}f^{A}$ and consequently proportional to $u^{\mu}$.
Thus there exists a positive quantity $n$, such that
\begin{equation}\label{ndef1}
n u^{\mu}:=\frac{1}{3 !} \epsilon^{\mu\nu\lambda\rho}v_{\nu\lambda\rho}
\end{equation}
\begin{definition}
We call $n$ particle number density.
\end{definition}
\begin{bem}
As was the four-velocity, the particle number density is a
functional of the configuration involving the configuration
gradient.
\end{bem}
The definition is justified by the following observation: the
vector-field $nu^{\mu}$ satisfies a continuity equation:
\begin{equation}\label{cont1}
\nabla_{\mu}(nu^{\mu})=0
\end{equation}
This can be verified using (\ref{ndef1}):
\begin{equation}
\nabla_{\mu}( \epsilon^{\mu\nu\lambda\rho}v_{\nu\lambda\rho} )=(\nabla_{\mu}\epsilon^{\mu\nu\lambda\rho})v_{\nu\lambda\rho} + \epsilon^{\mu\nu\lambda\rho}(\nabla_\mu v_{\nu\lambda\rho} )
\end{equation}
The first term vanishes, since the volume form on space-time is
covariant constant. The second is zero, since $V$ is closed and
whence also $v$ is closed.

Next we define the strain. Similar to a metric (as the name
suggests) it is a measure for the strain arising in the material as
a consequence of configuration.
\begin{definition}
The strain $h^{AB}$ corresponding to the configuration is defined as
\begin{equation}\label{sstrain}
h^{AB}:=\partial_{\mu}f^{A} g^{\mu\nu} \partial_{\nu}f^{B}.
\end{equation}
\end{definition}
Note that $h^{AB}$ does not in general give rise to a rank two
tensor on $\mathcal{B}$, since it depends on $x^{\mu}$, the
coordinates on space-time.
\begin{lemma}\label{bornlemma1}
A necessary and sufficient condition for the strain to define a
tensor on $\mathcal{B}$ is that the four-velocity $u^{\mu}$ is
Born-rigid, i.e. that the Lie-derivative of metric projection
orthogonal to $u^{\mu}$ in the $u^{\mu}$ direction vanishes:
\begin{equation}\label{borncond}
\mathcal{L}_{u}(g_{\mu\nu}+u_{\mu}u_{\nu})=0
\end{equation}
\end{lemma}
The strain as defined in (\ref{sstrain}) is no tensor, since $f$ is
no diffeomorphism. But if the strain is constant along the flow of
the four-velocity, then the strain is uniquely defined on the
quotient (of space-time along the four-velocity). But the latter
clearly is diffeomorphic to the body if the four-velocity $u^\mu$ is
geodesic (one just has to think in terms of initial data). This is
the case for solutions of the field equations as we will see. Thus
the strain can be viewed as a tensor field. If on the other hand the
strain really comes from a tensor field on $\mathcal{B}$, then it
must be constant along the flow of $u$. Both arguments come down to
\begin{equation}
\mathcal{L}_{u}h^{AB}=0
\end{equation}
Since the Lie-derivative of the configuration gradient (viewed as a
collection of one-forms on space-time) vanishes according to
\begin{equation}
\mathcal{L}_{u}\partial_\mu f^A=\partial_\mu (u^\nu \partial_\nu f^A)+ u^\nu \partial_{[\mu}\partial_{\nu]} f^A=0
\end{equation}
we find that this is equivalent to the statement that the
Lie-derivative of the inverse metric lies in the kernel of the
configuration gradient:
\begin{equation}
0=\mathcal{L}_{u}h^{AB}=f^A,_\mu f^B,_\nu \mathcal{L}_{u} g^{\mu\nu}
\end{equation}
Consequently there exists a scalar $A$ and a vector $v$ such that
\begin{equation}
\mathcal{L}_{u}g^{\mu\nu}= - A u^\mu u^\nu - 2 v^{(\mu}u^{\nu)}
\end{equation}
where without restriction of generality we can assume $v$ to be
orthogonal to $u$. (Otherwise just split $v$ into a parallel and an
orthogonal part and re-define $A$ and $v$ in the obvious way.) We
can lower the indices:
\begin{equation}\label{borncond1}
\mathcal{L}_{u}g_{\mu\nu}= A u_\mu u_\nu + 2 v_{(\mu}u_{\nu)}
\end{equation}
First note that $A=0$ according to
\begin{equation}
 A= u^\mu u^\nu \mathcal{L}_{u}g_{\mu\nu}= \mathcal{L}_{u}(g_{\mu\nu}u^\mu u^\nu)-2u_\mu \mathcal{L}_{u}u^\mu=0
\end{equation}
since $u^2=-1$ and $\mathcal{L}_{u}u=0$ for every vector field $u$.
To compute $v$ we contract (\ref{borncond1}) with $u$:
\begin{equation}
-v_\mu=u^\nu \mathcal{L}_{u}g_{\mu\nu}= \mathcal{L}_{u}(u_\mu)
\end{equation}
Thus we find that
\begin{equation}
\mathcal{L}_{u}g_{\mu\nu}=- 2 \mathcal{L}_{u}(u_{(\mu})  u_{\nu)}
\end{equation}
which is equivalent to the statement of the lemma.
\begin{bem}
Note that the Born-rigidity condition constitutes a restriction on
the space-time geometry, since the existence of a Born-rigid vector
field is not trivially given. An example, where a Born rigid
four-velocity may exist, is that of a space-time admitting a
time-like Killing vector field. The existence of a homothetic vector
field however turns out to be too weak.
\end{bem}
To see the last point of the remark, we insert
\begin{equation}
\mathcal{L}_{u}g_{\mu\nu}=Cg_{\mu\nu}
\end{equation}
into (\ref{borncond}). For nonzero $C$ we obtain
\begin{equation}
 g_{\mu\nu}=-2u_{(\mu}u_{\nu)}
\end{equation}
which contradicts the non-degeneracy of the metric. Note that
Born-rigidity does in turn not imply homotheticity, since
contracting (\ref{borncond}) twice with $f^A,_\mu$ leads to
\begin{equation}
f^{A\mu}f^{B\nu}\mathcal{L}_{u}g_{\mu\nu} =0
\end{equation}
independent of the values of $A$ and $B$. Thus $\mathcal{L}_{u}g\ne C g$ for non-zero $C$, since this would again lead to a contradiction. \\

Leaving behind the question of functional dependence and Born
rigidity we note some further observations regarding the strain:
\begin{itemize}
\item the strain is positive definite
\item therefore there exists an inverse, denoted by $h_{AB}$.
\item since the four-velocity is annihilated by the configuration gradient, we have that $h^{AB}=f^{A},_{\mu}f^{B},_{\nu}(g^{\mu\nu}+su^{\mu}u^{\nu} )$ for arbitrary real $s$.
\end{itemize}
The first statement comes from the fact, that the configuration
gradient in the definition of the strain projects out the (time
like) $u^{\mu}$-direction, leaving behind the positive definite
rest. The other points follow immediately.

We can use the strain to rewrite the particle number density as the
positive root of the determinant of the strain, taken w.r.t. the
volume form on the body:
\begin{equation}\label{ndeth}
6 n^2=\det_{\mathcal{B}}(h^{AB})=V_{ABC}V_{A'B'C'}h^{AA'}h^{BB'}h^{CC'}
\end{equation}
This can be seen by contracting (\ref{ndef1}) with itself and using
the standard formulas (see f.e. \cite{urbi2} p.83) to evaluate the
$\epsilon$-tensors.

\subsection{The space-time Lagrangian}

Our field equations will be the Euler-Lagrange equations associated
with an action principle connected with the action
\begin{equation}\label{staction0}
S_{\mbox{s}}[f]:=\int_{\mathcal{M}}\rho (f,\partial f ; g) \sqrt{-\det g}\mbox{d}^4 x
\end{equation}
The Lagrangian density $\rho$ is assumed to be a non-negative
function of the configuration, of the configuration gradient and of
the metric. An explicit $x^{\mu}$ dependence only enters via the
space-time metric. We subject $\rho$ to the following assumptions:
\begin{ann}
The Lagrangian is covariant under space-time diffeomorphisms, i.e.
it transforms as a scalar for given configurations.
\end{ann}
\begin{bem}
In non-relativistic elasticity this condition is called ``material
frame indifference" (see e.g. \cite{gur1} or \cite{gur2}).
\end{bem}

An equivalent formulation (see f.e. \cite{soper1}) of this
assumption is
\begin{equation}
\mathcal{L}_{\xi}\rho=\frac{\partial \rho}{\partial f^{A}}\mathcal{L}_{\xi}f^{A}+\frac{\partial \rho}{\partial f^{A},_{\mu}}\mathcal{L}_{\xi}f^{A},_{\mu}+\frac{\partial \rho}{\partial g^{\mu\nu}}\mathcal{L}_{\xi}g^{\mu\nu}
\end{equation}
for arbitrary vector fields $\xi^{\mu}\in T\mathcal{M}$. If we write
the Lie-derivatives in terms of partial derivatives we obtain
\begin{equation}\label{6754}
f^{A},_{\mu}\frac{\partial \rho }{\partial f^{A},_{\nu}}=2 g^{\nu\lambda} \frac{\partial \rho }{\partial g^{\mu\lambda}}
\end{equation}
By contraction with $u^{\mu}$ this implies
\begin{equation}\label{34567}
\frac{\partial \rho }{\partial g^{\mu\lambda}}u^{\lambda}=0
\end{equation}
This by the symmetry of the expression in turn implies
\begin{equation}
\frac{\partial \rho }{\partial
g^{\mu\nu}}=A_{AB}f^A,_{\mu}f^B,_{\nu}
\end{equation}
for some $A_{AB}$. From this the following is clear:
\begin{equation}
\rho (f^{A},f^{A},_{\mu};g^{\mu\nu})=\rho (f^{A},f^{A},_{\mu};g^{\mu\nu}+s u^{\mu}u^{\nu})
\end{equation}
for arbitrary real $s$. Let $s=1$. We claim that
\begin{equation}
g_{\mu\nu}+u_{\mu}u_{\nu}=f^{A},_{\mu}f^{B},_{\nu}h_{AB}
\end{equation}
The objects on both sides are symmetric. We verify the claim by
contracting with $u^{\mu}$ and $f^{C},_{\alpha}g^{\alpha\mu}$
respectively. For both cases it is easy to verify the claim. Since
those form an orthonormal basis of $T\mathcal{M}$, the claim is
proved.

The above formula then tells us that the Lagrangian can be rewritten
as a function of the configuration, the configuration gradient and
the strain:
\begin{equation}
\rho(f,\partial f ; g)=\sigma (f,\partial f ; h)
\end{equation}
Plugging this into equation (\ref{6754}) we see that
\begin{equation}
\frac{\partial \sigma}{\partial f^A,_\mu}f^A,_\nu=0
\end{equation}
Since the configuration gradient is assumed to be of maximal rank,
the first term has to vanish. Thus $\sigma$ is in fact explicitly
independent of the configuration gradient (Of course it still enters
implicitly via the strain). Keeping this in mind, we spit the
Lagrangian $\rho$ into
\begin{equation}
\rho=n \epsilon
\end{equation}
where by definition $\epsilon$ is a positive function depending
solely on the configuration and the strain
\begin{equation}\label{stinte1}
\epsilon=\epsilon (f, h)
\end{equation}
This factorization is justified, since the particle number density
$n$ can be written as the determinant of the strain (see
(\ref{ndeth})), thus involves only the strain and (via the volume
form on the body) the configuration. The action (\ref{staction0})
takes its final form
 \begin{equation}\label{staction1}
S_{\mbox{s}}[f]:=\int_{\mathcal{M}} n\epsilon(f, h) \sqrt{-\det g}\mbox{d}^4 x
\end{equation}
where the Lagrangian
\begin{equation}\label{stlagrangian1}
\mathcal{L}_{\mbox{s}}=n\epsilon(f, h) \sqrt{-\det g}
\end{equation}
depends solely on points of the base space (via the metric), on the
strain and on the configuration.
\begin{definition}
We call $\epsilon$ stored energy function or simply stored energy.
\end{definition}
The stored energy function contains all the information about the
material under consideration. Knowing it is equivalent to knowing
the equation of state.

We can derive some important quantities from the stored energy.
First of all it provides us with an intrinsic definition of stress
(depending only on the material):
\begin{definition}
The second Piola-Kirchhoff stress tensor is defined as the first variation of the stored energy w.r.t. the strain:
\begin{equation}\label{piola1}
\tau_{AB}:=\frac{\partial \epsilon}{\partial h^{AB}}
\end{equation}
\end{definition}
As was the case for the strain, the Piola-Kirchhoff stress is in
general no tensor on the body, since it depends on the space-time
coordinates $x^{\mu}$. We have the obvious symmetry property
\begin{equation}
\tau_{AB}=\tau_{BA}
\end{equation}
We will mainly use this second Piola-Kirchhoff stress tensor and
whenever we talk about the Piola stress we mean exactly this one
unless said otherwise. For completeness we stress that one can
define a first Piola-Kirchhoff tensor too. Let us denote it by
$\sigma_{\mbox{\small s}}$. In the space-time description it is
defined by
\begin{equation}
\sigma^\mu_{\mbox{\small s}A}:=\frac{\partial\epsilon }{\partial f^A,_\mu}
\end{equation}
One can check that the first and second Piola-Kirchhoff stress tensors are connected via
\begin{equation}
\sigma^\mu_{\mbox{\small s}A}=\tau_{CD}\frac{\partial h^{CD}}{\partial f^A,_\mu}=2\tau_{CD}\delta^{(C}_Af^{D)}_\nu g^{\mu\nu}
\end{equation}
Note that $\sigma^{\mu}_{\mbox{\small s}A} u_\mu =0$.

\subsection{Elasticity operator}\label{elasticity operator2222}

The second variation will turn out to be the most important quantity
for questions of local well-posedness in the neighborhood of
stress-free deformations:
\begin{definition}
The second variation of the stored energy w.r.t. the strain is
called elasticity operator. It is defined as
\begin{equation}\label{elastt1}
U_{ABCD}:=\frac{\partial ^2 \epsilon}{\partial h^{AB}\partial h^{CD}}
\end{equation}
\end{definition}
Again this does not in general define a tensor field on the body.
From the definition the following symmetries are evident:
\begin{equation}
U_{ABCD}=U_{BACD}=U_{ABDC}=U_{CDAB}
\end{equation}
\begin{bem}
Note that the second Piola stress and the elasticity operator are
introduced without referring directly to the configuration. This
fact will allow us to use the very same definitions in the material
setup.
\end{bem}
Before giving an important example for the elasticity tensor, we
give a short definition concerning symmetries of the material.
\begin{definition}
A material is called homogeneous, if the stored energy $\epsilon$
depends solely on the strain and does no more include the
configuration itself. This is equivalent to
\begin{equation}
\epsilon=\epsilon(h)
\end{equation}
We also require the volume form $V$ to be independent of $X$.
\end{definition}
Assume there is given a metric $G_{AB}$ on $\mathcal{B}$ (f.e.
coming from the strain of a special Born-rigid reference
configuration). Further assume that the stored energy is covariant
under diffeomorphisms of $\mathcal{B}$. Then it depends solely on
the invariants of the strain taken w.r.t. $G$. We cast this into the
following definition:
\begin{definition}
We call a material isotropic if the stored energy $\epsilon$ depends
on the configuration gradient only via its invariants, i.e.
\begin{equation}
\epsilon=\epsilon(f, \tr h ,\tr h^2, \det h)
\end{equation}
\end{definition}
An important special case of such materials are the so called
St.Vernard-Kirchhoff materials (see f.e. \cite{cia2}). They are
homogenous and isotropic and their elasticity tensor is given by
(remember that $G$ denotes some metric on $\mathcal{B}$)
\begin{equation}\label{stvk1}
U_{ABCD}=2\mu G_{A(C}G_{D)B}+\lambda G_{AB}G_{CD}
\end{equation}
where $\mu$ and $\lambda$ are constants.

One can formally develop the stored energy for an isotropic (not
necessarily homogeneous) material into a Taylor series around the
metric $G$ and find that the quadratic portion can always be written
in the form ($G^{AB}$ is the inverse metric)
\begin{equation}
\left( 2\mu  G_{A(C}G_{D)B}+\lambda G_{AB}G_{CD} \right) (h^{AB}-G^{AB})(h^{CD}-G^{CD})
\end{equation}
for $\mu$ and $\lambda$ being functions on $\mathcal{B}$. If the material is homogeneous, then they are constant.
\begin{definition}
The constants $\mu$ and $\lambda$ appearing in the quadratic portion
of the Taylor series of the stored energy for a homogenous isotropic
material are called the Lam\'e constants.
\end{definition}
This in particular means that the constants appearing in
(\ref{stvk1}) are the Lam\'e constants. For the special case of
St.Vernant-Kirchhoff materials they completely determine the
material.

To get used to the notation we apply the three positivity notions
introduced in section \ref{quasisec} to the St.Vernard-Kirchhoff
elasticity operator: What one obtains are conditions on the Lam\'e
constants:

For the Legendre-Hadamard condition we have to contract (\ref{stvk1}) with $a^Ab^Ba^Cb^D$. Then the condition reads
\begin{equation}
(\mu+\lambda )(a\cdot b)^2+\mu a^2 b^2 \ge \epsilon a^2 b^2
\end{equation}
By splitting $b$ in a part parallel and a part orthogonal to $a$, we
find that the above condition holds for all choices of $a$ and $b$,
if the Lam\'e constants satisfy the inequalities
\begin{equation}\label{spassimbuero}
\mu >0 \ \ \ \ \ \ 2\mu+\lambda >0
\end{equation}
These combinations, as we will see in section \ref{steom11111}, are
crucial entities for the hyperbolicity of the corresponding
equations of motion. We will also see, that these conditions
(\ref{spassimbuero}) are equivalent to the statement that plane
waves in the linearized theory propagate at real (and not imaginary)
speed.

Next we do this computation for strong point-wise stability
(\ref{sps1}): The inequality, we have to satisfy for arbitrary
$m^{AB}$ is (recall that all finite dimensional norms are
equivalent. See f.e. \cite{bbn1})
\begin{equation}
 ( 2\mu  G_{A(C}G_{D)B}+\lambda G_{AB}G_{CD}) m^{AB}m^{CD} \ge 2 \epsilon  G_{A(C}G_{D)B}  m^{AB}m^{CD}
\end{equation}
Evaluating both sides leads to
\begin{eqnarray}
( \mu -\epsilon )n^{(AB)}n_{(AB)}\ge 0  \\
(2\mu+3\lambda-\epsilon) m^A_A \ge 0
\end{eqnarray}
where $n$ denotes the trace-free portion of $m$. The above
conditions are satisfied for all possible choices of $m^{AB}$ if
\begin{equation}
\mu >0 \ \ \ \ \ \ 2\mu+3 \lambda >0
\end{equation}
We see that there is a tighter restriction on the possible values of $\lambda$.

Finally we come to rank-two-positivity. The corresponding inequality
reads
\begin{equation}
 ( 2\mu  G_{A(C}G_{D)B}+\lambda G_{AB}G_{CD}) m^{AB}m^{CD} \ge \epsilon  G_{AC}G_{DB}  m^{AB}m^{CD}
\end{equation}
At the first glance it may look similar to the case of
strong-point-wise-stability, but this is not true, since on the
right hand side antisymmetric terms do appear. But sine on the left
hand side only symmetric terms are involved, the antisymmetric terms
on the right hand side cannot be bounded. Thus there is no possible
way to satisfy rank-two-positivity of the elasticity operator for
St.Venant-Kirchhoff materials (This holds for general isotropic
materials. For this check the following remarks.) Note however that
it makes sense to apply the notion of rank-two positivity to certain
combinations involving the elasticity operator.
\begin{bem}
We stress that although we did the above example for the case of
St.Vernant-Kirchhoff materials, the analogous treatment works for
arbitrary isotropic materials.
\end{bem}
\begin{bem}
Note that $G$ entered in the conditions. It thus may well happen that the conditions are met for a certain choice of $G$ and cease to hold for a different choice. We have also seen, that some positivity conditions may just fail for certain types of materials.
\end{bem}

\subsection{Energy momentum}\label{sectem1}

Next we derive the energy-momentum tensor for this model. We can
derive it from the Lagrangian according to the formula (see f.e.
\cite{soper1} p.42 )
\begin{equation}\label{origem1}
T^{\mu}_{\nu}:=\frac{\partial \rho}{\partial
f^A,_{\mu}}f^A,_\nu-\rho \delta^{\mu}_{\nu}
\end{equation}
Recall that the Lagrangian $\rho$ did only depend on the
configuration and the strain. This allows us to write (recall the
definition of the strain (\ref{sstrain}))
\begin{equation}
\frac{\partial \rho}{\partial f^A,_{\mu}}f^A,_\nu=\frac{\partial \rho }{\partial h^{CD}}\frac{\partial h^{CD}}{\partial f^A,_\mu}f^A,_\nu=2\frac{\partial \rho }{\partial h^{CD}}f^C,_{\nu}f^D,_\alpha g^{\alpha\mu}
\end{equation}
On the other hand
\begin{equation}
\frac{\partial \rho}{\partial g^{\mu\nu}}=\frac{\partial \rho }{\partial h^{CD}}\frac{\partial h^{CD}}{\partial g^{\mu\nu}}=\frac{\partial \rho }{\partial h^{CD}}f^C,_\mu f^D,_\nu
\end{equation}
Beside a factor $2$ this is just the same term. Thus we can rewrite the energy-momentum tensor as
\begin{equation}\label{bla111}
T_{\mu\nu}=2\frac{\partial\rho}{\partial g^{\mu\nu}}-\rho g_{\mu\nu}
\end{equation}
Note that this expression is manifestly symmetric. Contracting with
the four-velocity and using (\ref{34567}) we find that
\begin{equation}
T_{\mu\nu}u^\nu=-\rho u_\mu
\end{equation}
Thus the energy-momentum tensor must be of the form
\begin{equation}
T_{\mu\nu}=\rho u_\mu u_\nu +S_{\mu\nu}
\end{equation}
where $S$ is symmetric and orthogonal to $u$, i.e. $S_{\mu\nu}u^\nu
=0$. ($S$ is sometimes called Cauchy-stress.) This in turn implies
that there exists an object $A_{AB}$ such that $S_{\mu\nu}=f^A,_\mu
f^B,_\nu A_{AB}$.
\begin{bem}
Note that this representation of the energy-momentum tensor allows a
good interpretation for the $\rho$ entering the Lagrangian. It plays
the role of the total energy density of the material in its rest
system.
\end{bem}
To compute the actual form of $A$, we have to use $\rho=n\epsilon$ to write out the right hand side in (\ref{bla111}):
\begin{equation}
T_{\mu\nu}= 2 \left( n\tau_{AB}+\frac{\partial n}{\partial h^{AB}}\epsilon \right) f^A,_\mu f^B,_\nu -\rho g_{\mu\nu}
\end{equation}
To compute the variation of the particle number density, we use a
well known result that connects the variation of the determinant
with the inverse tensor, namely (note that the strain is symmetric
and that $h_{AB}$ is the inverse strain as defined in section
\ref{mausi})
\begin{equation}\label{varnh}
\frac{\partial n}{\partial h^{AB}}=\frac{1}{2}nh_{AB}
\end{equation}
The expression entering the energy momentum tensor is thus given by
the symmetric tensor $h_{\mu\nu}:=h_{AB}f^A,_\mu f^B,_\nu$. A closer
look shows that
\begin{equation}
h_{\mu\nu}u^\mu=0
\end{equation}
as well as
\begin{equation}
h_{\mu\nu}f^{M \nu} =h_{AB}f^A,_\mu h^{BM}=f^{M},_{\mu}
\end{equation}
While the four-velocity is annihilated, $h_{\mu\nu}$ acts as
identity on the configuration gradient and hence also on the
orthogonal space. We conclude the formula
\begin{equation}
h_{\mu\nu}=g_{\mu\nu}+u_\mu u_\nu
\end{equation}
Insertion of this into the expression for the energy-momentum tensor yields
\begin{equation}\label{stemt111}
T_{\mu\nu}= \rho u_\mu u_\nu +2n\tau_{AB}f^A,_\mu f^B,_\nu
\end{equation}
This justifies to call the second Piola-Kirchhoff tensor $\tau_{AB}$
stress, since it is equivalent to the Cauchy stress. Note that there
are no currents involved in the energy-momentum tensor.

\begin{bem}
The energy-momentum tensor is divergence-free, if the configuration
is stress-free and the corresponding four-velocity is geodesic.
\end{bem}
This can be shown by direct computation:
\begin{equation}\label{divergencedesemtensors}
\nabla_\nu T^{\mu\nu}=\nabla_\nu (n  u^\nu) \epsilon u^\mu +n u^\nu\nabla_\nu (\epsilon u^\mu ) +2g^{\mu\mu'}\nabla^\nu (n\tau_{AB}f^A,_{\mu'} f^B,_\nu)
\end{equation}
The last term is zero, since the configuration is assumed to be
stress-free. The first term vanishes due to the continuity equation
(\ref{cont1}). The second term can be split into three parts:
\begin{equation}
n u^\nu \left(  \epsilon \nabla_\nu u^\mu + u^\mu\frac{\partial  \epsilon }{\partial f^A}f^A,_\nu     +   u^\mu\tau_{AB} h^{AB},\nu     \right)
\end{equation}
All these terms vanish. The last because the configuration is
assumed stress-free, the second because the four-velocity is
orthogonal to the configuration gradient and the first because the
four-velocity is geodesic.

\begin{bem}
From the proof of the last remark we can extract another interesting
observation: The divergence of the energy-momentum tensor (this time
for a general configuration) is annihilated by the material
velocity. Thus $\nabla_\nu T^{\mu\nu}=0$ are practically only three
equations, which are equivalent to $f^A,_\mu  \nabla_\nu
T^{\mu\nu}=0$
\end{bem}
This can be seen by contracting (\ref{divergencedesemtensors}) with
the four-velocity and then evaluating the resulting expression. We
will however proceed along a shorter route. Since by the Leibnitz
rule
\begin{equation}
u_\mu \nabla_\nu T^{\mu\nu}=\nabla_\nu (u_\mu T^{\mu\nu}) - T^{\mu\nu}\nabla_\nu u_\mu
\end{equation}
we find that (using that the four-velocity annihilates the
configuration gradient and is normalized)
\begin{equation}
u_\mu \nabla_\nu T^{\mu\nu}=-\nabla_\nu (\rho u^\nu )  2 n \tau_{AB}f^A,_{\mu} f^B,_\nu\nabla^\nu u^\mu
\end{equation}
We can split the total energy density according to $\rho=n\epsilon$. Then the first term becomes
\begin{equation}
-\nabla_\nu (\rho u^\nu )=-\epsilon \nabla_\nu (n u^\nu ) - n u^\nu \nabla_\nu \epsilon
\end{equation}
The first term vanished due to the continuity equation which is
identically satisfied in our theory. The second term gives
\begin{equation}
- n u^\nu \nabla_\nu \epsilon= - n u^\nu \frac{\partial \epsilon}{\partial f^A}f^A,_\nu - nu^\nu \tau_{AB} \nabla_\nu h^{AB}
\end{equation}
 The first is zero because the four-velocity annihilates the configuration gradient, the second term prevails. We finally obtain
\begin{equation}\label{ucompdesemtensors}
u_\mu \nabla_\nu T^{\mu\nu}=- nu^\nu \tau_{AB} \nabla_\nu h^{AB}-2 n \tau_{AB}f^A,_{\mu} f^B,_\nu\nabla^\nu u^\mu
\end{equation}
This is altogether zero. To see this we first substitute
\begin{equation}
u^\nu \nabla_\nu h^{AB}=2g^{\alpha\beta} u^\nu f^{(A},_\alpha\nabla_\nu f^{B)},_\beta
\end{equation}
Since the configuration is a collection of scalars, we can use that
the covariant derivative is torsion-free. This leads to
\begin{equation}
u^\nu \nabla_\nu h^{AB}=2g^{\alpha\beta} u^\nu f^{(A},_\alpha \nabla_\beta f^{B)},_\nu=-2f^{(A},_\alpha f^{B)},_\nu \nabla_\alpha u^\nu
\end{equation}
Thus the two terms in (\ref{ucompdesemtensors}) cancel and the statement of the remark is shown to be true.

\subsection{Reference state}\label{radenska1}

In the definition of isotropy we needed to assume the existence of a
metric $G$ on the body. We now assume that $G$ is generated by some
special reference configuration:
\begin{definition}
We call a configuration $\tilde f^A$ reference configuration
whenever it is stress-free, i.e.
\begin{equation}
\tau_{AB}|_{f=\tilde f}=0
\end{equation}
and the corresponding four-velocity is geodesic. Quantities
evaluated at this reference configuration will be denoted by an
upper tilde (like e.g. $\tilde f$). Relations, which hold only for
this reference configuration will be denoted by an $o$ above the
relation (f.e. $\stackrel{o}{=}$). We sometimes refer to the
reference configuration more generally as the reference state of the
material (this notation will be used in the material setup too).
\end{definition}
\begin{bem}
There is another possible way of defining the reference
configuration: one may assume it to be stress-free and in addition
to generate a divergence-free energy-momentum tensor. Then the
four-velocity is geodesic if the reference deformation solves the
field equations. This can be seen from the following arguments. We
shall however stick to our original definition.
\end{bem}
By the last remark of the foregoing section we know that the
energy-momentum tensor for the reference state is divergence-free.
We will see in section \ref{steom11111} that this is equivalent to
the statement that the reference configuration solves the field
equations (which will be the Euler-Lagrange equations corresponding
to the Lagrangian (\ref{stlagrangian1}).). But we would like to show
that also slight perturbations of this reference state give rise to
solutions. This will be shown to be the case (at least for the
unbounded setup), if the reference configuration satisfies a certain
minimality property, namely that the elasticity operator satisfies
the Legendre-Hadamard condition:
\begin{definition}
We call a reference configuration natural, whenever the elasticity
tensor for this reference state satisfies the Legendre-Hadamard
condition (\ref{LH1}). In formulas this reads
\begin{equation}
\tilde U_{ABCD}a^Aa^Cb^Bb^D \ge \epsilon |a|^2 |b|^2
\end{equation}
for all $a$ and $b$ and for a positive $\epsilon$.
\end{definition}
For most of the following work we will assume the existence of such
a natural configuration.
\begin{bem}
These definitions did not claim Born-rigidity for the reference
state's four-velocity. Thus the reference strain need not define a
tensor field on $\mathcal{B}$. This is a difference between the
formulation given here and the standard (non-relativistic)
formulations of elasticity, where a (flat) metric on the body is
assumed throughout.
\end{bem}
We have seen in section \ref{mausi} that the existence of a
Born-rigid vector field constitutes a restriction on the space-time
geometry. We will cast this statement into a more concrete form in
the following lemma:
\begin{lemma}
If the reference strain defines a tensor field on the body, then
space-time has to be stationary.
\end{lemma}
We have shown in lemma \ref{bornlemma1} that the strain defines a
tensor if and only if the Born rigidity condition holds. Thus the
problem is the following: we have to show that
\begin{equation}
\mathcal{L}_u g_{\mu\nu}=0
\end{equation}
for a four-velocity satisfying $u^2=-1$ as well as $u^\nu \nabla_\nu
u^\mu=0$ and the Born condition (\ref{borncond}). The latter implies
the Killing equation whenever
\begin{equation}
\mathcal{L}_u u_\mu =0
\end{equation}
This is equivalent to
\begin{equation}
0=2 u^\nu\nabla_{(\nu}u_{\mu)}=u^\nu\nabla_{\nu}u_{\mu}+ \frac{1}{2}\nabla_{\mu}u^2
\end{equation}
Both terms on the right hand side vanish due to the assumptions of
the lemma. This in turn implies the Killing equation to be valid for
the reference four-velocity. Consequently space-time has to admit a
time-like Killing vector field since otherwise one would have a
contradiction. Thus the metric has to be stationary at least.
\begin{bem}
This lemma implies that a non-stationary space-time can not admit a
Born-rigid reference four-velocity. In such a setup the reference
strain can never be a tensor on the body.
\end{bem}

We end this section with a slight adjustment of the original
definition of isotropy:
\begin{definition}
We call a material isotopic (w.r.t. the reference configuration)
whenever it is isotropic w.r.t. the reference strain.
\end{definition}
For convenience we slightly change notation in the following way: if
the material is isotropic, we redefine the Lam\'e moduli introduced
in (\ref{stvk1}) according to
\begin{equation}\label{isotropic elastopneu}
4 \tilde U_{ABCD}=\tilde \epsilon \left( 2\mu \tilde h_{A(C}\tilde
h_{D)B}+\lambda \tilde h_{AB}\tilde h_{CD} \right)
\end{equation}
This definition of isotropy will be used in the following.

\section{Material description}\label{eintelephonimwlad}
In this section we introduce the material description. The dependent
variables in this setup are time-dependent mappings from the body
onto space-like hyper-surfaces of space-time. Roughly speaking they
are the instantaneous inverse to the configurations. In this section
we will perform a $3+1$ decomposition of space-time, translate all
main objects into this new description and finally rewrite the
Lagrangian for this setup.

\subsection{Basic setup}

Assume there exists a foliation of space-time by space-like
hyper-surfaces. This means space-time is the product of a
three-dimensional space-like manifold $\Sigma$ with an interval
$I\subset \R$:
\begin{equation}
\mathcal{M}=I\times \Sigma
\end{equation}
Let the leafs of the foliation be the level sets of the function
$t$. Introducing coordinates $x^i$ ($i=1,2,3$) on $\Sigma$, we can
write the space-time metric (using $t=:x^0$ as fourth coordinate) as
\begin{equation}\label{ADMm1}
g_{\mu\nu} dx^\mu dx^\nu =-N^2dt^2+g_{ij}(dx^i +Y^i dt)(dx^j+Y^j dt)
\end{equation}
The inverse metric reads
\begin{equation}\label{invADMm1}
g^{\mu\nu}\partial_\mu \partial_\nu =-N^{-2}(\partial_t -Y^i\partial_i )^2 +g^{ij}\partial_i\partial_j
\end{equation}
All functions depend on $(t,x^i)$. Clearly we can write every metric
in the given form. This is the ADM-representation of the metric (see
\cite{mtw} for details). The function $N$ is assumed to be positive
and is called lapse while $Y^i$ is called shift. The geometrical
meaning of the quantities entering the ADM-representation is the
following: $g_{ij}(t,x)$ is the induced metric on the leaf
$\Sigma_t:= \{ t \}\times\Sigma$ of the foliation, $g^{ij}$ is its
inverse, i.e. $g^{ij}g_{jk}=\delta^i_k$. The norm of $dt$ is given
by $-N^{-2}$. Thus the unit co-normal is $n_{\mu}dx^\mu = \pm N dt$.
The (future-pointing) unit-normal is $n^\mu\partial_\mu =N^{-1}
(\partial_t -Y^i\partial_i )$.

\begin{bem}
One has to be careful about the notation: although $g_{ij}$ denote
the spatial components of $g_{\mu\nu}dx^\mu dx^\nu$, the analogous
statement fails to be true for the inverse metric. There the spatial
components involve both lapse and shift as well as $g^{ij}$.
\end{bem}
Given a configuration $f^A$. Since the corresponding material
velocity $u^\mu$ is assumed to be time-like, we have that $n_\mu
u^\mu \ne 0$. This means that $u^\mu$ is transversal to the
foliation. This implies that for each instant of time $t$, the
configuration is a diffeomorphism between the hyper-surface
$\Sigma_t$ and the body, since the configuration gradient is assumed
to be of maximal rank, thus generating a linear isomorphism between
$T\Sigma_t$ and $T\mathcal{B}$.

It thus makes sense to write down the inverse of the configuration
for every instant of time. We define
\begin{definition}
The deformation $F^i$ corresponding to the configuration $f^A$ is
defined by (remember that the coordinates on the body $\mathcal{B}$
were denoted by $X^A$)
\begin{equation}\label{conf1}
f^A(t,F^i(t, X))=X^A
\end{equation}
\end{definition}
We note that this definition involves the chosen foliation. We also
note that by the above statements the deformation is well-defined,
since we can solve (\ref{conf1}) for it by means of the implicit
function theorem. It is a collection of time dependent scalar maps
from the body onto the leafs of the foliation.
\begin{eqnarray}\nonumber
F: I \times \mathcal{B} &\to& \{ \cdot \} \times\Sigma  \subset \mathcal{M} \\
\nonumber (t,X^A) &\mapsto& F^{i}(t,X)
\end{eqnarray}
 \begin{bem}
One could easily construct a diffeomorphism between space-time and
$I\times \mathcal{B}$ by extending the configuration and the
deformation into four-dimensional mappings of the form $(t,f)$ and
$(t,F)$. We will however (mostly) stick to our original formulation.
\end{bem}
\begin{definition}
In accordance with the notion of configuration gradient, we call
$F^i,_A$ deformation gradient. Note that (in contrary to the
configuration gradient) the deformation gradient defined this way is
a linear isomorphism.
\end{definition}
From the definition of the deformation (\ref{conf1}) we can derive
the formal connection between the configuration gradient and the
deformation gradient: differentiating w.r.t. $X^B$ leads to
\begin{equation}\label{gradients1}
f^A,_i (t,F)F^i,_B=\delta^A_B
\end{equation}
Contracting with $F^j ,_{A}$ leads to
\begin{equation}
F^j,_A f^A,_i (t,F)F^i,_B=F^j ,_B
\end{equation}
which in turn gives
\begin{equation}\label{gradients2}
F^j ,_{A}f^A ,_{i}(t,F)=\delta^j _{i}
\end{equation}
We see that the configuration gradient can be obtained as the
inverse of the deformation gradient and vice versa.

Another consequence of the definition of $F^{i}$ is obtained by
differentiating (\ref{conf1}) with respect to $t$, namely (an upper
dot denotes differentiation w.r.t. $t$)
\begin{equation}\label{oo1}
\dot f^A (t,F(t,X)) +f^A ,_{i} (t,F(t,X)) \dot F^i (t,X)=0.
\end{equation}
This equation connects the time derivatives of the configuration and
the deformation.

\begin{bem}
From now on we will drop the arguments of the configuration
gradient. Whenever we write it down we understand it as the inverse
of the deformation gradient by means of (\ref{gradients1}),
(\ref{gradients2}) and (\ref{oo1}).
\end{bem}
For our purpose it turns out to be more convenient to introduce a
different normalization for the four-velocity which is in accordance
with the foliation. We introduce a new vector field $v^\mu$, which
is parallel to the four-velocity and satisfies $v^\mu (dt)_\mu =1$.
Thus
\begin{equation}
v^\mu\partial_\mu=\partial_t +v^i\partial_i
\end{equation}
Since the four-velocity is annihilated by the configuration
gradient, the same is true for $v$:
\begin{equation}
0=\dot f^A+v^i f^A,_i
\end{equation}
Subtracting this from equation (\ref{oo1}) we obtain
\begin{equation}
v^i(t,F(t,X)) f^A,_i(t,F(t,X))=f^A ,_{i} (t,F(t,X)) \dot F^i (t,X)
\end{equation} Since $f^A_i$ is non-degenerated, this implies
\begin{equation}
v^i(t,F(t,X)) =\dot F^i (t,X)
\end{equation}
Thus
\begin{equation}\label{v}
v^\mu\partial_\mu=\partial_t + \dot F^i (t,f(t,x))\partial_i
\end{equation}
\begin{definition}
From now on we use the term four-velocity for $v^\mu$ in the context
of the material setup. If we refer to $u^\mu$ we will speak of the
normalized four-velocity.
\end{definition}
The norm $g_{\mu\nu}v^\mu v^\nu$ of the four-velocity is given by
\begin{equation}
v^2=-N^2 +g_{ij}(Y^i+\dot F^i)(Y^j +\dot F^j)
\end{equation}
Introducing
\begin{equation}\label{W}
W^i:=\frac{1}{N}(\dot F^i+Y^i)
\end{equation}
we find that the four-velocity is time-like if and only if
\begin{equation}\label{W^2<1}
1-W^ig_{ij}W^j =:1-W^2>0
\end{equation}
\begin{ann}
Throughout the rest of this thesis we will assume that this
condition holds.
\end{ann}
This equation may be understood as a restriction on the possible
choices for the time flow. Roughly speaking lapse and shift have to
be chosen so that they balance the spatial evolution of the
material.

This quantity $W^i$ will appear throughout the following analysis.
We will also frequently encounter it in the following context:
\begin{equation}\label{gamma}
\gamma:=(1-W^2)^{-1}
\end{equation}
The condition (\ref{W^2<1}) is then equivalent to
\begin{equation}\label{gamma<infty}
1 \le \gamma<\infty
\end{equation}
We will find that this factor plays a similar role to the $\gamma$
factor of special relativity. (This can be seen clearly in the
Newtonian limit. We refer to the appendix for details.)

Next we deal with the strain. We can split the configuration
gradient into its spatial part and the rest. Then the strain becomes
\begin{equation}
h^{AB}=f^A,_if^B,_j (g^{ij} -\frac{1}{N^2}Y^iY^j )-
\frac{1}{N^2}\dot f^A\dot f^B + \frac{2}{N^2}f^{(A},_i\dot f^{B)}Y^i
\end{equation}
Recall that the first term inside the bracket is the spatial part of
the inverse metric (\ref{invADMm1}). Then we use (\ref{oo1}) to
replace the time derivatives of the configuration by those of the
deformation. One obtains
\begin{equation}\label{mstrain1}
h^{AB}=f^A,_i f^B,_j (g^{ij}-W^i W^j )
\end{equation}
If one understands the configuration gradient as being the inverse
of the deformation gradient, then this is an expression for the
strain only involving the deformation and the deformation gradient
besides the metric. We have thus successfully translated the strain
into the material description.
\begin{bem}
Abusing notation we keep the letter $h$ for the strain both in the
space-time and in the material description, although strictly
speaking those are different objects. The same remark applies to
most of the following cases. But since it is clear from the context,
which description is used, we stick to this simplified notation.
\end{bem}
The inverse strain is given by
\begin{equation}\label{minvstrain1}
h_{AB}=F^a,_AF^b,_Bg_{ia}g_{jb}(g^{ij}+\gamma W^iW^j )
\end{equation}
To prove this first note that $h_{AB}$ as defined is clearly
symmetric and positive definite, sice $\gamma$ is bounded from
below.

To show that it is indeed the inverse we contract with the strain:
\begin{equation}
h^{AB}h_{BC}=f^A,_a f^B,_b (g^{ab}-W^aW^b) F^c,_CF^l,_B g_{cj}g_{li}(g^{ij}+\gamma W^iW^j )
\end{equation}
Now we use (\ref{gradients1}) and (\ref{gradients2}) to evaluate the
contraction of the configuration gradient with the deformation
gradient. The result of this is
\begin{equation}
h^{AB}h_{BC}=\delta^A_C - f^A,_aW^a W^jg_{ij}F^j,_C(1-\gamma +\gamma W^2)
\end{equation}
The term in the bracket vanishes due to the definition of $\gamma$,
which proves that $h_{AB}$ is indeed the expression for the inverse
strain in the material setup.

\subsection{Material Lagrangian}

To translate the space-time action (\ref{staction1}) into the
material description, we have to find expressions for
\begin{itemize}
\item the particle number density (\ref{ndeth})
\item and the stored energy (\ref{stinte1}).
\item We also have to transform the volume element in the proper way.
\end{itemize}
We start with the stored energy. It is a scalar build from the
strain and the configuration, which themselves can be treated as a
collection of scalars. Thus for the material description we have
\begin{equation}\label{minte1}
\epsilon=\epsilon (X,h)
\end{equation}
It is an advantage of the material description that the dependence
of the stored energy becomes more clear: the energy stored in the
material depends on the strain acting at each point. Different
points of the (non-homogeneous) material react differently to
strain.
\begin{bem}
From this expression for the internal energy it is clear, that we
can simply use the definition given for the second Piola stress
(\ref{piola1}) and the elasticity operator (\ref{elastt1}). We just
have to replace the configuration $f^A$ by the image $X^A$.
\end{bem}
Only the first Piola tensor causes trouble, since it involves a
variation w.r.t. the configuration gradient. For the material setup
we introduce a new definition via the variation of the internal
energy w.r.t. the deformation gradient:
\begin{equation}
\sigma_{\mbox{\small m}i}^A :=\frac{\partial \epsilon}{\partial F^i,_A}
\end{equation}
But again this definition is given just for reasons of completeness.

Coming back to the original problem, namely that of translating the
Lagrangian, we next show how to deal with the particle density.
Recall that we could express it through the determinant of the
strain. For the strain we now have formula (\ref{mstrain1}). All we
need to do, is to compute the determinant of this expression w.r.t.
$V_{ABC}$, the volume form on the body.

First we introduce $k^{AB}$, the three-dimensional analogon of the
strain:
\begin{equation}\label{k1}
k^{AB}=f^A,_i f^B,_j g^{ij}
\end{equation}
This expression is exactly the non-relativistic version of the
strain (see f.e. \cite{gur1}). In this sense we can understand the
$W^i$ terms in (\ref{mstrain1}) as relativistic corrections to the
non-relativistic strain. The inverse of $k^{AB}$ is given by
\begin{equation}\label{invk1}
k^{-1}_{AB}=F^i,_A F^j,_Bg_{ij}
\end{equation}
which follows from basic linear algebra.

According to (\ref{ndeth}) the determinant of the strain (apart from
a factor $6$) is given by $h^{AA'}h^{BB'}h^{CC'}V_{ABC}V_{A'B'C'}$
Using (\ref{mstrain1}) we write this as
\begin{equation}\label{determinantedesstrains}
(g^{aa'}-W^aW^{a'})\dots f^A_a f^{A'}_{a'}\dots V_{ABC}V_{A'B'C'}
\end{equation}
Since $V_{ABC}$ is the volume form on the body, $f^A_af^B_bf^C_c
V_{ABC}$ is a volume form on the hyper-surface. Thus it must be
proportional to the three-dimensional metric-volume-form
$\epsilon_{ijk}$. We define a scalar $\kappa$ to be this
proportionality factor:
\begin{equation}\label{bodyspacevol}
f^A_af^B_bf^C_c V_{ABC}=:\kappa \epsilon_{abc}
\end{equation}
We will find according to (\ref{akppan}) that $\kappa$ is basically
given by the particle density. From the above equation we find that
(\ref{determinantedesstrains}) can be written as
\begin{equation}
(g^{aa'}-W^aW^{a'})(g^{bb'}-W^bW^{b'})(g^{cc'}-W^cW^{c'}) \kappa ^2  \epsilon_{abc} \epsilon_{a'b'c'}
\end{equation}
This (besides the factor $6 \kappa^2$) is the determinant of
$(g^{aa'}-W^aW^{a'})$. Determinants of such objects can be easily
computed using traces. The result is $(1-W^2)$. Thus the particle
density $n$ (which is the square root of the determinant of the
strain) becomes
\begin{equation}\label{akppan}
n=\kappa (1-W^2)^{1/2}=\kappa \gamma^{-\frac{1}{2}}
\end{equation}
This equation connects the particle density $n$ with the factor
$\kappa$.

 Finally we turn to the volume form. First we give its
expression in ADM variables, then we apply our transformation
(\ref{conf1}) on the spatial part. The result will be a volume form
on the body.

The space-time volume form can be written as
\begin{equation}
\sqrt{\det g_{\mu\nu}}d^4 x=N\sqrt{\det g_{ij}}dtd^3x
\end{equation}

Now (\ref{bodyspacevol}) implies that the volume-form on the
hyper-surfaces $\sqrt{\det g_{ij}}d^3x$ can be transformed into
$\kappa^{-1}V_{ABC}dX^AdX^BdX^C$, the volume form on $\mathcal{B}$
(modulo $\kappa$).

Collecting all pieces, we can write the action (\ref{staction1}) as
an integral over the material manifold times (an interval $I$ of)
$\R$:
\begin{equation}\label{maction1}
S_{\mbox{\small M}}[F]=\int_{I\times \mathcal{B} } N
\gamma^{-\frac{1}{2}}\epsilon V dt d^3X
\end{equation}
where $V$ is a scalar on the body defined by $V
d^3X:=V_{ABC}dX^AdX^BdX^C$. The corresponding Lagrangian is
obviously given by
\begin{equation}\label{mlagrangian1}
\mathcal{L}_{\mbox{\small M}}= N  \gamma^{-\frac{1}{2}}\epsilon V
\end{equation}
A final word on the functional dependence: $V$ is a scalar on the
body, $N$ depends on time $t$ and on the deformation. The stored
energy is a function on the strain bundle over the body, i.e. it
depends on the points of the body and on the strain. Finally we note
that $\gamma$ depends on time, on the deformation and on the
time-derivative of the deformation. This completes the listing of
the basic objects of the theory in the material description.

\subsection{Reference deformation}\label{refmat1}

This section gives a definition of reference deformation. We remain
as closely as possible to the definition used for the reference
state in the space-time description:
\begin{definition}
A deformation $\tilde F^i$ is called reference deformation, if it is
stress-free, i.e.
\begin{equation}
\tau_{AB}|_{F=\tilde F}=0
\end{equation}
and if the configuration corresponding to $\tilde F^i$ generates a
geodesic normalized four-velocity.

In analogy to the space-time description a reference deformation is
called natural, if the corresponding elasticity operator
(\ref{elastt1}) satisfies the Legendre-Hadamard condition
(\ref{LH1}).
\end{definition}
This is a rather pragmatic approach, since it involves
configurations in the definition. We have chosen it to keep the
definition simple and intuitive. The following lemma will give a
characterization of a reference state solely using deformations.
\begin{lemma}
A stress-free deformation $F^i$ is a reference deformation, if and
only if it satisfies the equation
\begin{equation}
\Gamma^{A}_{00}=0
\end{equation}
where $\Gamma^A_{00}$ are Christoffel symbols formed from the metric
(\ref{transformedmetric11}). The explicit form of this condition is
given by setting the right hand side of (\ref{noch hoffnungsloser})
equal to zero.
\end{lemma}

We prove the lemma by exploiting the relation between space-time and
$\R\times \mathcal{B}$: From what we have learned so far it is quite
clear that the mapping
\begin{eqnarray}
\mathcal{M}&\to&\R\times \mathcal{B}  \\
(t,x)&\mapsto&(t,f(t,x))
\end{eqnarray}
is a diffeomorphism with inverse
\begin{equation}
(t,X)\mapsto (t, F(t,X))
\end{equation}
We will apply this diffeomorphism to the geodesic equation for the
normalized four-velocity. The result will be the geodesic equation
for the transformed objects. This equation will be of a particular
simple form, so that the result can be readily read off.

For convenience we combine time $t$ and points $X^A$ on the body into a four-dimensional object
\begin{equation}
(X^\mu):=(X^0,X^A)=(t,X^A)
\end{equation}
this notation may seem misleading at the first sight, but it will
prove to be very practical. It will be clear from the context
whether Greek indices belong to space-time or $\R\times\mathcal{B}$.

Let the transformation of the normalized four-velocity be denoted by
$U^\mu$. Then by the standard transformation formula for vector
fields we have
\begin{equation}
U^{\mu}=\frac{\partial X^\mu}{\partial x^\nu}u^\nu
\end{equation}
(Recall the different meaning of $\mu$ and $\nu$ in this context).
For the normalized four-velocity this leads to the particular simple
expressions
\begin{eqnarray}
U^0(t,X)=u^0(t,F(t,X))=:u  \\
U^A=f^A,_\mu u^\mu =0
\end{eqnarray}
Then the geodesic equation takes the form
\begin{eqnarray}
0=u\partial_t u +\Gamma^0_{00}u^2 \\
0=\Gamma^A_{00}u^2
\end{eqnarray}
where the Christoffel symbols have to be taken w.r.t. the transformed metric
\begin{equation}\label{transformedmetric11}
G_{\mu\nu}=g_{\mu'\nu'}\frac{\partial x^{\mu'}}{\partial X^\mu}\frac{\partial x^{\nu'}}{\partial X^\nu}
\end{equation}
We obtain
\begin{equation}\label{mama}
(G_{\mu\nu}) =\left(\begin{array}{cc}-N^2(1- W^2) & N g_{ij} W^i F^j,_A \\ N g_{ij} W^i F^j,_B   & g_{ij}F^i,_A F^j,_B \end{array}\right)
\end{equation}
for the transformed metric where $W^2=g_{ij}W^iW^j$. The first thing
we can do is to determine the factor $u$. We find that $u^\mu u^\nu
g_{\mu\nu}=-1$ becomes
\begin{equation}
-1=G_{\mu\nu}U^\mu U^\nu =-u^2N^2(1 - W^2)
\end{equation}
From this we conclude that $u=N^{-1}\gamma^{\frac{1}{2}}$, so that we can write the geodesic equation as:
\begin{eqnarray}
0=N \gamma^{-\frac{1}{2}}\partial_t (N^{-1}\gamma^{\frac{1}{2}}) +\Gamma^0_{00} \\
0=\Gamma^A_{00}
\end{eqnarray}
Now we need the expressions for these Christoffel symbols. Note that
we can rewrite the transformed metric (\ref{mama}) in ADM form using
\begin{equation}
G_{AB}=g_{ij}F^i,_A F^j,_B
\end{equation}
as three-metric. The inverse is given by
\begin{equation}
G^{AB}=g^{ij}f^A,_i f^B,_j
\end{equation}
Then for the shift we take
\begin{equation}
Y^A=G^{AB}Y_B
\end{equation}
where
\begin{equation}
Y_B=N g_{ij} W^i F^j,_B
\end{equation}
The lapse finally is given by
\begin{equation}
X^2= Y^2 -G_{00}=N^2 (W^2+1-W^2)=N^2
\end{equation}
Note that the above calculation also implies
\begin{equation}
\gamma^{-1}=(1-\frac{Y^2}{N^2})
\end{equation}
Now we can apply the formulas for the Christoffel symbols in the
ADM-formalism from the appendix to rewrite the expressions entering
the equations as
\begin{eqnarray}\label{bloedsinnige rechnung}
\gamma\partial_t (N^2 \gamma^{-1}) -2N^2 \Gamma^0_{00}=W^2 \gamma  \partial_t (N^2\gamma^{-1}) -Y^A\left( (N^2\gamma^{-1}),_A +2 \dot Y_A \right) \\
\label{noch hoffnungsloser}
2N^2 \Gamma^A_{00}= - Y^A\partial_t (N^2 \gamma^{-1}) + \left[ N^2 G^{AB}-Y^AY^B\right] \left( (N^2 \gamma^{-1}),_B   +2 \dot Y_B \right)
\end{eqnarray}
From this we can show that it is sufficient to look at the second
equation, since it implies the first. This is seen by contraction of
(\ref{noch hoffnungsloser}) with $Y_A$:
\begin{equation}
2N^2\Gamma^A_{00}Y_A= -Y^2 \partial_t (N^2 \gamma^{-1}) + (N^2-Y^2)Y^B \left( (N^2 \gamma^{-1}),_B   +2 \dot Y_B \right)
\end{equation}
which is nothing but $N^2 \gamma^{-1}$ times the right hand side of
(\ref{bloedsinnige rechnung}). Thus (\ref{noch hoffnungsloser})
vanishes if and only if the normalized four-velocity is geodesic
which completes the proof on the lemma.

\chapter{Dynamics of Elastic Materials}\label{elvisdynamics}

This chapter deals with the dynamics of elastic materials. We will
derive the equations of motion governing the evolution of the
materials, which are the Euler-Lagrange equations arising from
variation of the actions given in the previous chapter.

The linearized equations (in the space-time picture; the same is
however true for the material picture, since the linearized
equations are the same in both descriptions) are shown to allow for
global solutions, which will be given in closed form. This will give
some insight into the propagation of elastic distortions. Then we
will show that the non-linear equations are hyperbolic in both
descriptions, i.e. that the Cauchy problem is well-posed.

After this we generalize this result in two directions: First the
self-interaction due to gravitational forces is taken into account
by analyzing the equations of elasticity coupled to the Einstein
equations. This will be done in the space-time description. On the
other hand we introduce boundaries, i.e. we specialize to materials
of finite extension. This will be done in the material picture,
since the boundary-conditions take a much simpler form in that
setup. For both cases (the self-gravitating as well as the
finite-volume) we show the Cauchy-problem to be well defined.

For all the well-posedness results in this chapter we employ the
theorems given in the chapter \ref{elvis1}.

\section{Equations of motion}\label{eav231}
This section is devoted to the derivation of the equations of
motion, which are the Euler-Lagrange equations arising in the
context of the Lagrangians (\ref{stlagrangian1}) and
(\ref{mlagrangian1}). Special attention will be given to the
principal part, since we know from chapter 2 that this part of the
equations is the relevant one for well-posedness.

Let $\mathcal{L}$ denote a Lagrangian depending on the field $\phi^A
(x^\mu)$ and its first order derivative. Then (under certain
regularity assumptions on the Lagrangian) $\phi^A$ is a critical
point of the action if and only if it satisfies the Euler-Lagrange
equations:
\begin{equation}\label{allgel1}
\frac{d}{dx^\mu}\frac{\partial \mathcal{L}}{\partial \phi^A,_\mu}=\frac{\partial \mathcal{L}}{\partial \phi^A}
\end{equation}
This is a quasi-linear system of second order for the unknowns
$\phi^A$. In the following we will derive these equations for both
the material and the space-time description.

\subsection{Space-time description}\label{steom11111}
We use the Lagrangian (\ref{stlagrangian1}) to obtain the equations
of motion for the space-time description. Recall the form of this
Lagrangian:
\begin{equation}
\mathcal{L}_{\mbox{s}}=n\epsilon(f, h) \sqrt{-\det g}
\end{equation}
Then the Euler-Lagrange equations can be written as
\begin{equation}
A^{\mu\nu}_{AB}f^B,_{\mu\nu}:=\frac{\partial^2 (n\epsilon)
}{\partial f^A,_\mu \partial f^B,_\nu}f^B,_{\mu\nu}=R_A
\end{equation}
where the lower order terms are given by
\begin{equation}
R_A:=\frac{\partial (n\epsilon) }{\partial f^A} - \frac{\partial
(n\epsilon)}{\partial f^A,_\mu f^B}f^B,_\mu -\frac{1}{\sqrt{-\det
g}}\partial_\mu \frac{\partial \mathcal{L}_{\mbox{s}}}{\partial
f^A,_\mu}
\end{equation}
Note that the differentiation in the last term on the right hand
side is a partial differentiation acting on those $x^\mu$ entering
explicitly (via the metric). They do not lead to second order
expressions.
\begin{bem} Note that we could as well use $A^{(\mu\nu
)}_{AB}$ instead of $A^{\mu\nu}$ in the principal part. But this
would lead to problems once boundary conditions are involved. (This
can be seen in section \ref{teeistgut}: there we require a certain
matching between principal part and boundary terms, which is
violated for the considered symmetry assumption) Therefore we stick
to the full principal part.
\end{bem}
It is interesting to note that the fact that the principal part
comes from a Lagrangian is reflected in the following symmetry of
the principal part:
\begin{equation}
A^{\mu\nu}_{AB}=A^{\nu\mu}_{BA}
\end{equation}
At this stage we pause to note a deep connection between the field
equations as given here and the energy-momentum tensor: In general
an alternative way to derive field equations is to claim
divergence-less-ness of the energy momentum tensor. We have seen in
section (\ref{sectem1}) that $u^\mu\nabla_\nu T^\nu_\mu =0$ is
identically satisfied due to the continuity equation. In the
following lemma we will see, that the vanishing of the remaining
components of the divergence is equivalent to the field equations
derived from the Lagrangian via variation.
\begin{lemma}\label{blabla345}
A configuration, which solves the Euler-Lagrange equation has a
divergence-free energy-momentum tensor and vice versa.
\end{lemma}
As mentioned before we only need to care about the components of the
divergence of the energy-momentum tensor orthogonal to the material
velocity. For this rest we first recall the definition of the
energy-momentum tensor from (\ref{origem1}):
\begin{equation}
T^{\mu}_{\nu}=\frac{\partial \rho}{\partial f^A,_{\mu}}f^A,_\nu-\rho \delta^{\mu}_{\nu}
\end{equation}
Then
\begin{equation}\label{eine unwichtige rechte seite}
\nabla_\mu T^{\mu}_{\nu}=T^{\mu}_{\nu},_\mu +\Gamma^\mu_{\mu\alpha}T^\alpha_\nu -\Gamma^\alpha_{\mu\nu}T^\mu_\alpha=T^{\mu}_{\nu},_\mu +\frac{1}{2}T^\alpha_\nu g^{\mu\beta}g_{\mu\beta},_\alpha-\frac{1}{2}T^{\mu\alpha}g_{\alpha\mu},_\nu
\end{equation}
where the last equality hods due to the symmetry of the
energy-momentum tensor. The first term on the right hand side can be
written as
\begin{equation}\label{schon sehr nah}
T^{\mu}_{\nu},_\mu =\left( \frac{\partial \rho }{\partial f^{A},_\mu}\right),_\mu f^A,_\nu -\frac{\partial\rho}{\partial f^A}f^A,_\nu-\partial_\nu \rho
\end{equation}
This is already very close to the desired result. As in section
\ref{sectem1} we can write
\begin{equation}
\partial_\nu \rho = \frac{\partial \rho }{\partial h^{AB}}f^A,_\alpha f^B,_\beta g^{\alpha\beta},_\nu
\end{equation}
since an explicit $x$-dependence only entered via the metric in the
strain. The other two terms on the right hand side of (\ref{eine
unwichtige rechte seite}) can be written as:
\begin{equation}\label{noch eine}
\frac{1}{2} \frac{\partial \rho}{\partial f^A,_\alpha}f^A,_\nu  g^{\mu\beta}g_{\mu\beta},_\alpha-\frac{1}{2}\frac{\partial \rho}{\partial f^A,_\mu}f^A,_\lambda g^{\lambda\alpha} g_{\alpha\mu},_\nu
\end{equation}
where the first term can be written as
\begin{equation}
 \frac{\partial \rho}{\partial f^A,_\alpha} \frac{(\sqrt{-\det g}),_\alpha}{\sqrt{-\det g}} f^A,_\nu
\end{equation}
This term together with the first two terms in (\ref{schon sehr
nah}) is the desired result. We have to show, that the second term
in (\ref{noch eine}) cancels the last term in (\ref{schon sehr
nah}). To see this recall from section \ref{sectem1} that
\begin{equation}
\frac{\partial \rho}{\partial f^A,_\mu}f^A,_\nu = 2\frac{\partial \rho}{\partial h^{AB}}f^A,_\alpha g^{\alpha \mu}f^B,_\nu
\end{equation}
Application of this to the two terms in question shows, that they
cancel. The remaining parts can be cast into the form
\begin{equation}
\nabla_\mu T^\mu_\nu= \frac{1}{\sqrt{-\det g}} \left[  \left( \sqrt{-\det g} \frac{\partial \rho }{\partial f^A,_\mu}\right),_\mu -\sqrt{-\det g}\frac{\partial \rho }{\partial f^A}          \right]f^A,_\nu
\end{equation}
This together with the continuity equation implies the lemma.

We need good information on the principal part in order to apply the
existence theorems from section \ref{quasisec}. Thus we will give a
more explicit form of $A^{\mu\nu}_{AB}$. In a first step we compute
\begin{equation}
\frac{\partial n\epsilon}{\partial f^A,_\mu}=\frac{n}{\partial f^A,_\mu}\epsilon +n \tau_{MN}\frac{\partial h^{MN}}{\partial f^A,_\mu}
\end{equation}
Thus
\begin{equation}\label{ststressedpp1}
A^{\mu\nu}_{AB}=\frac{\partial^2 n}{\partial f^A,_\mu \partial f^B,_\nu} \epsilon + n U_{MNKL}\frac{\partial h^{MN}}{\partial f^A,_\mu}\frac{\partial h^{KL}}{\partial f^B,_\nu} +\Phi^{\mu\nu}_{AB}
\end{equation}
There are basically three parts: The first involves the stored
energy, the second the elasticity tensor, and the last term is
linear homogeneous in the stress. More precisely the following holds
\begin{equation}
\Phi^{\mu\nu}_{AB}=\tau_{MN}\left( \frac{\partial n}{\partial f^A,_\mu}\frac{\partial h^{MN}}{\partial f^B,_\nu}+ \frac{\partial n}{\partial f^B,_\nu}\frac{\partial h^{MN}}{\partial f^A,_\mu} +n \frac{\partial^2 h^{MN}}{\partial f^A,_\mu \partial f^B,_\nu}   \right)
\end{equation}
For a stress-free state $\Phi^{\mu\nu}_{AB}|_{\tau=0}=0$ and the
principal part consists only of the first two terms in
(\ref{ststressedpp1}).

To obtain a higher level of detail, we have to compute the involving
partial derivatives. The easiest is the variation of the strain
w.r.t. the configuration gradient. It is straightforward to see that
\begin{equation}
\frac{\partial h^{MN}}{\partial f^A,_\mu}=2\delta^{(M}_A f^{N)},_\alpha g^{\mu\alpha}
\end{equation}
Then the second variation yields
\begin{equation}
\frac{\partial^2 h^{MN}}{\partial f^A,_\mu\partial f^B,_\nu}=2\delta^{(M}_A\delta^{N)}_Bg^{\mu\nu}
\end{equation}
The only missing ingredients are the variations of the particle
number density. We have seen in section \ref{sectem1} that
\begin{equation}
\frac{\partial n}{\partial h^{AB}}=\frac{1}{2}n h_{AB}
\end{equation}
with $h_{AB}$ being the inverse strain. Thus
\begin{equation}
\frac{\partial n}{\partial f^{A},_\mu}=n h_{AN} f^{N},_\alpha g^{\mu\alpha}
\end{equation}
From this expression we can compute the second variation
\begin{equation}\nonumber
\frac{\partial^2 n}{\partial f^{A},_\mu\partial f^B,_\nu}= n h_{BM} f^M,_\beta g^{\nu\beta} h_{AN} f^{N},_\alpha g^{\mu\alpha} + n \frac{\partial h_{AN}}{\partial f^B,_\nu} f^{N},_\alpha g^{\mu\alpha}+ n h_{AB}g^{\mu\nu}
\end{equation}
Since $h_{MN}$ is the inverse strain, we can use the formula
\begin{equation}
\frac{\partial  h_{AN}}{\partial f^B,_\nu}=-h_{AC}\frac{\partial h^{CD}}{\partial f^B,_\nu}h_{DN}
\end{equation}
This leads to
\begin{equation}
\frac{\partial  h_{AN}}{\partial f^B,_\nu}=-2 h_{B(A}h_{N)D}f^D,_\alpha g^{\alpha\nu}
\end{equation}
Inserting this into the formula for the second variation gives
\begin{equation}\label{secondn11}
2n h_{BM}h_{AN}f^M,_\alpha f^N,_\beta g^{\alpha [\nu}g^{\mu]\beta}+n h_{AB}\left(  g^{\mu\nu}-  h_{MN}f^N,_\alpha f^M,_\beta g^{\alpha\mu}g^{\beta \nu}  \right)
\end{equation}
This is where possible antisymmetric (in the upper index-pair)
portions of the principal part enter.

We can further simplify using two observations: first note that
\begin{equation}
\phi^\nu_B :=h_{BM}f^M,_\alpha g^{\alpha\nu}
\end{equation}
 is the inverse of the configuration gradient on the orthogonal space of the four-velocity. In formulas this means
\begin{equation}
\phi^\nu_B f^A,_\nu =\delta^A_B
\end{equation}
and
\begin{equation}
\phi^\nu_B f^B,_\mu =\delta_{\mu}^{\nu}+u_\mu u^\nu
\end{equation}
While the first equation is immediate from the definition of the
$\phi^\nu_B$, the second equality is shown by contracting
$\phi^\nu_B f^B,_\mu $ (which is symmetric w.r.t. the metric) with
the four-velocity and the configuration gradient respectively. The
first contraction obviously vanishes, while the second gives
\begin{equation}
\phi^\nu_B f^B,_\mu f^A,_\nu= f^A_\nu h_{BM}f^M,_\alpha g^{\alpha\nu} f^B,_\mu=f^A,_\mu
\end{equation}
This proves the given formulas.

On the other hand we can use this new information to simplify the
bracket-terms in (\ref{secondn11}). We can write them as
\begin{equation}
g^{\mu\nu}-  h_{MN}f^N,_\alpha f^M,_\beta g^{\alpha\mu}g^{\beta \nu}=-u^\mu u^\nu
\end{equation}
Thus the final version for the second variation of the particle density is
\begin{equation}\label{stsecvarn1}
\frac{\partial^2 n}{\partial f^{A},_\mu\partial f^B,_\nu}=2n \phi^{[\mu}_A\phi^{\nu]}_B - n h_{AB}u^\mu u^\nu
\end{equation}
With this expression at hand, we have all what we need to cast the principal part into its final appearance:
\begin{equation}\label{stressedpp2}
A^{\mu\nu}_{AB}=\left( 2n \phi^{[\mu}_A\phi^{\nu]}_B - n h_{AB}u^\mu u^\nu \right) \epsilon + 4n U_{ANBL}f^{N},_\alpha g^{\alpha\mu} f^L,_\beta g^{\beta\nu} +\Phi_{AB}^{\mu\nu}
\end{equation}
This will be the form of the principal part relevant in the next
sections. We have the following expression for $\Phi^{\mu\nu}_{AB}$:
\begin{equation}
\Phi^{\mu\nu}_{AB}=2 n\left(  \tau_{BN}f^N,_\alpha g^{\alpha\nu}\phi^\mu_A + \tau_{AN} f^N,_\alpha g^{\alpha\mu}\phi^\nu_B + \tau_{AB}g^{\mu\nu} \right)
\end{equation}
Note that besides the antisymmetric very first term the stress-free
part of the principal part is split into a term in $u$-direction and
a term orthogonal to the four-velocity. One can continue in this
direction by splitting the stress-terms in a similar manner.
Recalling the definition of the $\phi^\nu_B$, we find that we can
split the metric according to
\begin{equation}
g^{\mu\nu}=h_{MN}f^M,_\alpha g^{\alpha\mu}f^N,_\beta g^{\beta\nu}-u^\mu u^\nu
\end{equation}
We also find that the antisymmetric term in the principal part is
orthogonal to the four-velocity:
\begin{equation}
\phi^{[\mu}_A\phi^{\nu]}_B=h_{M[A}h_{B]N}f^M,_\alpha g^{\alpha\mu}f^N,_\beta g^{\beta\nu}
\end{equation}
Thus we can split the principal part into two orthogonal components:
\begin{equation}\label{hierweiss ich keinen namen1}
A^{\mu\nu}_{AB}= - n \left( \epsilon h_{AB} +2 \tau_{AB} \right)u^\mu u^\nu + 2 n M_{AMBN}f^M,_\alpha g^{\alpha\mu} f^N,_\beta g^{\beta\nu}
\end{equation}
where
\begin{equation}\label{hierweiss ich keinen namen2}
M_{AMBN}= 2 U_{AMBN} + \tau_{AB}h_{MN} +\tau_{BN}h_{AM} +\tau_{AM}h_{BN} + \epsilon h_{M[A}h_{B]N}
\end{equation}
Note the following symmetry of $M_{AMBN}$:
\begin{equation}
M_{AMBN}=M_{BNAM}
\end{equation}
The other symmetries of the elasticity operator (\ref{elasticity
operator2222}) are lost by virtue of the other terms.

We finally give the equations of motion in a convenient form
\begin{equation}\label{fuereisesser}
\left( -n \epsilon h_{AB}u^\mu u^\nu + 4nU_{AMBN}f^M,_\alpha g^{\alpha\mu} f^N,_\beta g^{\beta\nu}  +\Phi^{\mu\nu}_{AB} \right)\partial_\mu\partial_\nu f^B=R_A
\end{equation}
where $R_A$ again denotes the lower order terms.

\subsection{Material picture}\label{meom1111}

From the Lagrangian (\ref{mlagrangian1}) we derive the equations of
motion for a elastic medium in the material description. This
Lagrangian had the form
\begin{equation}
\mathcal{L}_{\mbox{\small M}}=N\gamma^{-\frac{1}{2}}\epsilon V
\end{equation}
In the above formula $N$ is the lapse, $V$ is the scalar volume
element and $\gamma^{-1}=(1-W^2)$ where $W$ was defined by
$NW^i:=\dot F^i+Y^i$, where $Y^i$ is the shift.

As in section \ref{refmat1} we combine time $t$ and points $X^A$ on the body into a four-dimensional object
\begin{equation}
(X^\mu):=(X^0,X^A)=(t,X^A)
\end{equation}
Whenever we use the material picture (and whenever the contrary is
not explicitly emphasized), we apply this notation. Greek indices
thus range from zero to three and denote points on
$\R\times\mathcal{B}$.

In this notation the Euler-Lagrange equations can be written as
\begin{equation}
A^{\mu\nu}_{ij}\partial_\mu\partial_\nu F^j = R_i
\end{equation}
where the lower order terms are given by
\begin{equation}
N R_i := \frac{\partial N\gamma^{-\frac{1}{2}}\epsilon}{\partial F^i}- \frac{\partial^2 N\gamma^{-\frac{1}{2}}\epsilon}{\partial F^i,_\mu \partial F^j} F^j,_\nu -\frac{1}{V}\partial_\nu \left( V \frac{ \partial N\gamma^{-\frac{1}{2}}\epsilon}{\partial F^i,_\mu}\right)
\end{equation}
As before in the space-time description we will no longer care about
this lower order terms. They are (as long as they stay smooth) of no
importance for the question of local well-posedness and only come
into play when the longtime-behavior is analyzed which is far beyond
the scope of this thesis.

The principal part reads
\begin{equation}
A^{\mu\nu}_{ij}=\frac{\partial^2 \gamma^{-\frac{1}{2}}\epsilon }{\partial F^i,_\mu \partial F^j,_\nu}
\end{equation}
Note that $\gamma$ only involves the time-derivative of the
deformation and none of its other derivatives. Thus the $F^i,_A$ can
only be found in the stored energy function (via the strain). With
this property of the Lagrangian variation looks easier than in the
space-time-description. But as we will see, this case is in fact the
more complicated one. This is because the deformation gradients
enter the Lagrangian in a less explicit way, which is a result of
the transformations involved in the transition between the
descriptions.

In analogy to the space-time-description the principal part is given
by
\begin{equation}\label{matpp1}
A^{\mu\nu}_{ij}=\frac{\partial^2 \gamma^{-\frac{1}{2}}}{\partial \dot F^i \partial \dot F^j }\delta^{\mu}_0 \delta^\nu_0\epsilon + \gamma^{-\frac{1}{2}}U_{MNAB}\frac{\partial h^{AB}}{\partial F^i,_\mu}\frac{\partial h^{MN}}{\partial F^j,_\nu} +\Phi^{\mu\nu}_{ij}
\end{equation}
where $\Phi^{\mu\nu}_{ij}$ is again linear homogeneous in the stress. It is of the form
\begin{equation}
\Phi^{\mu\nu}_{ij}=\tau_{AB}\left(  \gamma^{-\frac{1}{2}}\frac{\partial^2 h^{AB}}{\partial F^i,_\mu \partial F^j,_\nu} + \frac{\partial \gamma^{-\frac{1}{2}}}{\partial F^i,_\mu}\frac{\partial h^{AB}}{\partial F^j,_\nu} +  \frac{\partial \gamma^{-\frac{1}{2}}}{\partial F^j,_\nu}\frac{\partial h^{AB}}{\partial F^i,_\mu}   \right)
\end{equation}
As for the space-time-description we will derive more explicit
formulas by computing the variations. The expressions finally
obtained will be the counterparts of (\ref{stressedpp2}) and
(\ref{hierweiss ich keinen namen1}) in the material description.

In any case we need the variations of $\gamma^{-\frac{1}{2}}=(1-W^2)^{\frac{1}{2}}$ and of the strain. For the former we first compute
\begin{equation}
\frac{\partial \gamma^{-\frac{1}{2}}}{\partial \dot F^i}=-\frac{\gamma^{\frac{1}{2}}}{N}W^{a}g_{ai}
\end{equation}
 From this the second variation can be obtained
\begin{equation}
\frac{\partial ^2 \gamma^{-\frac{1}{2}}}{\partial \dot F^{i}\partial \dot F^{j}}=\left( -\frac{\gamma^{\frac{3}{2}}}{N^2}W^{a}W^{b}-\frac{\gamma^{\frac{1}{2}}}{N^2}g^{ab} \right) g_{ia}g_{jb}
\end{equation}
Now we turn to the variation of the strain: Note that the $f^A,_a$
entering the formula (\ref{mstrain1}) for the strain are by means of
(\ref{gradients1}) the inverse of the deformation gradient and thus
only depend on $F^a,_A$ and not on $\dot F^i$.

The first variations are given by
\begin{eqnarray}
\label{eiersalat}\frac{\partial h^{AB}}{\partial \dot F^{i}}=-\frac{2}{N}f^{(A},_{a}f^{B)},_{i}W^{a} \\
\frac{\partial h^{AB}}{\partial F^{i},_{I}}=-2f^{I},_{a}f^{(A},_{i}f^{B)},_{b}(g^{ab}-W^{a}W^{b})=-2f^{(A},_i h^{B)I}
\end{eqnarray}
where we have used the variation of (\ref{gradients1}) w.r.t. the
deformation gradient to compute the variation of $f^A,_a$. From the
above expressions for the first variations the second variations
follow:
\begin{eqnarray}
\frac{\partial ^2 h^{AB}}{\partial \dot F^{i} \partial \dot F^{j}}=-\frac{2}{N^2} f^{(A},_{i}f^{B)},_{j} \\
\frac{\partial ^2 h^{AB}}{\partial F^{i},_{I} \partial F^{j},_{J}}=2  f^{J},_{i} f^{(A},_{j} h^{B)I} +2 f^{(A},_{i}\left( f^{B)},_{j}h^{IJ} +h^{B)J} f^{I},_{j}  \right) \\
\frac{\partial^2 h^{AB}}{\partial F^i,_I\partial \dot F^j}=\frac{2}{N} f^{(A},_i[ f^{B)},_a f^I,_j + f^{B)},_j f^I,_a ]W^a
\end{eqnarray}
Using these formulas we can write down the principal part. Since the
resulting expressions are rather lengthy we split into
time-components, spatial components and mixed components:

\subsubsection{Time components}
We first compute the time-portion of the principal part (\ref{matpp1}):
\begin{equation}
A^{00}_{ij}=\frac{\partial^2 \gamma^{-\frac{1}{2}}}{\partial \dot F^i \partial \dot F^j }\epsilon + \gamma^{-\frac{1}{2}}U_{MNAB}\frac{\partial h^{AB}}{\partial \dot F^i}\frac{\partial h^{MN}}{\partial \dot F^j} +\Phi^{00}_{ij}
\end{equation}
From the formulas we have obtained for the variation of $\gamma$ and of the strain, we find that this can be written as
\begin{equation}\nonumber
A^{00}_{ij}=-\left( \gamma W_i W_j +g_{ij}\right) \frac{\gamma^{\frac{1}{2}}\epsilon}{N^2}+4 \frac{\gamma^{-\frac{1}{2}}}{N^2}U_{ABCD}f^A,_a W^a f^B,_i f^C,_cW^c f^D,_j + \Phi^{00}_{ij}
\end{equation}
where the index at $W$ was pulled with the induced three-metric:
$W_i=g_{ij}W^j$. For application it will be more convenient to
replace the original principal part $A^{\mu\nu}_{ij}$ by a new
object defined by
\begin{equation}
B^{\mu\nu}_{AB}:=A^{\mu\nu}_{ij}F^i,_A F^j,_B
\end{equation}
Since the deformation gradient $F^i,_A$ is an isomorphism, those two
objects can be viewed as equivalent. For example positivity of
$A^{\mu\nu}_{ij}$ implies positivity of $B^{\mu\nu}_{AB}$ and vice
versa. The reason for introducing $B^{\mu\nu}_{AB}$ will be clear,
once we write it down. Using the formula (\ref{minvstrain1}) for the
representation of the inverse strain we obtain
\begin{equation}
B^{00}_{AB}=-h_{AB}\frac{\gamma^{\frac{1}{2}}\epsilon}{N^2}+4 \frac{\gamma^{-\frac{1}{2}}}{N^2}U_{ACBD}f^C,_cf^D,_d W^cW^d +\Psi^{00}_{AB}
\end{equation}
where $\Psi^{\mu\nu}_{AB}:=\Phi^{\mu\nu}_{ij}F^i,_A F^j,_B$. To complete the time part we note that $\Phi^{00}_{ij}$ is given by
\begin{equation}
\Phi^{00}_{ij}=2 \frac{ \gamma^{\frac{1}{2}}}{N^2} \tau_{AB} \left(   - \gamma^{-1}f^A,_if^B,_j +2 W_{(i}f^A,_{j)}f^B,_bW^b  \right)
\end{equation}
Thus we obtain
\begin{equation}
\Psi^{00}_{AB}= 2 \frac{ \gamma^{\frac{1}{2}}}{N^2} \tau_{MN} \left( - \gamma^{-1} \delta^M_A\delta^N_B  +2  W_iF^i,_{(A}\delta^M_{B)}f^N,_bW^b  \right)
\end{equation}

\subsubsection{Spatial components}
Next we come to the spatial part. This is easily obtained, since $\gamma$ does not depend on $F^i,_A$. It is given by
\begin{equation}
A^{AB}_{ij}=\gamma^{-\frac{1}{2}}U_{MNKL}\frac{\partial h^{KL}}{\partial F^i,_A}\frac{\partial h^{MN}}{\partial F^j,_B} +\Phi^{AB}_{ij}
\end{equation}
This takes the simple form
\begin{equation}
A^{AB}_{ij}=4\gamma^{-\frac{1}{2}} U_{MNKL} f^M,_i h^{NA}f^K,_jh^{LB}+\Phi^{AB}_{ij}
\end{equation}
which becomes
\begin{equation}
B^{AB}_{IJ}=4\gamma^{-\frac{1}{2}} U_{IMJN}h^{MA}h^{NB}+\Psi^{AB}_{IJ}
\end{equation}
It remains to compute $\Phi^{AB}_{ij}$ and $\Psi^{AB}_{IJ}$. The former is given by
\begin{equation}
\Phi^{AB}_{ij}= 2 \gamma^{-\frac{1}{2}}\tau_{MN} \left( f^B,_if^M,_jh^{NA}+f^M,_i f^N,_jh^{AB} +f^M,_if^A,_jh^{NB}  \right)
\end{equation}
From this the latter is found to be
\begin{equation}
\Psi^{AB}_{IJ}=2 \gamma^{-\frac{1}{2}}\tau_{MN} \left(  \delta^B_I \delta^M_J h^{NA}+ \delta^M_I\delta^N_Jh^{AB}+\delta^M_I\delta^A_Jh^{NB}       \right)
\end{equation}
This completes the computation of the spatial terms.

\subsubsection{Mixed components}

Now we turn to the mixed components. They are obtained from
\begin{equation}
A^{0B}_{ij}=\gamma^{-\frac{1}{2}}U_{MNKL}\frac{\partial h^{KL}}{\partial \dot F^i}\frac{\partial h^{MN}}{\partial F^j,_B} +\Phi^{0B}_{ij}
\end{equation}
The other mixed components then follow from the symmetry of the
principal part:
\begin{equation}
A^{A0}_{ij}=A^{0A}_{ji}
\end{equation}
Applying our formulas to compute the included variations we obtain
\begin{equation}
A^{0B}_{ij}=\frac{4}{N}\gamma^{-\frac{1}{2}} U_{MNKL}f^M,_i h^{NB}f^K,_af^L,_jW^a   +\Phi^{0B}_{ij}
\end{equation}
This leads to
\begin{equation}
B^{0B}_{IJ}=\frac{4}{N}\gamma^{-\frac{1}{2}} U_{INJK}h^{NB}f^K,_aW^a  +\Psi^{0B}_{IJ}
\end{equation}
Finally we compute
\begin{equation}\nonumber
\Phi^{0B}_{ij}=\frac{2\gamma^{\frac{1}{2}}}{N}\tau_{MN}\left[ 2 \gamma^{-1} f^{(B},_if^{N)},_a W^a f^M,_j  + h^{BM}f^N,_j W_i  \right]
\end{equation}
and
\begin{equation}
\Psi^{0B}_{IJ}=\frac{2\gamma^{\frac{1}{2}}}{N}\tau_{MN}\left[  2 \gamma^{-1} \delta^{(B}_If^{N)},_aW^a\delta^M_J + h^{BM}\delta^N_J W_iF^i,_I          \right]
\end{equation}
With these formulas for the mixed components of the principal part we can proceed to the final task of this section:

\subsubsection{Full principal part}

We will add up the partial results from the proceeding subsections
and will derive the full principal part including the stress-terms
in analogy to (\ref{hierweiss ich keinen namen1}). Note that the key
point there was to split the principal part into a portion
proportional to the four-velocity and an orthogonal remainder. Here
we will proceed along the same lines. This means that first of all
we have to identify the analogous objects for the material setup.

The form of the principal part (\ref{matpp1}) suggests, that one
preferred direction will be $\partial_t$. To give an explanation for
this fact we recall our procedure in section \ref{refmat1}. There we
saw, that the transformation $(t,x)\leftrightarrow (t,X)$ given by
the diffeomorphism
\begin{equation}
(t,X)=(t,f(t,x)) \ \ \ \ \ \ (t,x)=(t,F(t,X))
\end{equation}
maps the normalized four-velocity to
\begin{equation}
u^\mu \mapsto u^0 \delta^\mu_0
\end{equation}
Thus $\partial_t$ is nothing but (a multiple of) the normalized
four-velocity mapped to $\R\times\mathcal{B}$. From this one can
guess that the transformation of $f^A,_\mu g^{\mu\nu}$ (viewed as a
collection of vector fields on space-time) will determine the
remainder. We find that
\begin{equation}
f^A,_\mu g^{\mu\nu}\mapsto -\delta^\mu_0 \dot f^A + \delta^\mu_Mh^{MA}=-\delta^\mu_0 \dot f^A +\delta^\mu_Mh^{MA}
\end{equation}
This leads to the definitions
\begin{eqnarray}
U^\mu:=\gamma^{\frac{1}{2}}\delta^\mu_0 \\
\phi^{C\mu}:=\delta^\mu_0 \dot F^a f^C,_a   +\delta^\mu_Mh^{MC}
\end{eqnarray}
The coefficient entering the definition of $U^\mu$ guarantees the
normalization. In order to keep the calculation as simple as
possible, we assume Gaussian coordinates on space-time. This choice
makes the computation a lot easier to overlook.

Even for this simple choice of coordinates the actual calculation is
rather lengthy, we will only catch the most important steps. We find
it more convenient to replace $A^{\mu\nu}_{ij}$ by
$B^{\mu\nu}_{AB}$. This new object is given by
\begin{eqnarray}
B^{\mu\nu}_{AB}=\delta^{\mu}_{0}\delta^{\nu}_{0}\left[-h_{AB}\gamma^{\frac{1}{2}}\epsilon+4\gamma^{-\frac{1}{2}}U_{AaBb}\dot F^{a}\dot F^{b}    - \right. \\
\left. 2\tau_{CD} \left( 2 \gamma^{\frac{1}{2}} \dot F^{c}f^{C},_{c}f^{D},_{(i}\dot F_{j)}F^{i},_{A}F^{j},_{B} - \gamma^{-\frac{1}{2}}\delta^{C}_{A}\delta^{D}_{B}   \right) \right] + \\
+ \delta^{\mu}_{I}\delta^{\nu}_{J} \left[ 4\gamma^{-\frac{1}{2}}U_{ACBD}h^{CI}h^{DJ}    \right.+ \\
\left. + 2\gamma^{-\frac{1}{2}}\tau_{CD}\left(     \delta^{J}_{A}\delta^{C}_{B}h^{DI}+\delta^{C}_{A}\delta^{D}_{B}h^{IJ}+\delta^{C}_{A}\delta^{I}_{B}h^{EJ}    \right) \right]+ \\
+\delta^{\mu}_{0}\delta^{\nu}_{J} \left[   4\gamma^{-\frac{1}{2}}U_{ACBD}\dot F^{a}f^{C},_{a}h^{DJ}          \right. + \\
+\left. 2\tau_{CB}\left(  \gamma^{\frac{1}{2}}\dot F^{k} g_{kn}F^{n},_{A} h^{CJ} +2 \gamma^{-\frac{1}{2}}\delta^{(C}_{A}f^{J)},_{m}\dot F^{m} \right)\right]          +\\
+\delta^{\nu}_{0}\delta^{\mu}_{I} \left[     4\gamma^{-\frac{1}{2}}U_{ACBD}\dot F^{a}f^{D},_{a}h^{CI}                   \right. + \\
+\left. 2\tau_{CA}\left(  \gamma^{\frac{1}{2}}\dot F^{k} g_{kn}F^{n},_{B} h^{CI} +2 \gamma^{-\frac{1}{2}}\delta^{(C}_{B}f^{I)},_{m}\dot F^{m}  \right) \right]
\end{eqnarray}
In analogy to (\ref{hierweiss ich keinen namen1}) we try to decompose this in the following way
\begin{equation}
B^{\mu\nu}_{AB}=-A_{AB}U^{\mu}U^{\nu}+4\gamma^{-\frac{1}{2}}U_{ACBD}\phi^{C\mu}\phi^{D\nu}+B_{ACBD}\phi^{C\mu}\phi^{D\nu}
\end{equation}
That this is indeed a valid ansatz will be seen in the course of the
computation: we will find that there are indeed no other terms than
the ones given above. In particular there will be no
$U^\mu\phi^{C\nu}$-terms.

There are in principle two ways to proceed. One may guess $A_{AB}$
and then compute $B_{ACBD}$, which is again a check of the validity
of the ansatz for $A_{AB}$. On the other hand one may try to
construct $B_{ACBD}$ by putting terms into the form $\phi^{C\mu}$.
The remaining terms form then $A_{AB}$. In both cases we have to
check that the coefficients of the elasticity tensor are of the
required form. This will be our first task. The parts involving
$U_{ACBD}$ are
\begin{eqnarray}
\delta^{\mu}_{0}\delta^{\nu}_{0}4\gamma^{-\frac{1}{2}}U_{AaBb}\dot F^{a}\dot F^{b}+\delta^{\mu}_{I}\delta^{\nu}_{J}4\gamma^{-\frac{1}{2}}U_{ACBD}h^{CI}h^{DJ} + \\
+\delta^{\mu}_{0}\delta^{\nu}_{J}  4\gamma^{-\frac{1}{2}}U_{ACBD}\dot F^{a}f^{C}_{a}h^{DJ}   +\delta^{\nu}_{0}\delta^{\mu}_{I}  4\gamma^{-\frac{1}{2}}U_{ACBD}\dot F^{a}f^{D}_{a}h^{CI}
\end{eqnarray}
Now it is straightforward to find that this is of the desired form.
To discuss the remaining terms we decide to make a proper ansatz for
$A_{AB}$ since this turns out to be the simplest route. In
reminiscence of the analogy for the part involving the elasticity
operator we pick
\begin{equation}
A_{AB}= \gamma^{-\frac{1}{2}}\left( h_{AB}\epsilon +2\tau_{AB} \right)
\end{equation}
Both these terms come from the variation twice w.r.t. $\dot F$. The
first term in the bracket is found immediately. The second requires
a little more care since it appears with the (at the first glance)
wrong factor. This is overcome by splitting $\gamma^{-1}=1-\dot
F^2$. The remaining terms then are
\begin{eqnarray}
\delta^{\mu}_{0}\delta^{\nu}_{0}\left[ 2\tau_{CD} \left( 2 \gamma^{\frac{1}{2}} \dot F^{c}f^{C}_{c}f^{D}_{(i}\dot F_{j)}F^{i}_{A}F^{j}_{B} +\gamma^{\frac{1}{2}}\delta^{C}_{A}\delta^{D}_{B} \dot F^2  \right) \right] + \\
+ \delta^{\mu}_{I}\delta^{\nu}_{J} \left[ 2\gamma^{-\frac{1}{2}}\tau_{CD}\left(     \delta^{J}_{A}\delta^{C}_{B}h^{DI}+\delta^{C}_{A}\delta^{D}_{B}h^{IJ}+\delta^{C}_{A}\delta^{I}_{B}h^{DJ}    \right) \right]+ \\
+\delta^{\mu}_{0}\delta^{\nu}_{J} \left[   2\tau_{CB}\left(  \gamma^{\frac{1}{2}}\dot F^{k} g_{kn}F^{n}_{A} h^{CJ} +2 \gamma^{-\frac{1}{2}}\delta^{(C}_{A}f^{J)}_{m}\dot F^{m} \right)\right]          +\\
+\delta^{\nu}_{0}\delta^{\mu}_{I} \left[   2\tau_{CA}\left(  \gamma^{\frac{1}{2}}\dot F^{k} g_{kn}F^{n}_{B} h^{CI} +2 \gamma^{-\frac{1}{2}}\delta^{(C}_{B}f^{I)}_{m}\dot F^{m}  \right) \right]
\end{eqnarray}
In the next step we will evaluate the Kronecker deltas and split
according to the indices of the stress. This leads to
\begin{eqnarray}
2 \tau_{AB}\left(\delta^{\mu}_{0}\delta^{\nu}_{0}\gamma^{\frac{1}{2}}\dot F^2+\delta^{\mu}_{I}\delta^{\nu}_{J}\gamma^{-\frac{1}{2}}h^{IJ} +\delta^{\mu}_{0}\delta^{\nu}_{J} \gamma^{-\frac{1}{2}} f^{J}_{m}\dot F^{m} + \delta^{\nu}_{0}\delta^{\mu}_{I} \gamma^{-\frac{1}{2}}f^{I}_{m}\dot F^{m}   \right) + \\
\label{26} + 2\tau_{CA}\left[  \delta^{\mu}_{0}\delta^{\nu}_{0} \gamma^{\frac{1}{2}} \dot F^{c}f^{C}_{c}  \dot F^{a}g_{ab}F^{b}_{B} +\delta^{\mu}_{I}\delta^{\nu}_{J}  \gamma^{-\frac{1}{2}}\delta^{I}_{B}h^{CJ}  + \right. \\
\left. +   \delta^{\nu}_{0}\delta^{\mu}_{I} \left(   \gamma^{\frac{1}{2}}\dot F^{k} g_{kn}F^{n}_{B} h^{CI} +   \gamma^{-\frac{1}{2}}\delta^{I}_{B}f^{C}_{m}\dot F^{m}  \right)       \right] + \\
+2\tau_{CB}\left[   \delta^{\mu}_{0}\delta^{\nu}_{0}  \gamma^{\frac{1}{2}}  \dot F^{c}f^{C}_{c}   \dot F^{a}g_{ab}F^{b}_{A} + \delta^{\mu}_{I}\delta^{\nu}_{J} \gamma^{-\frac{1}{2}}  \delta^{J}_{A}h^{CI}  +\right. \\
\left. +   \delta^{\mu}_{0}\delta^{\nu}_{J} \left(    \gamma^{\frac{1}{2}}\dot F^{k} g_{kn}F^{n}_{A} h^{CJ}+   \gamma^{-\frac{1}{2}}\delta^{J}_{A}f^{C}_{m}\dot F^{m} \right)         \right]
\end{eqnarray}
All left to do, is to simplify the terms in the brackets. The first
term is the easiest: it is proportional to
$\tau_{AB}h_{CD}\phi^{C\mu}\phi^{D\nu}$. The factor is obtained as
\begin{equation}
\tau_{AB}(\dots ) = \gamma^{-\frac{1}{2}}\tau_{AB} h_{CD}\phi^{C\mu}\phi^{D\nu}
\end{equation}
The other two terms are slightly more difficult but can also be
handled without much trouble. We do the calculation for the second
term, the third works just the same way. We want to show that
\begin{equation}
\tau_{CA}(\dots ) = \gamma^{-\frac{1}{2}}\tau_{AC} h_{BD}\phi^{C\mu}\phi^{D\nu}
\end{equation}
To prove the validity of this claim we expand the right hand side:
\begin{eqnarray}\nonumber
 h_{BD}\phi^{C\mu}\phi^{D\nu}=h_{BD}\dot f^{C}\dot f^{D}\delta^{\mu}_{0}\delta^{\nu}_{0}+\delta^{N}_{B}h^{MC}\delta^{\mu}_{M}\delta^{\nu}_{N}-\delta^{M}_{D}\dot f^{C}\delta^{\mu}_{M}\delta^{\nu}_{0}
- \delta^{\nu}_{0}\delta^{\mu}_{M}h_{BD}\dot f^{D} h^{CM}
\end{eqnarray}
The second and the third term match the second and last term in the
bracket in (\ref{26}). The other terms need some handling.
\begin{equation}
h_{BD}\dot f^{C}\dot f^{D}=F^{b}_{B}(g_{bd}+\gamma\dot F_{b}\dot F_{d})\dot F^{d}f^{C}_{c}\dot F^{c}=\gamma F^{b}_{B}g_{bd}\dot F^{d} f^{C}_{c}\dot F^{c}
\end{equation}
which shows that the first term equals the first term in (\ref{26}).
The remaining last term can be dealt with along the same lines.

The very same arguments show that
\begin{equation}
\tau_{CB}(\dots )=\gamma^{-\frac{1}{2}}\tau_{CB} h_{AD}\phi^{C\mu}\phi^{D\nu}
\end{equation}
So altogether we find that the principal part can be written as
\begin{equation}\label{fuertee1}
B^{\mu\nu}_{AB}=-\gamma^{-\frac{1}{2}}\left( h_{AB}\epsilon +2\tau_{AB} \right)U^\mu U^\nu + 2 \gamma^{-\frac{1}{2}}M_{ACBD}\phi^{C\mu}\phi^{D\nu}
\end{equation}
where
\begin{equation}\label{fuertee2}
M_{ACBD}=2 U_{ACBD} + \tau_{AB}h_{CD}+\tau_{BD}h_{AC} +\tau_{AC}h_{BD}
\end{equation}
We find that the principal part takes the very same form as
in(\ref{hierweiss ich keinen namen1}), apart from the missing
$h_{A[C}h_{D]B}$ term. Nevertheless we only have the symmetry
\begin{equation}
M_{ACBD}=M_{BDAC}
\end{equation}
\begin{bem}
The same result could be obtained including lapse and shift. One
would just have to replace $\dot F^i$ by $W^i$ and add the shift $N$
at the right places.
\end{bem}
As was the case in the space-time-description, the ``spatial''
portion of the principal part lacks the full symmetry of the
elasticity operator for general states. Only for special cases, f.e.
stress-free states of the material, this full set of symmetries is
given.

This ends the derivation of the field equations. We can now turn to
the discussion of local well-posedness for certain situations.
Before doing so, we make some comments on the existence of conserved
quantities.

\section{Conserved quantities}

In this section we discuss conserved quantities. By the Noether
theorem (see f.e. \cite{soper1}) we know that the existence of
symmetries implies the existence of conserved quantities. Under the
assumption of staticity we will derive such a conserved quantity,
which will be the energy of the system. We will then linearize this
energy at a stress-free reference solution. We show that the
linearized energy is of the form ``kinetic plus potential terms''.
We will also show that the linearized material Lagrangian is of the
form ``kinetic minus potential terms'' with the same expressions for
the kinetic and potential terms, which reflects the connection
between Lagrangian and Hamiltonian of classical mechanics. Finally
we rewrite the linearized potential terms using a covariant
derivative. This allows for a direct comparison with the
non-relativistic case. Note that the derived energy is conserved for
both finite and infinite matter distributions (this last statement
of course depends on the boundary conditions; in particular we will
see that it is true for natural boundary conditions as introduced in
section \ref{natbcchapter}).

\subsection{Symmetries-conserved quantities}

Given a Killing vector $\xi^{\mu}$, we can contract it with the
Energy-momentum tensor: $T^{\mu\nu}\xi_{\nu}$. This new object is
divergence free, if the Energy-momentum tensor itself is divergence
free, which we assume in the following. (We have seen that the
energy-momentum tensor is divergence-free for solutions of the
Euler-Lagrange equations) We have
\begin{equation}
\nabla_{\mu}\left( T^{\mu\nu}\xi_{\nu} \right)=0
\end{equation}
Integrating this over a region $\Omega$ in space-time, we obtain
\begin{equation}
0=\int\limits_{\Omega}\nabla_{\mu}\left( T^{\mu\nu}\xi_{\nu} \right) \mbox{vol}
\end{equation}
We can use Stoke's theorem to convert this into a surface integral
\begin{equation}
0=\int\limits_{\Omega}\nabla_{\mu}\left( T^{\mu\nu}\xi_{\nu} \right) \mbox{vol}=\int\limits_{\partial\Omega} T^{\mu\nu}\xi_{\nu}\mbox{vol}_{\mu}
\end{equation}
The volume element on the boundary can be written as
$\mbox{vol}_{\mu}=\sqrt{|\det g_{(i)}|}n_{\mu}d^3 x$. $n_{\mu}$ is
the outer unit normal, while $g_{(i)}$ denotes the induced metric on
$\partial\Omega$. We find that
\begin{equation}
0=\int\limits_{\partial\Omega} T^{\mu\nu}\xi_{\nu}n_{\mu}\sqrt{|\det g_{(i)}|}d^3 x
\end{equation}
Assume $\Omega$ to be a cylinder. Then $\partial\Omega$ consists of
three parts: The bottom and the top and the mantle. If one can show
that the integral over the mantle vanishes (due to boundary or
fall-off conditions for the fields), one finds (taking into account
the sign change in the normal) that the integral over the bottom
coincides with the integral over the top. A conserved quantity is
found.

\subsection{The conserved energy}

Assume the space-time to be static and that the Killing vector has
norm $-1$. Then we can introduce coordinates $(t,x^{i})$ such that
$t$ is the parameter along the flow of the Killing vector field and
the $x^{i}$ are coordinates on the hyper-surfaces $t=const$. The
metric can be written in the following form:
\begin{equation}
ds^2=-dt^2+g_{ij}(x)dx^{i}dx^{j}
\end{equation}
The Killing vector is then given by $\partial_{t}$. Integrating over
the world-tube of the material (i.e. the inverse of the body under
the configuration map), we assume that the integral over the mantle
of the tube vanishes. This is guaranteed by fall-off or by the
natural boundary conditions introduced in section \ref{teeistgut}.
Thus the integral of $ T^{\mu\nu}\xi_{\nu}n_{\mu}\sqrt{|\det
g_{(i)}|}dx$ over any space-like slice of the world-tube is
conserved. By choosing the slices to have constant $t$ we arrive at
an energy, which by construction is conserved in time:
\begin{equation}
E:=\int\limits_{f^{-1}(t)(\mathcal{B})}T^{\mu\nu}\xi_{\nu}\xi_{\mu}\sqrt{|\det g |}d^3x
\end{equation}
here $f^{-1}(t)(\mathcal{B})$ is the inverse of the body
$\mathcal{B}$ under the map $f$ at the instant $t$. Modulo the
determinant of the induced metric the integrand is given by
\begin{equation}
T^{\mu\nu}\xi_{\nu}\xi_{\mu}=\xi^{\mu}\xi^{\nu}(\rho u_{\mu}u_{\nu} +2 n \tau_{AB}f^{A},_{\mu}f^{B},_{\nu})
\end{equation}
For each value of $t$ this is a scalar on the hyper-surface
$t=const$. We can compute its pullback along the deformation map
$F^{i}$. We had the following relations between quantities in the
material and spatial picture:
\begin{eqnarray}
u^{\mu}=\frac{1}{\sqrt{-v^2}}v^{\mu} \\
v^{\mu}\partial_{\mu}=\partial_{t}+\dot F^{i}\partial_{i} \\
-v^2=1-\dot F^2
\end{eqnarray}
From this it follows that
\begin{equation}
u^{\mu}\xi_{\mu}=\frac{1}{\sqrt{1-\dot F^2}}
\end{equation}
and that we have
\begin{equation}
T^{\mu\nu}\xi_{\nu}\xi_{\mu}=\frac{\rho}{1-\dot F^2} +2n\tau_{AB}\Xi^{A}\Xi^{B}
\end{equation}
where
\begin{equation}
\Xi^{A}:=f^{A},_{\mu}\xi^{\mu}=\dot f^{A}=-f^{A},_{i}\dot F^{i}
\end{equation}
The pullback of the volume form on the $t=const$ hyper-surface is
given by (the volume-element on $\mathcal{B}$ is chosen such that
$V=1$)
\begin{equation}
\sqrt{\det g_{ij}}\det\left( \frac{\partial F^{k}}{\partial X^{A}} \right)dX=\frac{(1-\dot F^2 )^{\frac{1}{2}}}{n}d^3 X
\end{equation}
Now the energy can be rewritten as an integral over the body
$\mathcal{B}$: (The integral coincides with the integral of the
pull-back over the base space)
\begin{equation}
E=\int\limits_{\mathcal{B}}\left( \frac{\epsilon}{(1-\dot F^2 )^{\frac{1}{2}}}+\frac{2\tau_{AB}}{(1-\dot F^2 )^{\frac{1}{2}}} f^{A},_{i}f^{B},_{j}\dot F^{i}\dot F^{j}   \right) d^3 X
\end{equation}
Using the definition of $\gamma$ we can abbreviate this:
\begin{equation}\label{menergy1}
E=\int\limits_{B}\gamma^{\frac{1}{2}}\left(   \epsilon+2 \tau_{AB}  f^{A},_{i}f^{B},_{j}\dot F^{i}\dot F^{j}    \right) d^3 X
\end{equation}
This energy is conserved in time. It is built solely from material
quantities.

\subsection{Linearized energy}
In this section we will expand the energy in a neighborhood of a
static and stress-free natural deformation. By this we mean that the
reference configuration (the existence of which we take as given) is
assumed to satisfy
\begin{equation}\label{filmmuseum}
u^\mu\partial_\mu\stackrel{o}{=}\partial_t
\end{equation}
(An equivalent way of putting the above equations is to write
$\tilde u^\mu\partial_\mu =\partial_t$. It is however useful to
stick to the notation used in (\ref{filmmuseum}). For conventions
regarding the reference state see section \ref{radenska1}) By the
definition of the four-velocity (\ref{filmmuseum}) includes that the
reference configuration and hence also the reference deformation is
time-independent:
\begin{equation}
\dot F^i\stackrel{o}=0
\end{equation}

We will find that the first order variation of the energy vanishes
and that the second order terms are of the form kinetic energy plus
potential energy.

\begin{bem}
As will be seen in later sections this particular reference state is
a solution of the field equation satisfying the boundary conditions
(both fall-off and natural).
\end{bem}
As usual we will write $\stackrel{o}{=}$ for relations which only
hold for this natural deformation. The perturbation will be denoted
by $\delta F^{i}$.

To see that the first variation of the energy vanishes note that we
can easily compute the following partial derivatives:
\begin{eqnarray}
\frac{\partial \gamma^{\frac{1}{2}}}{\partial F^{i},_{\mu}}=\delta^{\mu}_{0}\gamma^{\frac{3}{2}} \dot F_{i}\stackrel{o}{=}0 \\
\frac{\partial \gamma^{\frac{1}{2}}}{\partial F^{i}}=\frac{1}{2}\gamma^{\frac{3}{2}}\dot F^{a}\dot F^{b}g_{ab},_{i}\stackrel{o}{=}0 \\
\frac{\partial \epsilon}{\partial F^{i},_{\mu}}=\tau_{AB}\frac{\partial h^{AB}}{\partial F^{i},_{\mu}}\stackrel{o}{=}0 \\
\frac{\partial \epsilon}{\partial F^{i}}=\tau_{AB}\frac{\partial h^{AB}}{\partial F^{i}}\stackrel{o}{=}0
\end{eqnarray}
The first variation of the second term in the bracket in
(\ref{menergy1}) vanishes, since it is quadratic in $\dot F^{i}$.
There is also the Piola stress tensor in the third term, so it only
contributes to third order. We may omit it in this analysis. We have
found that the expanded energy is at least quadratic in the
deformation field.

We now compute the second variation of the energy: Therefore we need
the second variations of $\gamma^{\frac{1}{2}}$ and the stored
energy $\epsilon$: For the former we obtain
\begin{equation}
\frac{\partial ^2 \gamma^{\frac{1}{2}}}{\partial F^{i},_{\mu}\partial F^{j},_{\nu}}=\delta^{\mu}_{0}\delta^{\nu}_{0}\left( 3\gamma^{\frac{5}{2}}\dot F_{i} \dot F_{j} +\gamma^{\frac{3}{2}}g_{ij}\right) \stackrel{o}{=}\delta^{\mu}_{0}\delta^{\nu}_{0}g_{ij}
\end{equation}
All other variations vanish for the obvious reason that they still
involve time-derivatives of the field, which are zero for the
natural reference deformation.

The second variation of the stored energy with respect to the
deformation gradient is
\begin{equation}
\frac{\partial^2\epsilon}{\partial F^{i},_{\mu} \partial F^{j},_{\nu}}=U_{ABCD} \frac{\partial h^{AB}}{\partial F^{i},_{\mu}}\frac{\partial h^{CD}}{\partial F^{j},_{\nu}} + \tau_{AB} (\dots ) \stackrel{o}{=}U_{ABCD} \frac{\partial h^{AB}}{\partial F^{i},_{\mu}}\frac{\partial h^{CD}}{\partial F^{j},_{\nu}}
\end{equation}
We can write out the variations of the strain and get
\begin{equation}
\frac{\partial^2\epsilon}{\partial F^{i},_{\mu}\partial F^{j},_{\nu}} \stackrel{o}{=}4U_{ABCD} h^{MB}f^{A},_{i}h^{ND}f^{C},_{j}\delta^{\mu}_{M}\delta^{\nu}_{N}
\end{equation}
The second variation with respect to the deformation reads
\begin{equation}
\frac{\partial^2\epsilon}{\partial F^{i}\partial F^{j}}=U_{ABCD} \frac{\partial h^{AB}}{\partial F^{i}}\frac{\partial h^{CD}}{\partial F^{j}} + \tau_{AB} (\dots ) \stackrel{o}{=}U_{ABCD} \frac{\partial h^{AB}}{\partial F^{i}}\frac{\partial h^{CD}}{\partial F^{j}}
\end{equation}
writing this in terms of the deformation we obtain
\begin{equation}
\frac{\partial^2\epsilon}{\partial F^{i}\partial F^{j}}\stackrel{o}{=}U_{ABCD}f^{A},_{a}f^{B},_{b}g^{ab},_{i}f^{C},_{c}f^{D},_{d}g^{cd},_{j}
\end{equation}
Finally we need to take care of the mixed term:
\begin{equation}
\frac{\partial^2\epsilon}{\partial F^{i},_{\mu}\partial F^{j}}=U_{ABCD}\frac{\partial h^{AB}}{\partial F^{j}}\frac{\partial h^{CD}}{\partial F^{i},_{\mu}}+\tau_{AB}(\dots )\stackrel{o}{=}U_{ABCD} \frac{\partial h^{AB}}{\partial F^{j}} \frac{\partial  h^{CD}}{\partial F^{i},_{\mu}}
\end{equation}
which gives
\begin{equation}
\frac{\partial^2\epsilon}{\partial F^{i},_{\mu}\partial F^{j}}\stackrel{o}{=}-2 U_{ABCD}f^{A},_{a}f^{B},_{b}g^{ab},_{j}h^{MC}f^{D},_{i}\delta^{\mu}_{M}
\end{equation}
These are all the terms entering the second variation of the energy.
All other vanish. Summing up we find that the lowest non-trivial
term in the expansion of the energy is given by the integral over
the following expression:
\begin{equation}\label{linenergy1}
 \epsilon g_{ij}\delta\dot F^{i}\delta \dot F^{j}+ 4 U_{ABCD}\left( X^{AB}+Y^{AB}\right) \left(X^{CD}+Y^{CD}\right)
\end{equation}
where
\begin{eqnarray}
X^{AB}= -\frac{1}{2}f^{A},_{a}f^{B},_{b}g^{ab},_{i}\delta F^{i} \\
Y^{AB}= h^{MA}f^{B},_{i}\delta F^{i},_{M}
\end{eqnarray}
To write this in a more convenient form, we introduce a covariant derivative:
\begin{equation}
\nabla_{A}V^{i}=\partial_{A}V^{i}+\Gamma^{i}_{jk}V^{k}F^{j}_{A}
\end{equation}
The $\Gamma^{i}_{jk}$ are the Christoffel symbols corresponding to
the induced three-metric $g_{ij}$. We will show, that we can combine
$X^{AB}$ and $Y^{AB}$ into this covariant derivative.

We first define
\begin{equation}
V^{MN}_{mn}:=4 U_{ABCD}h^{MA}h^{NC}f^{B},_{m}f^{D},_{n}
\end{equation}
Then the second term in (\ref{linenergy1}) becomes
\begin{equation}
V^{MN}_{mn} ( \delta F^{m},_{M} -\frac{1}{2}h_{MA}F^{i},_{B}f^{A},_{a}f^{B},_{b}g^{ab},_{j}\delta F^{j} ) \left( (\dots)^{n}_{N} \right)
\end{equation}
We will show that the second term in the brackets coincides with the
connection terms. A short computation indeed gives
\begin{equation}
h_{MA}F^{i},_{B}f^{A},_{a}f^{B},_{b}g^{ab},_{j}\delta F^{j}=-F^{a},_{M}g^{bm}g_{ab},_{j}\delta F^{j}
\end{equation}
All we have to show is that $g_{ab},_{j}$ is the only non-vanishing
term left from the Christoffel symbol $\Gamma_{baj}$. The rest,
which we will prove to vanish is $(-g_{aj},_{b}+g_{jb},_{a})$. This
is antisymmetric in $(a,b)$. We have to show that
\begin{equation}
F^{a},_{M}g^{bm}V^{MN}_{mn}=4 F^{a},_{M}g^{bm}U_{ABCD}h^{MA}h^{NC}f^{B},_{m}f^{D},_{n}
\end{equation}
is symmetric in this pair of indices. But this is nearly obvious
since the term on the right hand side becomes
\begin{equation}
4 h^{NC}f^{D},_{n}U_{ABCD}f^{B},_{i}f^{A},_{k}g^{ka}g^{ib}
\end{equation}
which of course has the desired symmetry by the symmetries of the
elasticity tensor. Thus the antisymmetric part of the Christoffel
symbol is annihilated and we end up with
\begin{equation}
\frac{1}{2}h_{MA}F^{i},_{B}f^{A},_{a}f^{B},_{b}g^{ab},_{j}\delta F^{j}=-F^{a},_{M}\Gamma^{m}_{aj}\delta F^{j}
\end{equation}
Together with the partial derivative we can combine this to the
covariant derivative. We finally have the following formula for the
linearized energy (besides the ground state energy $\tilde
\epsilon$, which is an additive constant):
\begin{equation}
E_{L}=\int\limits_{B}\left( \epsilon g_{ij}\delta\dot F^{i}\delta\dot F^{j} + V^{MN}_{mn}\nabla_{M}\delta F^{m}\nabla_{N}\delta F^{n} \right)d^3 X
\end{equation}
Note that $E_{L}$ is of the form kinetic energy plus potential
energy. Besides the covariant derivatives it coincides with the
corresponding non-relativistic quantity. Using the strong point-wise
stability condition (i.e. if the material's elasticity operator
satisfies this condition) on the elasticity tensor of the natural
deformation, we find that the linearized energy is non-negative. For
non-static, non-isometric deformations it is positive.

This very last remark needs some additional explanation.
\begin{definition}
We call a linearized deformation isometric, whenever it is a Killing
vector field w.r.t. the spatial metric $g_{ab}$.
\end{definition}
Note that it makes sense to speak of the linearized deformation
$\delta F^i$ as of a vector field, since by its nature it is an
element of the tangential bundle over the hyper-surfaces $t=const$.
Going back to the original form of the energy given in
(\ref{linenergy1}), we note that the part involving the elasticity
operator vanishes if and only if $X^{(AB)}+Y^{(AB)}=0$. We will show
that this is the case if and only if $\delta F^i$ is a Killing field
for the spatial metric. First note that $X^{(AB)}+Y^{(AB)}=0$ if and
only if
\begin{equation}
-\frac{1}{2}g^{ab},_i\delta F^i + f^M,_mg^{m(a}\delta F^{b)},_M=0
\end{equation}
Viewing $\delta F^i$ as a field on physical space $t=const$ rather
than on the body, we find that $\delta F^i,_m= f^M,_m\delta
F^{i},_M$. Lowering the indices in the above equation using the
spatial metric leads to
\begin{equation}
g_{ab},_i\delta F^i+ 2g_{m(a}\delta F^{m},_{b)}=0
\end{equation}
which is exactly the Killing equation $\mathcal{L}_{\delta F}g_{ab}=0$. We have the following lemma
\begin{lemma}
Assume that the strong point-wise stability condition holds. Then
the linearized energy is positive if and only if the linearized
deformation is non-static and non-isometric in the above sense.
\end{lemma}

Taking into account the energy of the natural state, which is the
first (i.e. zeroth order) term in the Taylor series, we find that
the energy increases, whenever one has a non-trivial (i.e.
non-isometric) perturbation of the natural state. This implies that
the assumed natural state (if it exists) minimizes the energy if the
strong-point-wise stability condition holds. This sheds some light
into the physical meaning of this condition.

\subsection{Linearized material Lagrangian}
Here we will derive the linearized material Lagrangian w.r.t. the given static natural state. We will see that it looks very similar to the linearized energy. Just the relative sign between the kinetic and the potential term changes. 
As was the case for the energy, we will find, that the first order perturbation of the Lagrangian vanishes and that the first non-vanishing order is the second.

Recall that the fully non-linear Lagrangian for the material picture is given by
\begin{equation}
S_{\mbox{\small M}}=\gamma^{-\frac{1}{2}}\epsilon
\end{equation}
Comparison with the energy shows that the first term differs only in
the sign of the exponent of $\gamma$. The first variation vanishes
due to well known reason: The variation of the internal energy gives
the Piola-stress tensor, while the variation of the $\gamma$-term
holds time-derivatives of the natural deformation.

The second variations become
\begin{eqnarray}
\frac{\partial^2 \mathcal{L}_{\mbox{\small M}}}{\partial F^{i}\partial F^{j}}\stackrel{o}{=} U_{ABCD}\frac{\partial h^{AB} }{\partial F^{i}}\frac{\partial h^{CD} }{\partial F^{j}}  \\
\frac{\partial^2 \mathcal{L}_{\mbox{\small M}}}{\partial F^{i},_{\mu}\partial F^{j}}\stackrel{o}{=} U_{ABCD}\frac{\partial h^{AB} }{\partial F^{i},_{\mu}}\frac{\partial h^{CD} }{\partial F^{j}} \\
\frac{\partial^2 \mathcal{L}_{\mbox{\small M}}}{\partial F^{i},_{\mu}\partial F^{j},_{\nu}}\stackrel{o}{=}-g_{ij}\epsilon \delta^{\mu}_{0}\delta^{\nu}_{0}+ U_{ABCD}\frac{\partial h^{AB} }{\partial F^{i},_{\mu}}\frac{\partial h^{CD} }{\partial F^{j},_{\nu}}
\end{eqnarray}
Besides the minus in the last line in front of the spatial metric,
this is just the same as we had for the linearization of the energy.
We thus can directly write down the linearized action
\begin{equation}\label{ooooo111}
S_{\mbox{\small M}L}=\int\limits_{B} \left( -\epsilon g_{ij}\delta \dot F^{i}\delta\dot F^{j}+ V^{MN}_{mn}\nabla_{M}\delta F^{m}\nabla_{N}\delta F^{n} \right)d^3 X
\end{equation}
This linearized Lagrangian is of the form potential minus kinetic
energy. From this linearized Lagrangian we can read off all the
information  needed for the local existence results proven in
section \ref{elvislebt}. If the metric is in addition flat, we can
introduce coordinates on space-time and on the body, such that the
configuration gradient and the strain coincide with the Kronecker
Delta. Then
\begin{equation}
V^{MN}_{mn} = 4
U_{ABCD}\delta^{MA}\delta^{NC}\delta^{B}_{m}\delta^{D}_{n}
\end{equation}
 so that $ V^{MN}_{mn}$ coincides with the elasticity tensor in these particular coordinate systems. But then the linearized relativistic Lagrangian coincides with the non-relativistic linearized Lagrangian.
\begin{bem}
We have seen that for a flat metric (i.e. elasticity on Minkowski
space) there is no way to distinguish between the relativistic and
the non-relativistic theory on the linearized level. Relativistic
effects only enter to third or higher order.
\end{bem}

\section{Linearized elasto-dynamics}

We linearize the space-time Lagrangian w.r.t. a stress-free static
reference solution. From this linearized Lagrangian we will derive
the corresponding linearized equations of motion. The actual
computation will be similar to the one done in the previous section.
There however we were working in the material description while here
the space-time picture is used. We do not know a priori that these
two descriptions lead to the same formulas in the linearized case.
That this is indeed true will be seen in the following computations.

Then in the second part of this section an explicit solution for
this linear system will be derived for isotropic and homogeneous
materials. For simplicity we assume space-time to be Minkowski
space.

\subsection{Linearized equations of motion}

The original nonlinear Lagrangian on Minkowski space is according to
(\ref{stlagrangian1}) given by
\begin{equation}
L=n\epsilon
\end{equation}
where the stored energy $\epsilon$ depends solely on the strain
$h^{AB}$ due to the assumption of homogeneity. The particle number
density $n$ depends again solely on the strain, if the medium is
assumed to be homogeneous. The strain itself is a function purely of
the configuration gradient and the space-time metric which is known.
Thus we can develop both the density and the internal energy into a
formal Taylor series:
\begin{eqnarray}
n=\tilde n +\frac{\partial n}{\partial f^{A},_{\mu}}|_{f=\tilde f}\delta f^A,_\mu +\frac{1}{2}\frac{\partial ^2 n}{\partial f^A,_\mu \partial f^B,_\nu }|_{f=\tilde f} \delta f^A,_\mu \delta f^B,_\nu +\mbox{h.o.t.}  \\
\epsilon= \tilde \epsilon  +\frac{\partial \epsilon}{\partial f^A,_\mu }|_{f=\tilde f} \delta f^A,_\mu+ \frac{1}{2}\frac{\partial ^2 \epsilon}{\partial f^A,_\mu \partial f^B,_\nu }|_{f=\tilde f}  \delta f^A,_\mu \delta f^B,_\nu +\mbox{h.o.t.}
\end{eqnarray}
We recall from section \ref{steom11111} that there always exist
$\Phi^\mu_A$ which form the inverse configuration gradient in the
sense that
\begin{equation}
f^A,_\mu \phi^\mu_B =\delta^A_B \ \mbox{and} \ \phi^\mu_B f^B,_\nu = h^\mu_\nu
\end{equation}
where
\begin{equation}
h^{\mu\nu}=\eta^{\mu\nu}+u^\mu u^\nu
\end{equation}
is the metric on the orthogonal space of the four-velocity. We
assume that we can choose coordinates in such a way that $\tilde
u=\partial_{t}$ in standard Minkowski coordinates. This means that
the normalized four-velocity is assumed to be a
hyper-surface-orthogonal time-like Killing field. Thus the
corresponding configuration $\tilde f^A$ is a time-independent
diffeomorphism between the body and the hyper-surfaces orthogonal to
$\tilde u^\mu$. In addition we assume $\tilde f^A$ to be
stress-free, i.e. $\tilde \tau_{AB}=0$.

Using $\phi^\mu_A$ and the formulas derived in section
\ref{steom11111} for the variation of the internal energy and the
particle number, we find that
\begin{eqnarray}
n=\tilde n \left( 1+\tilde \phi^i_{A} \delta f^A,_i -\frac{1}{2}\tilde h_{AB}\delta\dot f^A\delta\dot f^B         + \dots        \right)   \\
\epsilon= \tilde \epsilon   + 2 \tilde U_{AMBN}\tilde f^M,_m\delta^{mi}\tilde f^{N},_n\delta^{nj}\delta f^A,_i\delta f^B,_j + \dots
\end{eqnarray}
Thus the linearized Lagrangian (omitting the zeroth order term which
does not contribute to the variation) is given by
\begin{equation}
L_0 := - \tilde \epsilon \tilde h_{AB}\delta\dot f^A\delta\dot f^B    + 4 \tilde U_{AMBN}\tilde f^M,_m\delta^{mi}\tilde f^{N},_n\delta^{nj}\delta f^A,_i\delta f^B,_j
\end{equation}
This Lagrangian is quadratic in the linearized configuration
gradient $\delta f^A$. As in the foregoing section the Lagrangian is
of the form kinetic minus potential terms.

Once the linearized Lagrangian is given it is easy to obtain the
corresponding field equations which are the linearization of the
original Euler-Lagrange equations. We obtain
\begin{equation}
- \tilde \epsilon \tilde h_{AB}\partial_{t}\partial_{t}\delta f^B + 4\tilde U_{AMBN}\tilde f^M,_m\delta^{mi}\tilde f^{N},_n\delta^{nj}\partial_i\partial_j f^B =0
\end{equation}
Without restriction of generality we set $\tilde \epsilon=1$ (this
can be obtained by dividing by $\tilde \epsilon$ then redefining the
elasticity operator). We also use coordinates on the body, such that
$\tilde f^{M}_{i}=\delta^M_i$. Then the reference strain satisfies
$\tilde h_{AB}=\delta_{AB}$. We also specialize to the isotropic
case, where the internal energy depends on the strain only via its
invariants. Then the elasticity tensor is determined by two
constants, $\lambda$ and $\mu$, the so called Lam\'e constants (see
section \ref{elasticity operator2222}). In the chosen coordinates we
have
\begin{equation}\label{elip1}
4\tilde U_{AMBN}=\lambda \delta_{AM}\delta_{BN}+2\mu \delta_{A(B}\delta_{N)M}
\end{equation}
For convenience we write $g^B$ instead of $\delta f^B$. Then the equation reads
\begin{equation}\label{glg1}
-\ddot g^A +\mu \Delta g^A +(\mu+\lambda )\nabla^A (\nabla \cdot g) =0
\end{equation}
Note that this is hyperbolic if and only if the following two
conditions on the Lam\'e constants are satisfied:
\begin{equation}\label{lhc1}
c_2 ^2 := \mu >0 \ \ \mbox{and} \ \ c_1 ^2 := 2 \mu + \lambda >0
\end{equation}
In the following section we will derive an explicit solution for the
Cauchy problem.

\subsection{Explicit solution}

In this section we will derive a solution for equation (\ref{glg1}) for given initial data
\begin{equation}
g|_{t=0}=j \ \ \ \dot g|_{t=0}=h
\end{equation}
For explicit solutions in a more general setting (but only for the
case $j=0$) we refer to \cite{duff1}. There the requirement of
isotropy is dropped. The resulting system still has constant
coefficients but is coupled in a more complicated way.

\subsubsection{Zero initial configuration}

To keep the calculation clear and simple, we first assume that
\begin{equation}\label{id1}
g|_{t=0}=0 \ \ \ \dot g|_{t=0}=h
\end{equation}
The other case $h=0$ will be handled afterwards along similar lines.
By linearity the solution to the full problem is then the
superposition of these two solutions.

Let the Fourier transformation of $g$ be denoted by $\tilde g$. Then
the Fourier transformation (in space only) of (\ref{glg1}) is given
by
\begin{equation}\label{glg2}
 \ddot{\tilde g}^A + \mu k^2 \tilde g^A +(\mu+\lambda )k^A (\tilde g \cdot k)=0
\end{equation}
This equation is of the form
\begin{equation}
\ddot{\tilde g}^A=-C^A_B \tilde g^B
\end{equation}
where the (symmetric) matrix $C$ is given by
\begin{equation}
C_{AB}=(\mu +\lambda ) k_A k_B +\mu k^2 \delta_{AB}
\end{equation}
Indices are lowered and raised using the Kronecker delta (which
coincides with the reference strain in the chosen coordinates). The
solution of the above equation with the Fourier transform of
(\ref{id1}) as initial data is of the form
\begin{equation}
\tilde g^A=\left( \frac{\sin C^{1/2}t}{C^{1/2}}\right)^A_B\tilde h^B
\end{equation}
This can be easily verified by direct computation. Note that
$\frac{\sin C^{1/2}}{C^{1/2}}$ is a matrix, and that in general it
is not easy to actually compute this operator from the given $C$.
But one readily checks that $C$ is positive definite if the operator
(\ref{elip1}) is elliptic, i.e. as long as (\ref{lhc1}) holds. But
this is assumed anyway, so the root $C^{1/2}$ is well defined. For
the actual computation we split $C$ into projection operators:
\begin{equation}
C=(2\mu +\lambda ) k^2 P+\mu k^2 Q
\end{equation}
where the projectors
\begin{eqnarray}
P_{AB}=k^{-2}k_Ak_B \\
Q_{AB}= \delta_{AB}-k^{-2} k_Ak_B
\end{eqnarray}
satisfy the following relations
\begin{eqnarray}
 P^2 =P \ \ \ \ Q^2=Q \\
PQ=0 \ \ \ \ P+Q=\delta
\end{eqnarray}
Using this together with the definition of the sine via its Taylor
series and the uniqueness of the inverse one can show that (recall
the definition of the wave speeds $c_i$ according to (\ref{lhc1}))
\begin{eqnarray}
C^{1/2}=c_1 |k| P + c_2 |k| Q \\
C^{-1/2}=\frac{1}{c_1 |k|} P +\frac{1}{ c_2 |k|} Q \\
\sin C^{1/2}t=\sin c_1 |k|t P+ \sin  c_2 |k|t Q
\end{eqnarray}
Note that in the last expression the operators $P$ and $Q$ are not
affected by the sine. When taken together, the above statements
imply
\begin{equation}
\frac{\sin C^{1/2}t}{ C^{1/2}}=\frac{ \sin c_1 |k|t}{c_1 |k|} P+ \frac{\sin  c_2 |k|t}{c_2 |k|} Q
\end{equation}
Using that $P+Q=\delta$ we find that $\tilde g$ can be written as
\begin{equation}\label{allgformel1}
\tilde g^A= \left( \frac{ \sin c_1 |k|t}{c_1 |k|^3}- \frac{\sin  c_2 |k|t}{c_2 |k|^3}  \right)(\tilde h \cdot k)k^A +\frac{\sin  c_2 |k|t}{c_2 |k|} \tilde h^A
\end{equation}
The last term will provide a Dirac delta, the first terms become
Heaviside functions (and later on by partial integration generate
Kronecker deltas too). We first investigate the last integral, since
it is the easier one. We have to compute the following integral:
\begin{equation}
II^A :=\frac{1}{(2\pi)^{3/2}}\int  \frac{\sin  c_2 |k|t}{c_2 |k|} \tilde h^A e^{ik\cdot x} d^3 k
\end{equation}
By the standard Fourier formulas we can express this as the
convolution of $F$, the inverse Fourier transformation of
$(|k|c_2)^{-1}\sin c_2 |k| t$, with the initial data:
\begin{equation}\label{distribution1}
(2\pi)^{3/2}II=\int F(x-y)h(y) d^3y =\int F(y)h(x-y) d^3y
\end{equation}
We thus have to compute the integral kernel $F$ to obtain an
explicit statement. The usual Fourier transformation formula reads
\begin{equation}
F(x)=\frac{1}{(2\pi)^{3/2}}\int  \frac{\sin c_2 |k| t}{|k|c_2}e^{ik\cdot x }d^3k
\end{equation}
For convenience we introduce some notation: $\kappa = |k|$, $\xi
=|x|$ and $z=\cos \theta$ where $\theta$ is the angle between $x$
and $k$, so that $x\cdot k=\xi \kappa \cos \theta $. By a suitable
choice of coordinate axes we can introduce polar coordinates such
that the above formula becomes
\begin{equation}
 F(x)=\frac{1}{(2\pi)^{1/2}} \int\limits_{0}^{\infty}\int\limits_{-1}^{1} \frac{\sin c_2 \kappa t}{\kappa c_2}e^{i \kappa \xi z }\kappa ^2 dz d\kappa
\end{equation}
We can carry out the $z$ integration and obtain
\begin{equation}\label{inte1}
F(x)=\frac{1}{(2\pi)^{1/2}}\frac{2}{c_2  \xi} \int\limits_{0}^{\infty} \sin c_2 \kappa t \sin \kappa\xi d\kappa
\end{equation}
The integrand has a primitive, which is not defined at infinity.
Therefore we cannot hope at all to obtain a classical representation
of the above integral. But assuming that the initial datum $h$ is a
smooth function of compact support we can interpret
(\ref{distribution1}) as the action of the distribution $F$ on the
test function $h$. Thus we may as well search for distributional
$F$. To do so we note that we can rewrite $F$ using the Euler
formula:
\begin{equation}\nonumber
 F(x)= - \frac{1}{(2\pi)^{1/2}}\frac{1}{2 c_2  \xi} \int\limits_{0}^{\infty} \left(  e^{i \kappa (c_2 t +\xi )}+  e^{-i \kappa (c_2 t +\xi )} -   e^{i \kappa (c_2 t -\xi )}-  e^{-i \kappa (c_2 t -\xi )} \right) d\kappa
\end{equation}
Now we combine the first and the second term as well as the third and the last one into integrals over all of $\R$:
\begin{equation}
F(x)= - \frac{1}{(2\pi)^{1/2}}\frac{1}{2 c_2  \xi} \int\limits_{-\infty}^{\infty}  \left( e^{i \kappa (c_2 t +\xi )} -   e^{i \kappa (c_2 t -\xi )} \right) d\kappa
\end{equation}
The integral is basically the well-known formula for the Fourier
transform of the Dirac delta distribution. We thus finally obtain
\begin{equation}
F(x)= -\left( \frac{\pi}{2}\right)^{1/2}\frac{1}{c_2  \xi}\left[ \delta (c_2 t +\xi) - \delta (c_2 t - \xi) \right]
\end{equation}
This allows a great simplification in (\ref{distribution1}). Since
we are interested in the solution for positive $t$, which
corresponds to solving to the future, the first Dirac delta does not
contribute (recall that $\xi =|x| \ge 0$). The solution is then
given by
\begin{equation}
II= \frac{1}{4\pi}\frac{1}{c_2 } \int  \delta (c_2 t - |x-y|)   \frac{h(y)}{|x-y|}  d^3y
\end{equation}
We can use the delta to transform the nominator and get it in front of the integral:
\begin{equation}
II= \frac{1}{4\pi}\frac{1}{c_2^2 t}\int  \delta (c_2 t - |x-y| )h(y) d^3y
\end{equation}
One clearly sees that the initial disturbance travels along the
sound-cone given by $c_2 t - |x-y|=0$. Next we introduce spherical
coordinates. Then we can directly carry out the radial integration.
This allows us to rewrite $g$ as a surface integral over a ball of
radius $c_2 t$ with center $x$.
\begin{equation}
II= \frac{1}{4\pi}\frac{1}{c_2^2 t}\int_{\partial B(x,c_2 t)}h(y)
dS_y
\end{equation}
which is $t$ times the average integral of the initial datum over
the sphere $\partial B(x,c_2 t)$. Now we turn to the first two terms
in (\ref{allgformel1}). They are a result of the coupling in the
equation (\ref{glg1}) and are a little more complicated to handle.
We have to compute
\begin{equation}
I^A := \frac{1}{(2\pi)^{3/2}}\int \left( \frac{ \sin c_1 |k|t}{c_1 |k|^3}- \frac{\sin  c_2 |k|t}{c_2 |k|^3}  \right)(\tilde h \cdot k)k^A e^{ik\cdot x} d^3 k
\end{equation}
Since this integral consists of two identical parts we will only
compute one of them and then add the analogous expression for the
other one. First we note that
\begin{equation}
\int\frac{\sin c_1 |k|t}{c_1 |k|^3}(\tilde h \cdot k)k^A e^{ik\cdot x} d^3 k=\frac{1}{i}\frac{\partial }{\partial x^{A}}\int\frac{\sin c_1 |k|t}{c_1 |k|^3}(\tilde h \cdot k)e^{ik\cdot x} d^3 k
\end{equation}
The remaining integral will again be written as convolution of a
function derived from the initial data and an integral kernel
obtained from the sine term.

Let $G$ denote the inverse Fourier transform of $\frac{\sin c_1
|k|t}{c_1 |k|^3}$ and let $H$ denote the inverse Fourier transform
of $\tilde h \cdot k$. Then
\begin{equation}
\int\frac{\sin c_1 |k|t}{c_1 |k|^3}(\tilde h \cdot k)e^{ik\cdot x} d^3 k=\int G(x-y)H(y)d^3y
\end{equation}
and consequently
\begin{equation}
\int \frac{ \sin c_1 |k|t}{c_1 |k|^3}(\tilde h \cdot k)k^A e^{ik\cdot x} d^3 k=\frac{1}{i}\int \frac{\partial }{\partial x^{A}}G(x-y)H(y)d^3y
\end{equation}
We first determine $H$. We claim
\begin{equation}
H(y)=\frac{1}{i}\partial_{A}h^{A}(y)
\end{equation}
To prove this we show that by partial integration (recall that $h$ is assumed to be of compact support)
\begin{equation}
\int \partial_{A} h^{A} e^{-ik\cdot x} d^3 x=i \int h^{A}k_A  e^{-ik\cdot x} d^3 x
\end{equation}
The $k$ can be put in front of the integral. This proves that the
Fourier transform of $\partial_{A}h^{A}$ is given by $i k_A\tilde
h^A$, which implies the claim.

It thus only remains to compute the integral kernel $G$. We proceed
similar to what we did with the last term of (\ref{allgformel1}).
Along the same paths we find that
\begin{equation}
G=-\frac{1}{(2\pi)^{1/2}}\frac{1}{2c_1 \xi }\int\limits_{-\infty}^{\infty} \frac{1}{\kappa ^2} \left( e^{i\kappa (c_1 t +\xi )} -e^{i\kappa (c_1 t-\xi)}\right) d\kappa
\end{equation}
The integrals are nothing but one-dimensional Fourier transformations of $\kappa^{-2}$. Thus we obtain
\begin{equation}
G=\sqrt{\frac{\pi}{2}}\frac{1}{2c_1 \xi }\left[  (c_1 t +\xi )\sgn (c_1 t+\xi) -( c_1 t -\xi )\sgn ( c_1 t -\xi)            \right]
\end{equation}
we can give $g$ a better known appearance by exchanging the $\sgn=2
\Theta -1$ by the Heaviside step function $\Theta$. This gives
\begin{equation}
G=-\sqrt{\frac{\pi}{2}}\frac{1}{c_1 }-\sqrt{\frac{\pi}{2}}\frac{1}{c_1 \xi }\left[ ( c_1 t -\xi ) \Theta ( c_1 t -\xi )-   (c_1 t+\xi) \Theta (c_1 t+\xi)\right]
\end{equation}
Next we have to take the derivative with respect to $x$.
\begin{equation}
\frac{\partial G}{\partial x}=\sqrt{\frac{\pi}{2}}\frac{x}{c_1 \xi^3
}\left[ \dots \right]+\sqrt{\frac{\pi}{2}}\frac{1}{c_1 \xi }\left[
\frac{x}{\xi} \Theta ( c_1 t -\xi )+\frac{x}{\xi}  \Theta (c_1
t+\xi) \right]
\end{equation}
There are also Dirac deltas, but they only appear together with the
argument (i.e. as $x\delta(x)$) and therefore do not contribute.
Taking into account the terms in the first bracket we end up with
\begin{equation}
\frac{\partial G}{\partial x} =-\sqrt{\frac{\pi}{2}}\frac{xt}{\xi
^3}[\Theta (c_1 t+\xi) -\Theta (c_1t -\xi) ]
\end{equation}
Note that the constant $c_1$ only appears as an argument in the
Heaviside functions. Having this expression for $G$, one half of $I$
is determined by
\begin{equation}\nonumber
\frac{1}{i}\int \frac{\partial G}{\partial
x}(x-y)H(y)d^3y=\sqrt{\frac{\pi}{2}}\int \frac{(x-y)t}{|x-y|
^3}[\Theta (c_1 t+|x-y|) -\Theta (c_1t -|x-y|)
]\partial_{A}h^{A}(y)d^3y
\end{equation}
The second half of $I$ is just the negative of this expression only
with $c_1$ replaced by $c_2$. At this stage we assume that $c_1
>c_2$. When these two expressions are added up, we find that the
Heaviside functions add up in way, which only leaves $B(x,c_1
t)\setminus B(x,c_2 t)$ as integration domain. (If $c_2>c_1$ the
balls would simply be interchanged.) Thus we have
\begin{equation}
I^A=-\frac{1}{4\pi}\int_{B(x,c_1 t)\setminus B(x,c_2 t)} \frac{(x-y)^A t}{|x-y| ^3}(\partial \cdot h)(y)d^3y
\end{equation}
This is possible, since (as already mentioned) the wave speeds $c_i$
only appear as arguments of the Heaviside function. Now we can add
$II$ to obtain the full solution $g$:
\begin{equation}\nonumber
g^A= \frac{1}{4\pi} \frac{1}{c_2^2 t}\int_{\partial B(x,c_2 t)}h^A(y) dS_y  - \frac{t}{4\pi}\int_{B(x,c_1 t)\setminus B(x,c_2 t)} \frac{(x-y)^A }{|x-y| ^3}(\partial \cdot h)(y)d^3y
\end{equation}
This formula still needs some cosmetics, since there appears a
divergence of the initial data. This problem is easily overcome by
partial integration:
\begin{equation}\label{partint1}
-\int_V \frac{\partial}{\partial y^A} h^A \phi^B = \int_V h^A \frac{\partial}{\partial y^A}\phi^B - \int_{\partial V}h^A N_A \phi^B
\end{equation}
Note that in contrary to what was done before, we must not omit the
boundary terms, since integration takes place over a finite domain.
In our case $V=B(x,c_1 t)\setminus B(x,c_2 t)$. Thus the co-normal
$N$ is given by $|x-y|N=\pm (y-x)$ where the sign depends whether we
are at the outer or inner boundary. The derivative of
$\phi=(x-y)|x-y|^{-3}$ with respect to $y$ is given by
\begin{equation}
\frac{\partial}{\partial y^A }\phi^B=\frac{1}{|x-y|^3}\left(-\delta^B_A+3\frac{(x-y)^B}{|x-y|}\frac{(x-y)_A}{|x-y|} \right)
\end{equation}
For convenience we introduce
\begin{equation}
n^A=\frac{(y-x)^A}{|x-y|}
\end{equation}
At the outer boundary this $n$ coincides with the unit normal and on
the inner boundary it is just the anti-normal. We write
\begin{equation}
\frac{\partial}{\partial y^A }\phi^B=\frac{1}{|x-y|^3}\left(-\delta^B_A+ 3n^Bn_A \right)
\end{equation}
This means that the volume term on the right hand side of (\ref{partint1}) becomes
\begin{equation}
\int_{B(x,c_1 t)\setminus B(x,c_2 t)} \frac{1}{|x-y|^3}(3 n_A n^B -\delta_A^B )  h^A      d^3y
\end{equation}
The surface integral over $\partial V$ splits into two integrals
over spheres of radius $c_i t$ with center $x$. They are given by
\begin{eqnarray}
-\frac{1}{c_1^2 t^2}\int_{\partial B(x,c_1 t)}n^Bn_A h^A(y) dS_y  \\
\frac{1}{c_2^2 t^2}\int_{\partial B(x,c_2 t)}n^Bn_A h^A(y) dS_y
\end{eqnarray}
Adding the last three integrals up according to (\ref{partint1}) and
combining the result with the original formula, we arrive at the
final form for the unknown $g$:
\begin{eqnarray}\label{hirter111}
g^B=\frac{t}{4\pi}\int_{B(x,c_1 t)\setminus B(x,c_2 t)} \frac{1}{|x-y|^3}(3 n_A n^B -\delta_A^B )  h^A(y)      d^3y + \\ \nonumber + \frac{1}{4\pi} \frac{1}{c_2^2 t}\int_{\partial B(x,c_2 t)}(\delta_A^B -n_An^B )h^A(y) dS_y + \\ \nonumber +\frac{1}{4\pi} \frac{1}{c_1^2 t}\int_{\partial B(x,c_1 t)}n^Bn_A h^A(y) dS_y
\end{eqnarray}
According to this formula, a given initial disturbance at a point
$y$ is noticed by an observer sitting at $x$ in the following way:
there will at first be a sharp signal followed by a continuous wave,
which finally ceases with another sharp signal. There is also a
grave difference in the nature of the sharp signals. One ($c_1$)
wave is transversal, while the other one ($c_2$) is longitudinal.
Usually the longitudinal waves are called pressure waves and the
transversal waves are known as shear waves (see f.e. \cite{gur1} or
another introductory textbook).

\subsubsection{Zero initial momentum}

In order to complete the picture we investigate the behavior of
solutions to (\ref{glg1}) with the following initial data (compare
to (\ref{id1})):
\begin{equation}\label{id2}
g|_{t=0}=j \ \ \ \dot g|_{t=0}=0
\end{equation}
Due to the linearity the solution for general initial data is then
the sum of the two solutions obtained for the initial data
(\ref{id1}) and (\ref{id2}). For the above data, the solution of the
Fourier transformed equations (\ref{glg2}) is
\begin{equation}
\tilde g =\cos C^{1/2}t \tilde j
\end{equation}
The $C$ can again be split into the projection operators $P$ and $Q$ which allows us to explicitly compute $\cos  C^{1/2}t$.  We finally obtain
\begin{equation}\label{hihihi1}
\tilde g=\cos c_2 |k| t \tilde j +\left( \frac{\cos c_1 |k|t}{|k|^2} -\frac{\cos c_2 |k|t}{|k|^2} \right) (\tilde j \cdot k)k
\end{equation}
The solution $g$ is the inverse Fourier transform of this expression. We split into two parts, the first of them being
\begin{equation}
I:=\frac{1}{(2\pi)^{3/2}}\int \tilde j \cos c_2 |k| t  e^{ik\cdot x} d^3 k
\end{equation}
Thus
\begin{equation}
(2\pi)^{3/2}I=\int E(y)j(x-y) d^3 y
\end{equation}
where $E$ is the inverse Fourier transform of $\cos c_2 |k| t$.
Introducing spherical coordinates we again do the integration over
the angles. We obtain
\begin{equation}
E=\frac{\pi}{i \xi(2\pi)^{3/2}}\int\kappa\left(e^{i\kappa (c_2 t+\xi)} +e^{i\kappa(\xi-c_2 t)} \right) d\kappa
\end{equation}
Writing this as
\begin{equation}
E=-\frac{\pi}{c_2 \xi(2\pi)^{3/2}}\partial_{t}\int \left(e^{i\kappa (c_2 t+\xi)} -e^{i\kappa(\xi-c_2 t)} \right) d\kappa
\end{equation}
we obtain
\begin{equation}
E(x)=-\left( \frac{\pi}{2}\right)^{1/2}\frac{1}{c_2 \xi}\partial_{t} \left[ \delta ( c_2 t+\xi)-\delta(\xi- c_2 t)     \right]
\end{equation}
This allows us to evaluate $I$: We can write the $t$-derivative in
front of the convolution. We also omit the first Dirac delta, since
$(c_2 t+\xi)>0$. The remaining terms read
\begin{equation}\label{Iort1}
I(x)=\frac{1}{4\pi}\partial_{t}\int\frac{1}{c_2^2 t}\delta(|x-y|- c_2 t)j(y)d^3y
\end{equation}
We introduce spherical coordinates and use the Dirac delta to do the radial integration. This leads to
\begin{equation}
I=\frac{1}{4\pi}\partial_{t}\left(\frac{1}{c_2^2t} \int_{\partial B(x,c_2t)}j(y)dS_y \right)
\end{equation}
To actually carry out the differentiation we transform the integral in one over $\partial B(0,1)$:
\begin{equation}
I=\frac{1}{4\pi}\partial_{t}\left( t\int_{\partial B(0,1)}j(x+c_2t y)dS_y \right)
\end{equation}
Now we can interchange the integration with the differentiation, since the domain $\partial B(0,1)$ is independent of $t$. Then we transform the domain back. This finally gives
\begin{equation}
I=\frac{1}{4\pi}\left(  \frac{1}{c_2^2 t^2} \int_{\partial B(x,c_2t)}j dS_y +\frac{1}{c_2 t}  \int_{\partial B(x,c_2t)}\frac{\partial j}{\partial N}  dS_y  \right)
\end{equation}
where $\frac{\partial j}{\partial N}$ denotes the outer normal derivative.

Next we compute the inverse Fourier transform of the remaining terms in (\ref{hihihi1}).We start by computing
\begin{equation}
II_1:= \frac{1}{(2\pi)^{3/2}}\int\frac{\cos c_1 |k|t}{|k|^2} (\tilde j \cdot k)k e^{ik\cdot x}d^3k
\end{equation}
Using the previous result we obtain
\begin{equation}
(2\pi)^{3/2}II_1^A(x)=-\int \frac{\partial }{\partial x^A} D_1(x-y)\partial \cdot j d^3y
\end{equation}
where
\begin{equation}
D_1(x)=\frac{1}{(2\pi)^{3/2}}\int\frac{\cos c_1 |k|t}{|k|^2}e^{ik\cdot x}d^3k
\end{equation}
Along the meanwhile well known paths we convert this into a one-dimensional Fourier integral:
\begin{equation}\nonumber
D_1(x)=\frac{1}{(2\pi)^{1/2}}\frac{2}{\xi}\int\limits_0^\infty \frac{1}{\kappa} \cos c_1\kappa t \sin \kappa \xi d\kappa=\frac{1}{2i\xi(2\pi)^{1/2}}\int\limits_{-\infty}^\infty \frac{1}{\kappa}\left( e^{i\kappa (c_1 t +\xi)}+e^{i\kappa (\xi-c_1 t)}\right) d\kappa
\end{equation}
The integrals give (modulo constants) the $sign$ distributions. Thus $D_1$ becomes
\begin{equation}
D_1(x)=\sqrt{\frac{\pi}{2}}\frac{1}{\xi}\left[ \Theta(c_1t+\xi)+\Theta(\xi-c_1 t)-1 \right]
\end{equation}
Consequently (since $t>0$ we have $\Theta(c_1t+\xi)=1$ and $\delta(\xi+c_1 t)=0$)
\begin{equation}
\partial D_1(x)=\sqrt{\frac{\pi}{2}}\left( -\frac{x}{\xi^3} \Theta(\xi-c_1 t) +\frac{x}{\xi^2}\delta(\xi-c_1 t)\right)
\end{equation}
This means that $II:=II_1-II_2$ is given by
\begin{eqnarray}\nonumber
4\pi II=-\int_{B(x,c_1t)\setminus B(x,c_2 t)} \frac{(x-y)}{|x-y|^3}\partial\cdot j d^3y -\frac{1}{c_1^2t^2}\int_{\partial B(x,c_1t)}(x-y)\partial\cdot jdS_y + \\ \nonumber + \frac{1}{c_2^2t^2}\int_{\partial B(x,c_2t)}(x-y)\partial\cdot jdS_y
\end{eqnarray}
We know from the previous case of initial data (\ref{id1}) how to
deal with the volume integral. By partial integration we can split
in up into a volume integral involving the undifferentiated initial
data plus some surface terms. Rearranging finally yields
\begin{eqnarray}
4\pi g^B=\int_{B(x,c_1t)\setminus B(x,c_2 t)}\frac{1}{|x-y|^3}(3 n_A n^B -\delta_A^B )  j^A(y)      d^3y + \\ \nonumber  +\frac{1}{c_2^2 t^2}\int_{\partial B(x,c_2 t)}(\delta_A^B -n_An^B )j^A(y) dS_y + \\ \nonumber +\frac{1}{c_1^2 t^2}\int_{\partial B(x,c_1 t)}n^Bn_A j^A(y) dS_y  + \\ \nonumber + \frac{1}{c_2 t} \int_{\partial B(x,c_2t)}\left[ n^A\partial_A j^B -n^B\partial_A j^A \right] dS_y +\\ \nonumber + \frac{1}{c_1t}\int_{\partial B(x,c_1t)}n^B\partial_A j^A dS_y
\end{eqnarray}
Then the full solution is the superposition of this solution with
(\ref{hirter111}).

Note that properly speaking we have not proved the existence of a
solution but have rather derived the form of the solution, if
existence is assumed. In principle we would have to prove that the
representation formulas we derived in this section really solve the
system (\ref{glg1}) and satisfy the initial conditions. But since
this is a rather trivial task, once the formulas are given, we omit
it at this stage.

\section{Unbounded setup}\label{elvislebt}

The minimum requirement a mathematical theory should fulfill to be
viewed as physical reasonable is that the equations of motion
describing the dynamics are well-posed in a certain sense. In our
case the simplest non-linear setup is that of an infinitely extended
elastic material on a given background space-time, since in that
case we do not have to worry about effects arising from
self-gravitation nor have to care about boundary terms. It is thus a
crucial test for our matter model to provide this well-posedness for
the field equations arising in the said setup.

In this section we thus take a closer look at the equations of
motion derived in the sections \ref{steom11111} and \ref{meom1111}.
We will compute the principal symbol, and check hyperbolicity of the
equations of motion. Finally we apply the basic existence theorem
from section \ref{loceubd11111} to obtain local well-posedness for
the Cauchy problem for data close to those describing natural states
of the material.

Beside the results on well-posedness we will also derive some
results on the underlying geometric structure of the equations.

Throughout the following analysis we assume the internal energy to
be a smooth function of its entries (i.e. of the configuration and
the strain in the space-time description and of the body point and
the strain in the material picture).

We finally note that some of the results of this section can be
found in \cite{bs1}, \cite{beig1} or \cite{bewp1}.

\begin{bem}
Note that from now on we again deal with the fully non-linear case.
\end{bem}

\subsection{Space-time description}\label{uhrzeit}

The equations of motion for the configuration took the form
\begin{equation}\label{stiegl1111}
\left( -n \epsilon h_{AB}u^\mu u^\nu + 4nU_{AMBN}f^M,_\alpha g^{\alpha\mu} f^N,_\beta g^{\beta\nu}  +\Phi^{\mu\nu}_{AB} \right)\partial_\mu\partial_\nu f^B=R_A
\end{equation}
This is a quasi-linear second order system for the configuration
$f^A$. $R_A$ denoted the lower order terms and $\Phi^{\mu\nu}_{AB}$
was linear homogeneous in the stress.

\subsubsection{Second order well-posedness}
We will show that the principal part is regular hyperbolic in the
neighborhood of a natural configuration. The existence of this
natural configuration will be assumed in the following.

To prove regular hyperbolicity for configurations close to the
natural one (in a proper topology of course) it suffices to show
that the natural configuration itself has the required property. By
continuity (f.e. in the $C^1$ topology) the same is then true for
all configurations sufficiently close. Remember that regular
hyperbolicity implied weak hyperbolicity as was shown in section
\ref{loceubd11111}. The principal symbol for the natural
configuration reads
\begin{equation}
A_{AB}(k):= -\tilde n \tilde \epsilon \tilde h_{AB} (\tilde u^\mu k_\mu )^2 + 4 \tilde n \tilde U_{AMBN} \tilde f^M,_\alpha k^\alpha \tilde f^N,_\beta k^\beta
\end{equation}
This form of the principal symbol strongly suggest to try to prove
regular hyperbolicity w.r.t the pair $(\tilde u^\mu, -\tilde u_\nu
)$. Indeed one finds that
\begin{equation}
A_{AB}(-\tilde u)=- \tilde n \tilde \epsilon \tilde h_{AB}
\end{equation}
which is negative definite. On the other hand for all $l_\mu$
satisfying $l_\mu u^\mu =0$ we have that
\begin{equation}
A_{AB}(l)= 4 \tilde n \tilde U_{AMBN} \tilde f^M,_\alpha l^\alpha \tilde f^N,_\beta l^\beta
\end{equation}
which is positive definite by virtue of the Legendre-Hadamard
condition implied in the definition of the natural configuration.
Finally one checks the final requirement: $u^\mu (-u_\mu)=1>0$. Thus
we have the following result:
\begin{lemma}
The system (\ref{stiegl1111}) is regular hyperbolic w.r.t. the pair $(\tilde u^\mu,-\tilde u_\nu )$.
\end{lemma}
Then the corollary from section \ref{loceubd11111} immediately
implies the following result.
\begin{theorem}
The Cauchy problem for (\ref{stiegl1111}) is well-posed in the
following sense: Given a smooth natural configuration. Then there
exists a foliation of space-time, such that the initial value
problem for initial data $(f_0, f_1)$ being close to the data
induced by the natural configuration $(\tilde f_0, \tilde f_1)$ in
$H^{s+1}\times H^s$ (for $s\ge 3$) give rise to a unique local (both
in space and time) solution, which depends continuously on the
initial data.
\end{theorem}
\begin{bem}
The problem with this result is that we only get existence and
uniqueness locally in space. This can be improved if the we add an
additional assumption. If the characteristic speeds of the material
are smaller than the speed of light, one may obtain regular
hyperbolicity w.r.t. $(\partial_t, dt)$ all over the leafs of the
foliation and not only in certain neighborhoods. For the case of an
isotopic material this is explicitly carried out for the material
setup in the following section in lemma \ref{exlicittdt}, but it
would work just the same wy in the given context of the space-time
description.
\end{bem}
Note that the connection between the Legendre-Hadamard condition and
regular hyperbolicity is even deeper than we already know. The
following holds
\begin{lemma}
The system (\ref{stiegl1111}) is regular hyperbolic w.r.t. $(u^\mu,
-u^\nu)$ for a reference configuration if and only if it is natural.
If the material is isotropic w.r.t. the reference strain, then
(\ref{stiegl1111}) is regular hyperbolic if and only if the
reference configuration is natural.
\end{lemma}
The first part of the lemma is easy to prove: One part of the
statement we have already checked. The other part is obvious from
the structure of the principal part: suppose that the system is
regular hyperbolic w.r.t $(u^\mu,-u_\nu)$. Then the
Legendre-Hadamard condition for the elasticity operator follows
directly from the definition.

The second part of the lemma is a little bit harder. 
We will see that regular hyperbolicity w.r.t. any $(X^\mu,\omega_\nu)$ implies the Legendre-Hadamard condition. We have
\begin{equation}
0< A_{AB}(k)m^Am^B \le 4 n U_{AMBN}  f^M,_\alpha k^\alpha  f^N,_\beta k^\beta m^A m^B
\end{equation}
for all $k_\mu X^\mu =0$. Since we assume isotropy, we can write this as
\begin{equation}
0< \mu m^2 \kappa^2  + (\mu+\lambda ) (m\cdot \kappa )^2
\end{equation}
where $\kappa^A:=f^A,_\alpha k^\alpha$ and contractions are taken
w.r.t. the strain. We can always find a $k$, which gives rise to a
non-vanishing $\kappa$. Then $m=\kappa$ includes $2\mu+\lambda >0$,
while choosing $m$ orthogonal to $\kappa$ gives rise to $\mu>0$. But
those are just the conditions on the Lam\'e moduli which establish
the Legendre-Hadamard condition. The statement of the lemma is
proven.

\begin{bem}
This lemma does not generalize to non-isotropic materials.
\end{bem}
The reason for this lies in the following fact: suppose the system
is regular hyperbolic w.r.t. a pair $(X^\mu,\omega_\nu)$ where
$X^\mu$ is not parallel to the four-velocity. Then the hyperbolicity
conditions would imply the Legendre-Hadamard condition only on the
subspace, which is the intersection between the orthogonal spaces of
the four-velocity and the given $X^\mu$.

 To recover the full Legendre-Hadamard condition one would in addition
 require $0<U_{AMBN}  f^M,_\alpha X^\alpha  f^N,_\beta X^\beta m^A
 m^B$ to be true. But in general there is no way to obtain this
 inequality. One could of course try to derive a necessary condition
 on the elasticity operator which would result in the above inequality and thus would imply the statement of
 the lemma for such materials but this will not be done here.

We make a further comment on possible choices for $(X^\mu,
\omega_\nu)$ for isotropic materials:
\begin{lemma}
Let the material be isotropic. Then the system (\ref{stiegl1111}) is
regular hyperbolic w.r.t. the pair $(X^\mu,\omega_\nu)$ if and only
if the following inequalities
\begin{equation}
\left( g^{\mu\nu} + \left(   1-\frac{1}{\bar c^2} \right)u^\mu u^\nu
\right)\omega_\mu \omega_\nu <0
\end{equation}
as well as
\begin{equation}
\left( g_{\mu\nu} + \left(   1-\tilde c^2 \right)u_\mu u_\nu
\right)X^\mu X^\nu <0
\end{equation}
hold for $\bar c$ being the maximum of $c_1=\sqrt{2\mu+\lambda}$ and
$c_2=\sqrt{\mu}$ while $\tilde c$ is the minimum. For this lemma we
also assume that $1>\bar c \ge \tilde c$, i.e. that those constants
are bounded from above by the speed of light (the physical meaning
is that the sound waves travel with physical velocities). $u^\mu$ is
the reference four-velocity.
\end{lemma}
\begin{bem}
This lemma sheds some light on the geometric structure of the
problem. It tells us that $X^\mu$ as well as $\omega^\nu$ have to
lie inside certain sound-cones. Recall that we have found in the
last section (at least for the linearized setup) that the elastic
perturbations travel along these cones.
\end{bem}
To prove the lemma we first of all have to know the isotropic
principal symbol. In principle (that means beside positive
multiples) it is given by
\begin{equation}\label{isotropic principal symbol}
A(k)_{AB}m^Am^B=-m^2 (u^\mu k_\mu)^2+\left(   \mu m^2 \kappa^2  +
(\mu+\lambda ) (m\cdot \kappa )^2  \right)
\end{equation}
where $\kappa^A=f^A,_\mu k^\mu$ and contractions are taken w.r.t.
the reference strain. Without restriction of generality we assume
$m^2=1$.


Recall the definition: a system is regular hyperbolic w.r.t.
$(X^\mu, \omega_\nu)$ if and only if $A(\omega)_{AB}$ is negative
definite, while $A(l)_{AB}$ has to be positive definite for all
$l_\mu X^\mu=0$. By continuity small changes in the argument $k$ do
not change the positivity properties of $A(k)_{AB}$. A change of
these positivity properties would reflect itselve in the condition
$\det A(k)_{AB}=0$.

In general this determinant can be computed to be (see f.e.
\cite{bs1})
\begin{equation}
\det A(k)_{AB}=(\epsilon n)^3(g_{1}^{\mu\nu}k_\mu k_\nu )
(g_{2}^{\mu\nu}k_\mu k_\nu )
\end{equation}
where
\begin{equation}
g_{i}^{\mu\nu}= g^{\mu\nu} + \left(   1-\frac{1}{c_i^2} \right)u^\mu
u^\nu
\end{equation}
(later on we will use $\bar g^{\mu\nu}$ and $\tilde g^{\mu\nu}$ with
the same meaning) The determinant vanishes if and only if
$g_{1}^{\mu\nu}k_\mu k_\nu=0$ or $g_{2}^{\mu\nu}k_\mu k_\nu =0$.
Since the $g_i^{\mu\nu}$ are non-degenerate these conditions define
cones in the co-tangent space.

The principal part taken w.r.t. $-u_\mu$ is negative definite. Since
\begin{equation}
\left( g^{\mu\nu} +\left( 1-\frac{1}{c_i^2} \right) u^\mu u^\nu
\right)u_\mu u_\nu =-\frac{1}{c_i^2} <0
\end{equation}
the principal part is negative definite w.r.t. all $\omega_\nu$
satisfying the condition $\bar g^{\mu\nu}\omega_\mu \omega_\nu <0$.
Thus all co-vectors lying inside cone $\bar g^{\mu\nu}k_\mu k_\nu
=0$ are sub-characteristic.

The proof for $X^\mu$ is slightly more difficult: We need the
following result
\begin{proposition}
$\tilde g^{\mu\nu}k_\mu k_\nu >0 $ for all $k_\mu X^\mu=0$ if and
only if $\tilde g^{-1}_{\mu\nu}X^\mu X^\nu<0$
\end{proposition}
This proposition however is trivially true since $\tilde g_{\mu\nu}$
is a Lorentz-metric for $1>\tilde c^2$.

The proposition gives the second inequality of the lemma. At first
glance it might seem strange that one inequality involves $\tilde c$
while the other one contains $\bar c$. This comes from the fact,
that in one case we approach the co-sound-cone from above and in the
other case from below.

Note that (strictly speaking) up to now we have only proven the
"if"-part of the lemma. To complete the proof we simply note that
positive (respectively negative) definite operators have positive
(respectively negative) determinant. This observation gives the
"only if"-part and thus completes the proof of the lemma.

\begin{bem}
Note that the inner and outer cone interchange when viewed in
tangent and co-tangent space respectively. Thus $X^\mu$ lies in the
inner cone, while $\omega_\nu$ lies in the inner co-cone.
\end{bem}

\subsubsection{Rank two vs rank one positivity}

Usually people in the relativity community try to prove
well-posedness by casting the equations of motion into a symmetric
hyperbolic first order system. We will not give a definition (which
can e.g. be found in \cite{fima1}). Just note that there exists a
complete local-in space-theory for such systems. We could of course
try to apply this theory for the given setup as is done f.e. in
\cite{bs1}. But in general there may be a prize to pay for the
transition to the first order system. In the following we shall
briefly sketch, how this problem looks like and how it (at least in
the special case of an isotropic material) can be overcome:

If one prefers to cast the original equations into a first order
system, one would need to contract the principal part not only with
rank one objects but with general tensors with two indices to obtain
hyperbolicity and hence well-posedness. Hence the original
Legendre-Hadamard condition may in general fail to be sufficient,
since it only covers rank one positivity.

Note that the principal symbol and even the original second order system is unchanged if we exchange the elasticity tensor by
\begin{equation}\label{gaugetrafo}
U_{ABCD}\to \tilde U_{ABCD}:=U_{ABCD}+\Lambda_{ABCD}
\end{equation}
if the additional term has the following symmetries:
\begin{equation}\label{symm111}
\Lambda_{ABCD}=-\Lambda_{ADCB}=-\Lambda_{CBAD}=\Lambda_{CDAB}
\end{equation}
We will see (at least in the case of an isotropic material) that a
gauge transformation of the kind introduced in (\ref{gaugetrafo})
can be used to obtain hyperbolicity of the first order system by
just using the Legendre-Hadamard condition.

For an isotropic material we have seen that the elasticity operator
can be written as (omitting the upper tilde for the reference
strain)
\begin{equation}
4 U_{ABCD}= \epsilon \left( 2\mu h_{A(C}h_{D)B}+\lambda h_{AB}h_{CD}
\right)
\end{equation}
\begin{bem}
For convenience we will omit the factor $4$ on the left hand side as
well as the stored energy $\epsilon$ from now on for the rest of
this subsection.
\end{bem}
Since we do not assume the material to be homogeneous they need not
be constant and may still be functions on the body. $h_{AB}$ is the
inverse strain at the reference deformation. On flat background we
may use coordinates such that $h_{AB}=\delta_{AB}$ but we will stick
to this more general notation. We know that the Legendre-Hadamard
condition is equivalent to the following condition on the Lam\'e
moduli (compare section \ref{elasticity operator2222}):
\begin{equation}\label{LH2}
\mu >0 \ \ \ \ 2\mu +\lambda >0
\end{equation}
Our goal is to establish a stronger positivity result for the
transformed elasticity tensor $\tilde U_{ABCD}$ using these
conditions on the Lam\'e moduli. Namely we will show positivity of
\begin{equation}\label{lalala}
\tilde U_{ABCD}\phi^{AB}\phi^{CD}= U_{ABCD}\phi^{AB}\phi^{CD}+\Lambda_{ABCD}\phi^{AB}\phi^{CD}
\end{equation}
Splitting into its antisymmetric ($\omega^{AB}$) and symmetric but trace-free ($\chi^{AB}$) parts we obtain
\begin{equation}\label{splitting1}
\phi^{AB}=\omega^{AB}+\chi^{AB}+\frac{1}{3}\tr \phi h^{AB}
\end{equation}
where $\tr \phi =h_{AB}\phi^{AB}$ is the trace of $\phi^{AB}$ w.r.t. the reference strain. The first term in (\ref{lalala}) can then be written as
\begin{equation}
\left( \frac{2\mu}{3}+\lambda \right) (\tr\phi )^2 +2\mu \chi_{CB}\chi^{CB}
\end{equation}
The indices were lowered using the reference strain. This is
nonnegative if and only if $\mu>0$ and the bulk modulus $\kappa:=
3\lambda +2\mu >0$. \begin{bem} We note two things:
\begin{itemize}
\item those conditions are stronger than (\ref{LH2})
\item they do not include the antisymmetric part of $\phi^{AB}$
\end{itemize}
\end{bem}
We cannot use this expression to bound the contracted original elasticity tensor away from zero, since the left hand side vanishes for antisymmetric $\phi^{AB}$. Thus the gauge term $\Lambda_{ABCD}$ must give rise to antisymmetric terms thus breaking the symmetry (isotropy) of the setup.

The easiest and at the same time most natural choice for a gauge term is
\begin{equation}
\Lambda_{ABCD}=\alpha h_{A[B}h_{D]C}
\end{equation}
This term obeys the required symmetries as stated in (\ref{symm111}). The free function $\alpha$ will be specified in the following. Computing the second term in (\ref{lalala}) using this gauge term we obtain
\begin{equation}
\Lambda_{ABCD}\phi^{AB}\phi^{CD}= \alpha \left( \frac{2}{3}(\tr \phi )^2-\chi^{AB}\chi_{AB}+\omega^{AB}\omega_{AB} \right)
\end{equation}
Adding this to the terms from the original elasticity tensor gives
\begin{equation}
\tilde U(\phi,\phi)=\left( \frac{2}{3}(\mu+\alpha )+\lambda \right) (\tr \phi )^2 +(2\mu-\alpha )\chi^{AB}\chi_{AB}+\alpha \omega_{AB}\omega^{AB}
\end{equation}
If we would chose $\alpha=2\mu$ then we could use the
Legendre-Hadamard condition to bound the first and the last term but
the second term would vanish. This however would lead to problems as
indicated in the remark above. We thus need a more refined argument.
Assume the Legendre-Hadamard condition holds. Then by continuity
there exists a $\delta >0$ such that
\begin{eqnarray}
2\mu +\lambda - \frac{2}{3}\delta >0 \\
2\mu-\delta >0
\end{eqnarray}
Then we set $\alpha=2\mu-\delta$ and obtain
\begin{equation}
\tilde U(\phi,\phi)=\left(  2 \mu +\lambda - \frac{2}{3}\delta  \right) (\tr \phi )^2 + \delta \chi^{AB}\chi_{AB}+(2\mu-\delta ) \omega_{AB}\omega^{AB}
\end{equation}
Now all coefficients are positive by virtue of the Legendre-Hadamard condition and all summands are nonnegative.

We have shown how to construct a gauge term which leaves the second
order system unchanged but allows for the Legendre-Hadamard
condition to be sufficient for positivity and hence for
hyperbolicity in the context of the first order system.

\subsection{Material description}\label{fuertee3}

Now we turn to the material setup. In this section we will show that
the equations of motion derived from the material Lagrangian are
regular hyperbolic close to a natural reference state. This will
include well-posedness for the second order system.

We could also try to cast the system into a symmetric hyperbolic
system. Then in general stronger assumptions on the elasticity
operator (and thus on the material) would be needed. For isotropic
materials the last part of the previous section could be used to
overcome this discrepancy. But (besides this remark) we will not go
any deeper into the first order formulation, since the second order
formulation works out well.

As in the space-time setting we will find that there exists a
preferred pair $(X^\mu, \omega_\nu)$, with respect to which the
systems unfolds its hyperbolicity properties in the most natural
way. Finally the connection between this very choice and the pair
$(u^\mu, -u_\nu)$ used in the previous sections will be discussed.

The equations of motion were derived in the form (see section
\ref{meom1111} for the derivation)
\begin{equation}\label{puntigammer111}
A^{\mu\nu}_{ij}\partial_\mu \partial_\nu F^j =R_i
\end{equation}
This is a second order quasi-linear system for the deformation
$F^j$. As before $R_i$ denotes lower order terms.
\begin{bem}
Note that Greek indices are not space-time indices but rather denote
points of $\R \times \mathcal{B}$.
\end{bem}
It turns out to be most convenient not to use the principal part
directly to check the various positivity requirements but rather
work with
\begin{equation}
B^{\mu\nu}_{AB}:=A^{\mu\nu}_{ij}F^i,_A F^j,_B
\end{equation}
Since $F^i,_A$ is an isomorphism, this newly defined
$B^{\mu\nu}_{AB}$ has the very same positivity properties as the
original $A^{\mu\nu}_{ij}$ had. The advantage is that the new object
can be given in the more convenient form (see again section
\ref{meom1111})
\begin{equation}
B^{\mu\nu}_{AB}=\gamma^{-\frac{1}{2}}\left( -\epsilon h_{AB}U^\mu
U^\nu  +   4 U_{ACBD} \phi^{C\mu}\phi^{D\nu}\right)
+\Phi^{\mu\nu}_{AB}
\end{equation}
where $\Phi^{\mu\nu}_{AB}$ is linear homogeneous in the stress. We
have seen in section \ref{meom1111} how to obtain $U^\mu$ and
$\phi^{A\mu}$ for Gaussian coordinates on space-time. It is
straightforward to see that for general coordinates (i.e. for
non-vanishing lapse and shift) we obtain very similar expressions
\begin{eqnarray}
U^\mu = N^{-1}\gamma^{\frac{1}{2}}\delta^{\mu}_0 \\
\phi^{C\mu}=N^{-1}\delta^\mu_0 W^a f^C,_a + \delta^\mu_M h^{CM}
\end{eqnarray}
For the rest of this section we will stick to this more general
formulation. Note that $\gamma$ is thus again given by the more
general expression $\gamma=(1-W^2)^{-1}$

Then for a reference deformation the principal part reads
\begin{equation}
\tilde \gamma^{\frac{1}{2}}\tilde B^{\mu\nu}_{AB}=- \tilde \epsilon
\tilde h_{AB} \tilde U^\mu \tilde U^\nu  +   4 \tilde U_{ACBD}
\tilde\phi^{C\mu}\tilde\phi^{D\nu}
\end{equation}
We now try to find a pair $(X^\mu, \omega_\nu)$ w.r.t. which the
system is regular hyperbolic. Note that the principal part consists
of two portions. The first is non-positive, while the second is
non-negative once the Legendre-Hadamard condition is assumed for the
elasticity operator. Thus a good choice for each element of the pair
$(X^\mu, \omega_\nu)$ will annihilate one part of the principal
symbol. For the reference deformation this is in general given by
\begin{equation}
\tilde \gamma^{\frac{1}{2}} B(k)_{AB}=- \tilde \epsilon \tilde
h_{AB} (\tilde U^\mu k_\mu )^2 +4 \tilde U_{ACBD}
\tilde\phi^{C\mu}k_\mu \tilde \phi^{D\nu}k_\nu
\end{equation}
From the form of the principal symbol it is clear that the choice
for $X^\mu$ has to be
\begin{equation}
X^\mu=\tilde U^\mu
\end{equation}
because then for all $l_\mu X^\mu=0$ we have that
\begin{equation}
\tilde \gamma^{\frac{1}{2}} B(l)_{AB}=4 \tilde U_{ACBD} \tilde
\phi^{C\mu}l_\mu \tilde \phi^{D\nu}l_\nu
\end{equation}
which is positive by the Legendre-Hadamard condition unless
$\tilde\phi^{C\mu}l_\mu=0$. This however can be excluded since
\begin{equation}
\tilde\phi^{C\mu}l_\mu=\tilde h^{CM}l_M \ne 0
\end{equation}
which is non-zero by the non-degeneracy of the strain.

In order to obtain a negative principal part for $\omega_\nu$ we
have to assure that the part involving the elasticity operator does
not enter the symbol. This is guaranteed only by the following
condition:
\begin{equation}
\tilde\phi^{C\mu}\omega_\mu=0
\end{equation}
We conclude
\begin{equation}
\omega_M=- \tilde N^{-1} \omega_0 \tilde h_{MC} \tilde W^a \tilde
f^C,_a
\end{equation}
Thus for fixing $\omega_0$ through the (arbitrary) condition
$X^\mu\omega_\mu=1$ we obtain
\begin{equation}
\omega=\tilde N \tilde \gamma^{-\frac{1}{2}}\left( dt- \tilde N^{-1}
\tilde h_{MC} \tilde W^a \tilde f^C,_a dX^M \right)
\end{equation}
Together this gives the following result
\begin{lemma}\label{das andere lemma}
The system (\ref{puntigammer111}) is regular hyperbolic w.r.t. the
pair
\begin{equation}
(\tilde N^{-1}\tilde
\gamma^{\frac{1}{2}}\partial_t,\tilde N \tilde
\gamma^{-\frac{1}{2}}( dt- \tilde N^{-1} \tilde h_{MC} \tilde W^a
\tilde f^C,_a dX^M ) )
\end{equation}
 in a neighborhood of a natural
deformation.
\end{lemma}
\begin{bem}
Needless to say that the system is also regular hyperbolic w.r.t.
all pairs sufficiently close to the one given above. Nevertheless
the system (\ref{puntigammer111}) satisfies the conditions for
regular hyperbolicity in this optimal way only for the above choice.
\end{bem}
We can thus obtain the analogous result as in the previous section
stating that the system (\ref{puntigammer111}) is regular hyperbolic
near a natural deformation if and only if it is regular hyperbolic
w.r.t. the above choices for $X^\mu$ and $\omega_\nu$. The form of
the corresponding lemmas can be taken word by word from the last
section. Instead of spending too much time on this, we shall rather
clarify the connection between the above choice for $(X^\mu,
\omega_\nu)$ and the four-velocity.

For this purpose we make use of the transformation introduced in
section \ref{meom1111}, namely the diffeomorphism
\begin{equation}
(t,X)=(t,f(t,x)) \ \ \ \ \ \ (t,x)=(t,F(t,X))
\end{equation}
We have seen in the section mentioned above, that the material
velocity maps to $U^\mu$ which is a multiple of $\partial_t$. We
have also seen that
\begin{equation}
f^A,_\mu g^{\mu\nu} \mapsto \phi^{A\nu}
\end{equation}
From this one may guess that $-u_\mu \mapsto \omega_\mu$. To see
this we use the following result from section \ref{refmat1}:
\begin{equation}
(g_{\mu\nu})\mapsto (G_{\mu\nu}) =\left(\begin{array}{cc}-N^2(1-
W^2) & N g_{ij} W^i F^j,_A \\ N g_{ij} W^i F^j,_B   & g_{ij}F^i,_A
F^j,_B
\end{array}\right)
\end{equation}
From this it is easy to obtain that
\begin{equation}
u_\nu \mapsto U^\mu G_{\mu\nu}=N^{-1}
\gamma^{\frac{1}{2}}G_{0\nu}=-N\gamma^{-\frac{1}{2}}\delta^0_\nu +
\gamma^{\frac{1}{2}}g_{ij}W^i F^j,_A \delta^A_\nu
\end{equation}
To complete the argument we have to show that the spatial parts
coincide (The time-part already fits). This can be seen by recalling
the formula for the inverse strain derived in (\ref{minvstrain1}):
\begin{equation}
h_{MC}f^C,_aW^a=(g_{ij} + \gamma W_i W_j ) F^i,_M W^j = \gamma W_i
F^i,_M
\end{equation}
Using this calculation on the spatial art of $\omega$ we immediately
find that $u_\mu \mapsto - \omega_\mu$ thus proving the claim.
\begin{bem}
The above computation shows how the geometric importance of the
four-velocity $u^\mu$ carries over to the material description.
\end{bem}

From our observations up to now the following lemma connecting the
material and the space-time description is immediate:
\begin{lemma}
The material system close to a natural deformation is regular
hyperbolic w.r.t. $(X^\mu,\omega_\nu)$ as defined above if and only
if the system in the space-time picture is regular hyperbolic w.r.t.
$(\tilde u^\mu, -\tilde u_\nu)$ where $\tilde u^\mu$ is the natural
normalized four-velocity. Moreover $(X^\mu,\omega_\nu)$ and $(\tilde
u^\mu, -\tilde u_\nu)$ are connected via the diffeomorphism
introduced in section \ref{refmat1} as indicated above.
\end{lemma}
\begin{bem}
To prove this one just has to use that from regular hyperbolicity
w.r.t. one of the above pairs the Legendre-Hadamard condition
follows, which gives the hyperbolicity in the other description.
\end{bem}
We carry on with a local existence result. As in the space-time
description we can use the basic theorem given in section
\ref{loceubd11111} to obtain local well-posedness:
\begin{theorem}
The Cauchy problem for (\ref{puntigammer111}) is well-posed in the
following sense: Given a natural deformation. Then there exists a
(local) foliation of $\R \times \mathcal{B}$, such that for initial
data $(F^i_0, F^j_1)$ close to the ones induced by the natural
deformation on the leafs of the foliation in the $H^{s+1}\times H^s
$ topology ($s\ge 3$) there exists a unique regular (i.e. at least
$C^2$) local (in space and time) solution which depends continuously
on the initial data.
\end{theorem}
\begin{bem}
The actual function spaces involved can be taken from the original
theorem from section \ref{loceubd11111}.
\end{bem}
\begin{bem}
The reason why we only get local in space solutions in general lies
in the fact that $u_\mu$ (and hence $\omega_\mu$) is not closed in
general.
\end{bem}
There is another natural choice for the pair $(X^\mu,\omega_\nu)$,
given by the foliation $\R\times \mathcal{B}$. This pair is given by
$(\partial_t, dt)$. In the following lemma we will see how to deal
with this setup (at least for isotropic materials)
\begin{lemma}\label{exlicittdt}
Let the material be isotropic w.r.t a natural deformation. Then the
system (\ref{puntigammer111}) evaluated in a neighborhood of the
natural deformation is regular hyperbolic w.r.t $(\partial_t, dt)$
if and only if the characteristic speeds $c_1^2 :=2\mu +\lambda $
and $c_2^2 = \mu $ are bounded from above in the following way:
$W^{-2}>c_1^2>0$ as well as $W^{-2}>c_2^2>0$.
\end{lemma}
The following corollary follows immediately from the $1 > W^2 \ge
0$:
\begin{corollary}
The isotropic system is regular hyperbolic w.r.t. $(\partial_t, dt)$
close to the natural deformation if the characteristic speeds $c_i$
are bounded by the speed of light, i.e. $1>c_i>0$.
\end{corollary}
\begin{bem}
It is interesting that due to the lemma one may obtain regular
hyperbolicity w.r.t. $(\partial_t , dt)$ even if $c_1\ge 1$.
\end{bem}
\begin{bem}
An analogous result can be obtained for the space-time setup for a
suitable foliation.
\end{bem}
For the proof of the lemma we note that the Legendre-Hadamard
condition guarantees that the principal symbol is positive definite
w.r.t. all $l_\mu$ satisfying $l_0=0$. This has already been
mentioned in the proof of lemma \ref{das andere lemma}. The new part
consists in showing that $dt$ is sub-characteristic. We have to show
that the principal symbol (for convenience we omit the tilde) taken
w.r.t. $dt$ is negative definite, i.e.
\begin{equation}
N^2 \gamma^{\frac{1}{2}} B(k)_{AB}=-  \epsilon  h_{AB} \gamma +4
U_{ACBD}f^C,_c W^c f^D,_d W^d <0
\end{equation}
Since the coefficient on the left hand side is positive (recall that
$1\ge \gamma >0$) we only have to care about the right hand side.
Using the expression for the isotropic elasticity operator, the
question comes down to
\begin{equation}
- m^2\gamma + \left( (\mu +\lambda) (m\cdot w) ^2 + \mu m^2 w^2
\right) <0
\end{equation}
for non-vanishing $m^A$ where products were taken w.r.t. the strain
$h_{AB}$, while $w^A:=f^A,_aW^a$. The coefficient in front of
$(m\cdot w) ^2$ lacks a clear sign (the Legendre-Hadamard condition
only leads to the conditions given in the lemma). We thus first have
to rewrite the above expression. This is done by splitting $m^A$
into a part parallel to $w^A$ and an orthogonal rest:
\begin{equation}
m^A= \frac{m\cdot w}{w^2} w^A + m_{\perp}^A
\end{equation}
This leads to
\begin{equation}
m^2 = \frac{(m \cdot w)^2}{w^2} + m_{\perp}^2
\end{equation}
Thus besides positive multiples the principal symbol twice
contracted with $m^A$ is given through
\begin{equation}
- \left( m_{\perp}^2 + \frac{(m\cdot w)^2}{w^2} \right) \gamma +  (2
\mu +\lambda) (m\cdot w) ^2 + \mu m_{\perp}^2 w^2
\end{equation}
now the Legendre-Hadamard condition can be used to guarantee
positivity of the coefficients. Since $m_{\perp}^2$ and $(m\cdot
w)^2$ can be chosen independently, the symbol is negative definite
if and only if the following two inequalities hold:
\begin{eqnarray}
-m_{\perp}^2 \gamma +  \mu m_{\perp}^2 w^2 <0  \\
-(m\cdot w)^2 \gamma + (2 \mu +\lambda) (m\cdot w)^2 w^2 <0
\end{eqnarray}
Assuming the non-trivial case, those two inequalities are
essentially of the type
\begin{equation}
-\gamma + c^2 w^2 <0
\end{equation}
It remains to compute $w^2=W^a W^b f^A,_a f^B,_b h_{AB}$ and compare
the result to $\gamma$. From the formula for the inverse strain
given by equation (\ref{minvstrain1}) we obtain
\begin{equation}
w^2=\gamma W^2
\end{equation}
Thus the inequality we need becomes
\begin{equation}
-1+c^2 W^2 <0
\end{equation}
which is the claimed inequality.

\begin{bem}
Note that we can use the lemma to establish a local existence result
on all of $\mathcal{B}$ getting rid of the locality in space. The
same argument works for the space-time description. There we can use
it to obtain valid data on all of $t=const$.
\end{bem}

\section{Gravito-elastodynamics}\label{eisauch}
Once the initial value problem for the unbounded setup on a fixed
background is shown to be well-posed, there are two natural ways to
proceed. One can consider finite material bodies or one can
investigate unbounded self-gravitating elastic materials. While the
former task will be undertaken in the following section
\ref{teeistgut} we shall concentrate here on the latter challenge.
We will show that the initial value problem for a self-gravitating
unbounded elastic medium is well-posed under certain (mild)
restrictions on the material, already well-known to us from the
previous sections (namely the Legendre-Hadamard condition).

We will first discuss the initial value problem for the Einstein
field equations. Following the exposition of Friedrich and Rendall
\cite{frra1} we will use the harmonic gauge to cast them into a
quasi-linear wave equation, which is then shown to be regular
hyperbolic. Then we will discuss issues like gauge propagation and
proper initial data.

Then in the second part we will show that the coupled system
consisting of the Einstein(-matter field) equations (in harmonic
gauge) and the elastic equations of motion in space-time description
is regular hyperbolic and discuss how the arguments given for the
un-coupled case carry over.

\subsection{Einstein field equations}

Let space-time be denoted by $(M,g_{\mu\nu})$. The Einstein field
equations determine the space-time geometry (encoded in the metric
$g_{\mu\nu}$) through the space-time entities (matter). They are
obtained from the action principle ($L$ denotes the matter
Lagrangian)
\begin{equation}\label{eisaction}
\int_{M} ( R\sqrt{-\det g_{\mu\nu}}  +   2 \kappa L ) d^4 x
\end{equation}
and read
\begin{equation}\label{efe2}
G_{\mu\nu}=\kappa T_{\mu\nu}
\end{equation}
where $G_{\mu\nu}$ is the Einstein tensor formed from the Ricci
tensor $R_{\mu\nu}$, which depends on the metric and its first and
second derivatives while $T_{\mu\nu}$ is the energy-momentum tensor
while the constant $\kappa=\frac{8\pi G}{c^4}$ ($G$ being the
gravitational constant). An equivalent version reads
\begin{equation}\label{efe}
R_{\mu\nu}=\kappa \left( T_{\mu\nu}- \frac{1}{2}
T_{\alpha\beta}g^{\alpha\beta} g_{\mu\nu} \right)=:M_{\mu\nu}
\end{equation}
For vacuum the energy-momentum tensor vanishes and the Einstein
field equations simplify to
\begin{equation}
R_{\mu\nu}=0
\end{equation}
\begin{bem}
For the moment we do not specify a particular matter model. In this
spirit we do not give an explicit expression for the energy-momentum
tensor. We rather assume it to be a functional of the matter fields
to keep this discussion on a general level. Later on it will of
course be the one derived for elastic matter (\ref{stemt111}). It is
however important to assume that it does not contribute to the
principal part.
\end{bem}
When written in terms of the metric equation (\ref{efe}) reads
\begin{equation}\label{eisevfe}
 -\frac{1}{2}g^{\lambda\rho}g_{\mu \nu,\lambda\rho}+\nabla_{(\mu}\Gamma_{\nu)}+\Gamma^{\eta}_{\lambda\mu}g_{\eta\delta}g^{\lambda\rho}\Gamma^{\delta}_{\rho\nu}+2\Gamma^{\lambda}_{\delta\eta } g^{\delta\rho }g_{\lambda (\mu}\Gamma^{\eta}_{\nu )
 \rho}=M_{\mu\nu}
\end{equation}
where the contracted Christoffel symbols are defined by
$\Gamma^\mu:=\Gamma^\mu_{\alpha\beta} g^{\alpha\beta}$. Since the
Christoffel symbols involve first derivatives of the metric, the
principal part is given by
\begin{equation}\label{eisevpp}
P_{\mu\nu}^{\alpha\beta\gamma\delta}\partial_\gamma\partial_\delta
g_{\alpha\beta} :=\frac{1}{2}g^{\lambda\rho}g_{\mu
\nu,\lambda\rho}+\partial_{(\mu}\Gamma_{\nu)}
\end{equation}
\begin{bem}
We could exchange the covariant derivative by an ordinary one,
because the difference is of lower order in the metric.
\end{bem}
Note that this is a quasi-linear expression, since the inverse
metric depends on the metric itself but not on higher order terms.
One can show (see e.g. \cite{frra1}) that every hyper-surface is
characteristic in the sense that for every choice of $k_{\mu}$ the
characteristic polynomial (see chapter \ref{quasisec} for the
definitions) vanishes:
\begin{equation}
P(k)=0
\end{equation}
This in particular implies that there exists neither a
sub-characteristic co-vector nor can the equation in the form
(\ref{eisevfe}) be hyperbolic in the sense of any of our
definitions. This defect can however be shown to be a result of the
diffeomorphism invariance (i.e. of coordinate freedom) of the field
equations.

In order to obtain a reasonable system of differential equations we
have to make a suitable choice of coordinates. The form of the
principal part (\ref{eisevpp}) suggest that a good coordinate system
would allow us to get rid of the term $\partial_{(\mu}
\Gamma_{\nu)}$, since we then would be left with a wave-equation.
This is possible, since the Christoffel symbols are no tensorial
objects. We can thus hope to find coordinates for which
\begin{equation}
\Gamma_\mu =0
\end{equation}
More precisely we have the following:
\begin{lemma}
Given a space-like hyper-surface. Let $x^a$ be coordinates in some
open neighborhood $U$ in this hyper-surface. Then there exists
coordinates $y^\mu$ on some neighborhood of $U$ in $M$ with the
property $(y^\mu)|_U = (0,x^i)$ such that in these new coordinates
\begin{equation}\label{eisgamma}
\Gamma^\mu=0
\end{equation}
\end{lemma}
\begin{definition}
The coordinates introduced in the fore-going lemma are called
"harmonic coordinates".
\end{definition}
We can prove the lemma by rewriting (\ref{eisgamma}) (we construct
harmonic coordinates):
\begin{equation}
\Box_x y^\mu (x)=\Box_y y^\mu= -\Gamma^\mu=0
\end{equation}
where the second Christoffel symbol corresponds to the new
coordinate system $(y^\mu)$. The first equality holds since
coordinates are scalars on space-time. Using proper initial
conditions one can obtain the solution claimed in the lemma (a way
of seeing this is by application of the theorem in section
\ref{loceubd11111}).

In harmonic coordinates (which will from now on be denoted by
$x^\mu$) the principal part (\ref{eisevpp}) becomes
\begin{equation}\label{eisharmonicpp}
A_{\mu\nu}^{\alpha\beta\gamma\delta}\partial_\gamma\partial_\delta
g_{\alpha\beta} :=\frac{1}{2}g^{\lambda\rho}g_{\mu \nu,\lambda\rho}
\end{equation}
At the given stage it is easy to show that the principal part (after
raising the indices $(\mu\nu)$ in the above equation using the
metric) in harmonic coordinates is regular hyperbolic:
\begin{lemma}\label{eislemma1}
The principal part (\ref{eisharmonicpp}) is regular hyperbolic
w.r.t. every pair $(X^\mu, \omega_\nu)$ where both $X^\mu$ and
$\omega_\nu$ are time-like and satisfy $X^\mu\omega_\mu >0$.
\end{lemma}
Given proper initial data we can thus apply the existence theorem
from section \ref{loceubd11111} to obtain local well-posedness for
the Einstein equations in harmonic gauge (assuming proper
well-posedness of the matter equations).
\begin{bem}
The procedure of reducing the Einstein field equations to a
hyperbolic system by a suitable choice of coordinates is called
"hyperbolic reduction".
\end{bem}
The story however does not end at this point. Two (three) questions
remain open:
\begin{itemize}
\item What are proper initial data? The related question: Are there any constraints or can the data be
given freely? If there are constraints, do they propagate?
\item Does the gauge condition hold off the initial surface? (Is it
preserved under the evolution?)
\end{itemize}
These question will be answered in separate subsections.

\subsubsection{Initial data and constraints}

One may at first guess that the metric and a suitable transversal
derivative given on a suitable initial surface play the role of
proper initial data. This however is not the case. We will see that
an initial data set consist of a three-dimensional Riemannian
manifold $(S, g_{ij})$, which in addition is equipped with a
symmetric covariant two-tensor $k_{ij}$ plus a diffeomorphism
identifying $S$ with a space-like hyper-surface in $M=\R \times S$.
Then $g_{ij}$ plays the role of first fundamental form (induced
metric) on this initial hyper-surface, while $k_{ij}$ is the second
fundamental form (external curvature). If matter is present, the
initial data set has to be extended to include the matter data.

While lapse and shift can be prescribed freely (each choice
corresponds to a certain coordinate choice on $M$), $g_{ij}$ and
$k_{ij}$ are subjected to certain constraint equations. From these
"constrained" initial data one can construct initial data for
(\ref{eisevfe}).

To actually see this we have to carry out a 3+1 decomposition of the
Einstein field equations (\ref{efe}). Assuming a foliation of
space-time by space-like hyper-surfaces $\{ t=const \}$ we write the
metric in its ADM representation
\begin{equation}
g_{\mu\nu}dx^\mu dx^\nu =-N^2dt^2 +g_{ij} (dx^{i} +Y^i dt) (dx^j +
Y^j dt)
\end{equation}
Then the projector onto the tangent space of the hyper-surface $\{
t=const \}$ is $h^{\mu\nu}= g^{\mu\nu} + n^\mu n^\nu$ where the
unit-normal $n^\mu$ is given by $n^\mu=N^{-1}(
\delta^\mu_0-\delta^\mu_i Y^i )$.

Then $g_{ij}$ is the induced metric on $\{ t=const \}$, while the
second fundamental form is given by
\begin{equation}\label{2.ff}
k_{ij}=\frac{1}{2N }\left( \partial_t   - \mathcal{L}_{Y}
\right)g_{ij}
\end{equation}
One finds that certain components of the Einstein equations
(\ref{efe2}) beside material quantities only depend on the first and
second fundamental form and do not involve higher order transversal
derivatives of the metric. First note the following trivial
consequences of (\ref{efe2}):
\begin{eqnarray}
0= n^\mu n^\nu \left( G_{\mu\nu} -\kappa T_{\mu\nu} \right) \\
0= n^\mu h^\nu_\lambda\left( G_{\mu\nu} -\kappa T_{\mu\nu} \right)
\end{eqnarray}
The interesting point about these equations is that they can be
written as
\begin{eqnarray}\label{c1}
0&=&r-k_{ab}k^{ab} + k^2 - 2\kappa j \\
\label{c2} 0&=& 2D_{[a}k_{b]}^a + \kappa j_b
\end{eqnarray}
where
\begin{equation}
T^{\mu\nu}n_\mu n_\nu=:j  \ \ \ \ \ T^{\mu\nu}n_\mu
h_{\nu\alpha}=-j_\alpha   \ \ \ \ \ \ k=k_{ij}g^{ij}
\end{equation}
Here $r$ denotes the Ricci scalar corresponding to the induced
metric (one must use the Gauss and the Codazzi equation to arrive at
the stated result. See \cite{eisenhart} for details) while $D_a$ is
the covariant derivative associated with $g_{ij}$.

Note that equations (\ref{c1},\ref{c2}) form an under-determined
system for first and second fundamental form, once the material
quantities are known. In this sense (\ref{c1},\ref{c2}) constitute
necessary restrictions on possible initial data and are therefore
called constraint equations. If the coupled system is considered,
these equations also constrain the material initial data.

It remains to show how to construct the initial data $g_{\mu\nu}$
and $\partial_t g_{\mu\nu}$ for (\ref{eisevfe}) on $\{ t=0 \}$ once
$(g_{ij}, k_{ij})$ (or rather the corresponding objects on the
initial hyper-surface) are given.

From the 3+1 decomposition of the Christoffel symbols one can derive
the following equations for lapse and shift (see \cite{frra1} for
the derivation)
\begin{eqnarray}
\label{eisbier1}\partial_t N - N,_i Y^i &=& N^2 (k - n_\mu \Gamma^\mu ) \\
\label{eisbier2}\partial_t Y^i - Y^i,_j Y^j &=& N^2 (\gamma^i -
D^i\ln N - h^i_\nu \Gamma^\nu )
\end{eqnarray}
Here $\gamma^i$ denote the contracted Christoffel symbols
corresponding to the induced metric.

Given lapse and shift and assuming harmonic gauge, the above
equations give the time derivatives of lapse and shift in terms of
$(g_{ij}, k_{ij}, N, Y^i)$. What we still lack is an equation for
$\partial_t g_{ij}$ in terms of the data. For this we use the
definition of the second fundamental form (\ref{2.ff}). We can
rewrite it in the form ($\mathcal{L}_v$ denotes the Lie-derivative
w.r.t. the vector field $v^i$)
\begin{equation}
\label{eisbier3} \partial_t g_{ij}= 2 N k_{ij} +
\mathcal{L}_{Y}g_{ij}
\end{equation}
This equation gives the sought time-derivative of the induced
metric.

We sum up: Given initial data $(S, g_{ij}, k_{ij})$ (we suppress the
material data for the moment) which obey the constraints
(\ref{c1},\ref{c2}). These data induce corresponding objects on the
initial hyper-surface, which is the isomorphic embedding of
$(S,g_{ij})$ into space-time. Then for any sufficiently smooth
choice of lapse and shift (this choice is pretty arbitrary, it just
fixes the time flow off the initial surface) we can construct
$g_{\mu\nu}$ on the initial surface. The components of $\partial_t
g_{\mu\nu}$ can be obtained using equations
(\ref{eisbier1},\ref{eisbier2},\ref{eisbier3}) under the restriction
$\Gamma^\mu =0$ (coming from the requirement for harmonic
coordinates) as indicated above.

We have seen that the relevant initial data are given by $(S,
g_{ij}, k_{ij})$ together with an isometric embedding into
space-time. It has been shown how these data can be used to obtain
suitable data for the problem as formulated in (\ref{eisevfe}).

\subsubsection{Propagation of gauge}
Assume we have proper initial data on a hyper-surface, such that the
theorem from section \ref{loceubd11111} provides us with local
existence and uniqueness. (The accurate form of this statement will
be presented in the last section.) Then we still have to show that
this solution (obtained from the reduced system) actually is a
solution to the full Einstein field equation (\ref{efe}). This is
done by showing that if the contracted Christoffel symbols vanish
initially, then they also vanish off the initial surface.

For the solution of the reduced system $g_{\mu\nu}$ we construct the
un-contracted and contracted Christoffel symbols (which for the
moment we do not assume to vanish). For these symbols we derive the
Einstein tensor $G_{\mu\nu}$. Subtracting the energy-momentum tensor
and the reduced field equations (which hold due to the assumptions)
leads to ($\Gamma^\mu$ is assumed to be computed from $g_{\mu\nu}$)
\begin{equation}\nonumber
G_{\mu\nu}-\kappa T_{\mu\nu}=G_{\mu\nu}-\kappa T_{\mu\nu}
-(\mbox{reduced equations})=
\nabla_{(\mu}\Gamma_{\nu)}-\frac{1}{2}g_{\mu\nu}\nabla_\alpha\Gamma^\alpha
\end{equation}
(The second term arises when we express
$g^{\alpha\beta}g_{\mu\nu},_{\alpha\beta}$ in terms of covariant
derivatives.) Now we employ the contracted Bianchi identity (see
f.e. \cite{wald} or \cite{urbi1}), which hold for purely geometrical
reasons independently of the field equations:
\begin{equation}
\nabla^\mu G_{\mu\nu}=0
\end{equation}
If the energy-momentum tensor $T_{\mu\nu}$ is divergence-free (which
is the case for elastic materials according to lemma
\ref{blabla345}) this leads to a homogeneous wave equation for the
contracted Christoffel symbols:
\begin{equation}\label{gaugeprop1}
\Box \Gamma_\mu +R^\nu_\mu \Gamma_\nu =0
\end{equation}
If we can show that $\Gamma_\mu $ and $\partial_t \Gamma_\mu $ both
vanish on the initial surface, we may employ the basic existence and
uniqueness result from section \ref{loceubd11111}: obviously
$\Gamma_\mu =0$ is a solution for trivial initial data. The
uniqueness part then assures that this is indeed the only solution.
This means that $g_{\mu\nu}$ found from this harmonic coordinate
ansatz is a solution to the Einstein equations.

To see that the initial data for (\ref{gaugeprop1}) are indeed
trivial, we recall equations (\ref{eisbier1},\ref{eisbier2}). We
used the harmonic gauge to obtain $\partial_t N$ and $\partial_t
Y^i$ in terms of the given data. Reverting this argument implies
that the initial data have been chosen in a way such that
$\Gamma_\mu $ vanishes initially.

The argument for $\partial_t \Gamma_\mu$ works in a similar way, but
is more complicated, since the expression under consideration is
second order in the metric. By this it is not sufficient anymore to
argue in terms of initial data only. We have to make use of the
field equations. These enter in the form introduced in the beginning
of this sub-section. Contracting with the unit-normal $n^\mu$ leads
to
\begin{equation}
n^\nu ( G_{\mu\nu}-\kappa T_{\mu\nu} ) = n^\nu \left(
\nabla_{(\mu}\Gamma_{\nu)}-\frac{1}{2}g_{\mu\nu}\nabla_\alpha\Gamma^\alpha
\right)
\end{equation}
Recall that the left hand side only involves initial data when
evaluated on the initial surface (this was shown in the last
sub-section). But the initial data satisfy the constraint equations,
so that we can conclude
\begin{equation}
0= n^\nu \nabla_{(\mu}\Gamma_{\nu)}-\frac{1}{2}n_\mu
\nabla_\alpha\Gamma^\alpha
\end{equation}
Since we already know that $\Gamma_\mu$ vanishes on the initial
surface this can be written as (we stress that this equality only
holds on the initial surface as do the following equations. We
include no statements off the initial surface.)
\begin{equation}
0= n^\nu \partial_{(\mu}\Gamma_{\nu)}-\frac{1}{2}n_\mu
\partial_\alpha\Gamma^\alpha
\end{equation}
Note that we can omit spatial derivatives of the contracted
Christoffel symbols, since the latter vanish identically on all of
$t=0$. It is then a simple matter of inserting the definitions of
the unit-normal together with the ADM-representation of the metric
to obtain the desired result. We can conclude that $\partial_t
\Gamma_\mu$ has to vanish on the initial surface.

Thus we have found that the data for (\ref{gaugeprop1}) are trivial
so that the only solution is trivial as well. We conclude that the
metric obtained from the reduced equations is indeed a solution to
the Einstein equations.


\begin{bem}
Note that the only restriction on the matter model was the
divergence-less-ness of the energy-momentum tensor. For elastic
materials which satisfy the equations of motion this condition
holds. This is true independently of the equations of motion for the
metric.
\end{bem}

\subsubsection{Propagation of constraints}
We have seen in the previous section that the constraints
(\ref{c1},\ref{c2}) naturally arise in the context of 3+1
decomposition. If we would have used the Einstein equations in their
ADM-version (i.e. as evolution equation for the induced metric), we
would have to assure that the constraints hold off the initial
surface.

In the given context however this requires no further work, since a
solution to the initial value problem connected to the equation
(\ref{eisevfe}) gives rise to a solution $g_{\mu\nu}$ of the full
Einstein equations (\ref{efe}). Thus at each instant of time the
constraints are trivially satisfied and require no further
argumentation.

This property still holds once matter fields (and the corresponding
equations of motion) are added to the system.

\subsubsection{Hyperbolicity for the ADM variables in harmonic
gauge}\label{bringtdochwas?}

Although this is somehow "off-route" we will show how the harmonic
gauge can be used to obtain a well-posed system for the
ADM-variables. Therefore we first define
\begin{equation}
g_{\mu\nu}=:Z_{\mu\nu}(N ,Y^{k},g_{ij})
\end{equation}
and introduce
\begin{equation}
(\phi^{A}):=(N,Y^{i},g_{jk})
\end{equation}
In this notation we have
\begin{equation}
g_{\mu\nu}=Z_{\mu\nu}(\phi^{A})
\end{equation}
This gives for the derivatives
\begin{equation}
g_{\mu\nu,\rho}=Z_{\mu\nu,A}\phi^{A}_{,\rho}
\end{equation}
and
\begin{equation}
g_{\mu\nu,\rho\tau}=Z_{\mu\nu,AB}\phi^{A}_{,\rho}\phi^{B}_{,\tau}+Z_{\mu\nu,A}\phi^{A}_{,\rho\tau}
\end{equation}
so the new principal part derived from (\ref{eisharmonicpp}) is
given by
\begin{equation}\label{ppneu}
Z^{\rho\tau}Z_{\mu\nu,A}\phi^{A}_{,\rho\tau}.
\end{equation}
Note that the mapping $Z:\phi^{A}\mapsto Z_{\mu\nu}(\phi^{A})$ is a
diffeomorphism, so that $Z_{\mu\nu},_{A}$ is a linear isomorphism.

The symbol corresponding to (\ref{ppneu}) reads
\begin{equation}\label{3}
k^2 Z_{\mu\nu,A}
\end{equation}
In order to apply theorem1 we have to modify this. The most natural
way is by use of $Z_{\mu\nu ,A}$. For this purpose we first raise
the free Greek indices in (\ref{3}) by a non-degenerated
contra-variant two tensor $G^{\mu \mu '}$, which we will be specify
later. It may depend on the ADM variables and their first but not on
their second derivatives. The new symbol reads
\begin{equation}\label{neupp}
k^2 Z_{\mu '\nu '},_{B}G^{\mu \mu '} G^{\nu \nu '}Z_{\mu\nu,A}
\end{equation}
In order to apply our definition of regular hyperbolicity we have to
modify this. The most natural way is by use of $Z_{\mu\nu ,A}$. For
this purpose we first raise the free Greek indices in (\ref{3}) by a
non-degenerated contra-variant two tensor $G^{\mu \mu '}$, which we
will be specify later. It may depend on the ADM variables and their
first but not on their second derivatives. The new symbol reads
\begin{equation}\label{neupp}
k^2 Z_{\mu '\nu '},_{B}G^{\mu \mu '} G^{\nu \nu '}Z_{\mu\nu,A}
\end{equation}
The proper choice of $G^{\mu\nu}$ is crucial for the whole argument to work. 
A natural candidate is the following:
\begin{equation}
G^{\mu\nu}=Z^{\mu\nu}+2n^{\mu}n^{\nu}
\end{equation}
where $n^{\mu}=N^{-1}(-\delta^{\mu}_{0}+Y^{i}\delta^{\mu}_{i})$ is
the unit normal to the $t=const$ hyper-surface. If we contract
(\ref{neupp}) with $\delta \phi^{A}\delta\phi^{B}$, we obtain
\begin{equation}
k^2N^{-4}\left( 4N^2 (\delta N)^2 +2 g_{ij}\delta Y^{i} \delta Y^{j}
+\delta g_{ik}\delta g_{jl} g^{jk}g^{il} \right)
\end{equation}
Since the terms inside the bracket is positive, the symbol is
positive (negative) definite for space-like (time-like) choice of
$k_{\mu}$. Therefore the symbol is regular hyperbolic with respect
to every pair $(X^{\mu},\omega_{\nu})$ where both $X^\mu$ and
$\omega^\nu$ are time-like and satisfy $X^{\mu}\omega_{\mu} >0$.

\subsection{Coupled}

Here we will collect the partial results from the fore-going
sections to add them up into a local well-posedness result for the
equations of gravito-elasto-dynamics. We know that the Einstein
equations in harmonic coordinates form a regular hyperbolic second
order system. We also know that the solution of this system is
indeed a solution to the Einstein field equations (\ref{efe}) if the
energy-momentum tensor is divergence-free (which in our case is
trivial since the latter condition is equivalent to the matter
equations of motion, see lemma \ref{blabla345}). Finally we know
which restrictions (namely the constraints) have to be imposed on
the initial data and that these constraints propagate in the proper
sense.

We know from section \ref{uhrzeit} that equations governing the
dynamics of an elastic material on a fixed background are regular
hyperbolic close to a natural reference state (i.e. a stress-free
state, which satisfies the Legendre-Hadamard condition).

What we lack up to now is knowledge on the coupled system. For this
purpose we take a closer look at the action describing the coupled
scenario. For a self-gravitating elastic materials we rewrite
(\ref{eisaction}) as (one could include a coupling constant in front
of one of the two terms, but we set it equal to one, since we can
f.e. absorb it into the stored energy)
\begin{equation}
\int_M \left(   R + n \epsilon    \right) \sqrt{-\det g_{\mu\nu}}d^4
x
\end{equation}
At this stage we point out the following important detail: the
elastic Lagrangian does not depend on derivatives of the metric
while the Ricci scalar is independent of the configuration at all.
Thus the principal part decouples and in a harmonic coordinate
system we find that we can write the system as
\begin{equation}\label{cpp1}
\left( \begin{array}{cc} g^{\mu\nu}\delta^{\lambda '}_{\lambda}\delta^{\rho '}_{\rho} &  0  \\
0 &  A^{\mu\nu}_{IJ}  \end{array}
\right)\partial_{\mu}\partial_{\nu}\left( \begin{array}{c}
g_{\lambda '\rho '}  \\ f^{J} \end{array}\right)=\mbox{l.o.t.}
\end{equation}
where
\begin{equation}
A^{\mu\nu}_{IJ}= \frac{\partial ^2 n\epsilon}{\partial f^I,_\mu
\partial f^J,_\nu}
\end{equation}
is the principal part for the material equations as given in
(\ref{stressedpp2}) (We have divided the material principal part by
$\sqrt{-\det g_{\mu\nu}}$). As already indicated previous to lemma
\ref{eislemma1} it is more convenient to raise the indices in the
portion of the principal part involving the second order derivatives
of the metric. (There appear extra terms, but they are omitted here
because they are at most of first order.) The new principal part
reads
\begin{equation}\label{pphirter}
\left( \begin{array}{cc}  g^{\mu\nu}\delta_{\lambda '}^{\lambda}\delta_{\rho '}^{\rho} &  0  \\
0 &   A^{\mu\nu}_{IJ}
\end{array} \right)\partial_{\mu}\partial_{\nu}\left(
\begin{array}{c} g^{\lambda '\rho '}  \\ f^{J} \end{array}\right)
\end{equation}
Next we recall the material principal part for a reference
configuration (recall that all quantities corresponding to the
reference configuration $\tilde f^A$ were labeled by an upper tilde;
see (\ref{fuereisesser}) for the principal part)
\begin{equation}
\tilde A^{\mu\nu}_{IJ}=-\tilde n \tilde \epsilon \tilde h_{AB}\tilde
u^\mu \tilde u^\nu + 4\tilde n \tilde U_{AMBN}\tilde f^M,_\alpha
g^{\alpha\mu} \tilde f^N,_\beta g^{\beta\nu}
\end{equation}
We know from the section \ref{uhrzeit} that this is regular
hyperbolic w.r.t the pair $(\tilde u^\mu, - \tilde u_\nu)$. Since
the four-velocity $\tilde u^\mu$ is time-like by construction, we
find that the principal part (\ref{pphirter}) evaluated at a natural
configuration is regular hyperbolic w.r.t. $(\tilde u^\mu, - \tilde
u_\nu)$. This is due to the diagonal form of the principal part
which itself is a result of the uncoupling of the equations in the
highest order. By continuity (the stored energy is assumed to be
sufficiently smooth) the same is then true for all configurations
sufficiently close. We thus have the following statement:
\begin{lemma}
The principal part (\ref{pphirter}) is regular hyperbolic w.r.t. the
pair $(u^\mu, - u_\nu)$ for all pairs $(g_{\mu\nu}, f^A)$ where
$f^A$ is close enough to a natural configuration.
\end{lemma}
\begin{bem}
Note that the definition of natural configuration did involve the
Legendre-Hadamard condition.
\end{bem}
We can thus apply the existence theorem (or the corollary, depending
on whether $\tilde u_\mu$ is closed) from section \ref{loceubd11111}
to obtain local well-posedness.

There remains just one small obstacle regarding proper initial data:
we have to identify the material quantities used in the constraint
equations (\ref{c1},\ref{c2}) in terms of the configuration. For
this we have to chose a proper initial surface. Let us for
convenience assume that $\tilde u_{\mu}$ is closed.
\begin{bem}
Carefully note that this very last assumption implies the existence
of a reference metric $\tilde g_{\mu\nu}$ with respect to which the
above statement is true. This particular tilde here does not stand
for the reference configuration but for the reference metric.
\end{bem}
Then redefine $\tilde u_\mu := \tilde g_{\mu\nu} \tilde u^\mu$. This
one-form then by definition constitutes a foliation of space-time by
space-like hyper-surfaces $S_{t}$ (which we assume to be isomorphic
to $\R^3$). For those hyper-surfaces the four-velocity $\tilde
u^\mu$ is the unit-normal w.r.t. $\tilde g_{\mu\nu}$. Then the
constraints (\ref{c1},\ref{c2}) can be written as (note that in this
particular foliation $\tilde u_b=0$)
\begin{eqnarray}\label{c11}
0&=&r-k_{ab}k^{ab} + k^2 - 2\kappa T^{\mu\nu}\tilde u_\mu \tilde u_\nu \\
\label{c22} 0&=& 2D_{[a}k_{b]}^a + \kappa T^{\mu\nu}\tilde u_\mu
\tilde g_{\nu b}
\end{eqnarray}

Now we can state the main result of this section:

\begin{theorem}
Let the stored energy depend smoothly on its arguments. Given a
reference state $(\tilde g_{\mu\nu}, \tilde f^A)$, where $\tilde
f^A$ is a natural configuration such that $\tilde u_\nu:=\tilde
g_{\mu\nu} \tilde u^\mu$ is closed. Then this reference state
defines a foliation of space-time by space-like hyper-surfaces
$S_t$. Given an initial data set $(S, g_{ij}, k_{ij}, f^A_0, f^A_1)$
satisfying the constraints (\ref{c11},\ref{c22}) together with a
diffeomorphism mapping $S$ onto the initial hyper-surface $S_0$
(defined using the reference state $(\tilde g_{\mu\nu}, \tilde
f^A)$). Then the embedded data give rise to data $(g_{\mu\nu},
\partial_t g_{\mu\nu}, f^A, \partial_t f^A)|_{t=0}$ for the system
(\ref{cpp1}). If the latter are close to the data induced by the
reference state $(\tilde g_{\mu\nu}, \tilde f^A)$ in the topology
$H^{s+1}(\R^3) \times H^s(\R^3) \times H^{s+1}(\R ^3) \times
H^{s}(\R^3)$ for $s\ge 3$ they generate a unique local solution
$(g_{\mu\nu}, f^A)$ to the coupled system (\ref{efe},
\ref{stiegl1111}) which depends continuously on the initial data.
\end{theorem}
For the actual function space for the solution we refer to the
existence theorem from section \ref{loceubd11111}.
\begin{bem}
Once the theorem is given is becomes clear what we meant by the
phrase "close to": namely that the corresponding data on a suitable
hyper-surface would be sufficiently close in the topology given in
the theorem. An example would be states which are close to the
reference state in the supremum-norm.
\end{bem}
We could have tried to obtain a similar result using the corollary
from section \ref{loceubd11111}: If $\tilde u_\mu$ is not assumed to
be closed, then one obtains local well-posedness. We shall formulate
this the following way:
\begin{lemma}
Given a reference state $(\tilde g_{\mu\nu}, \tilde f^A)$, where
$\tilde f^A$ is a natural configuration. Then there exists a (local
in space and time) foliation of space-time such that all data close
to the the ones induces by the reference state give rise to a unique
local (in space and time) solution for the Cauchy problem.
\end{lemma}

\section{Bounded domains}\label{teeistgut}

In the context of matter models most naturally the question arises
whether the equations of motion admit solutions subjected to certain
boundary conditions. In this context it is again the most natural
choice to consider so called natural boundary condition. This means
that the normal stress is assumed to vanish. An elastic body like a
planet (neglecting the atmosphere) would be described by such a
model. Nevertheless one should stress that a planet like earth is
still well described by the Newtonian version of the equations given
here. On the other hand the next bigger objects, where relativistic
corrections come into play, are stars, which are described well
enough by fluid dynamics.

Fluid dynamics can by kinematically treated as a simple subcase of
an elastic material (the stored energy depends only on the particle
density, see e.g. \cite{bs1}). Nevertheless the techniques used in
this work are not suitable to discuss the dynamics of fluids. This
leaves us with the only physical example, which seems to be realized
in our universe: neutron stars. Those extremely dense objects are
known to admit solid crusts (\cite{glitches} or \cite{hae1}) and it
may well be the case that the core is solid as well. This provides a
physical motivation for the given problem. Of course we do not treat
the self-gravitating case but the setup discussed here has to be
rather seen as a natural first step towards this challenge.

There is another reason to talk about finite elastic materials,
which is a rather pragmatic one: there is a small number of physical
reasonable theories where a well-posed initial-boundary-value
problem can be considered without too much trouble. From this point
of view one can argue that it is reasonable to consider the given
problem for the sole reason that we can do so.

But instead of giving further arguments to justify the following
analysis we shall go in medias res. We finally note that the results
of this section can be too found in \cite{bewp1}.

We will use the equations of motion in the material picture. The
reader will find that in comparison to the foregoing sections we
will place stronger restrictions on the space-time geometry. The
reason for this is technical and will be discussed "on the way".

This section is organized as follows: First we recall the basic
setup including the relevant equations of motion. Then we discuss
suitable boundary conditions, which in our case happen to be natural
ones. This discussion is then followed by some remarks concerning
the connection between the Garding inequality and strong point-wise
stability as introduced in section \ref{quasisec}. Finally we will
show that the corollary to the basic existence theorem due to Koch
\cite{koch1} given in section \ref{kochsect1} applies to the system
under consideration, which gives local well-posedness of the
initial-boundary value problem.

\subsection{Basics and equations of motion}\label{teeouzo}

It is convenient for the problem at hand to use the equations of
motion cast in the material description. The reason lies in the
boundary conditions. We will find in the following section that
while the boundary in space-time is free, thus the problem in that
setup is that of solving a free-boundary value problem, it is fixed
in the material picture and given through the boundary of the
material manifold.

In the given scenario the basic unknowns are the deformation
mappings $F$, which are time-dependent mappings from a compact
manifold $\bar{\mathcal{B}}=\mathcal{B}\cup \partial \mathcal{B}$
with smooth boundary $\partial \mathcal{B}$ onto a bounded region
$\mathcal{S}$ of the standard leaf of the foliation, i.e.
\begin{eqnarray}
F: \bar{\mathcal{B}}\times \R & \to & \mathcal{S} \\
(X^A,t) & \mapsto & F^i(X,t)
\end{eqnarray}
As in the unbounded case the body is equipped with a volume-form
$V_{ABC}$, which is smooth on $\bar{\mathcal{B}}$.

For this setup we can define the basic kinematic objects just the
same way as for an unbounded medium. We thus use the definitions as
given in section \ref{eintelephonimwlad}.

For technical reasons we restrict ourselves to so called
"ultra-static" space-times. This means we assume the existence of a
hyper-surface-orthogonal normalized time-like Killing vector field.
This restriction may seem harsh but we stress that the sole
existence of a reference deformation, which admits a Born-rigid
normalized four-velocity, already provides a quite strong
restriction on the space-time geometry. If this reference
deformation ought to be a solution of the field equations, we found
in section \ref{steom11111} that this necessarily implies the
existence of a normalized time-like Killing vector field.

For technical reasons, which will be explained at a later stage it
is very convenient to assume this Killing field additionally to be
hyper-surface-orthogonal.

Under this assumption a suitable choice of coordinates allows us to
write the metric as
\begin{equation}\label{teemetric}
ds^2= -dt^2 + g_{ij}(x)dx^i dx^j
\end{equation}

Note that this form of the metric already leads to a notable
simplification of the formalism: Since both the lapse and the shift
take their most trivial form $N=1$, $Y^i=0$, we find that
\begin{equation}
W^i=\dot F^i
\end{equation}
which includes
\begin{equation}
h^{AB}=f^A,_a f^B,_b ( g^{ab} - \dot F^a \dot F^b )
\end{equation}
as well as
\begin{equation}
\gamma=(1-\dot F^2 )^{-1}
\end{equation}
\begin{bem}
Remember that $f^A,_a$ are to be understood as the inverse of
$F^a,_A$.
\end{bem}
We finally recall the Lagrangian
\begin{equation}
L=\gamma^{-\frac{1}{2}} \epsilon V
\end{equation}
where the stored energy $\epsilon$ is a smooth function of $X^A$ and
of the strain.

Then the field equations are
\begin{equation}\label{teeallgglg1}
\frac{d}{d X^\mu}\left( \frac{\partial L}{\partial F^i,_\mu}
\right)=\frac{\partial L}{\partial F^i}
\end{equation}
Thus the principal part is determined by
\begin{equation}\label{teeohnerum}
A^{\mu\nu}_{ij}=\frac{\partial ^2 L}{\partial F^i,_\mu
\partial F^j,_\nu}
\end{equation}

\subsection{Boundary conditions}\label{natbcchapter}
We now turn to boundary conditions. In non-relativistic elasticity
theory there are essentially two kinds of boundary conditions:
\begin{itemize}
\item boundary conditions of place: the boundary of the material
in space-time (i.e. the set $f^{-1}(\partial \mathcal{B})$ ) is
prescribed.

\item traction boundary conditions: the normal traction, i.e. the
normal components of the stress are prescribed.

\end{itemize}
When the normal traction is zero, we speak of "natural" boundary
conditions. For obvious reasons they are the ones appropriate for
modeling a freely floating finite object like a star or a planet.
This reason together with a second one strongly suggests this
choice: all other boundary conditions but the natural ones become
inconsistent once one couples to the Einstein equations.

For the space-time description these natural boundary conditions are
given by
\begin{equation}
T^\nu_\mu n_\nu |_{f^{-1}(\partial \mathcal{B})}=0
\end{equation}
where $T^{\mu\nu}$ is the energy-momentum tensor as given in
(\ref{stemt111}) while $n_\mu$ denotes the co-normal of
$f^{-1}(\partial \mathcal{B})$. To see this note that the material
velocity is tangential to the inverse images of points of
$\mathcal{B}$ under the configuration map. Thus it is annihilated by
the co-normal. Hence the above equation becomes (recall that the
configuration gradient is assumed to be of maximal rank)
\begin{equation}\label{teebc1}
\tau_{AB}f^B,_\nu g^{\mu\nu}n_\nu|_{f^{-1}(\partial \mathcal{B})}=0
\end{equation}
which is just the statement that the normal stress vanishes.

There is however a huge disadvantage about these boundary
conditions: the boundary $f^{-1}(\partial \mathcal{B})$ is free,
i.e. it is not determined a priori, but it is rather a part of the
problem to find the actual boundary.

This problem is overcome by switching to the material description.
Then the boundary conditions are evaluated on $\partial
\mathcal{B}\times \R$, which is fixed. To actually obtain these new
boundary conditions we have to translate (\ref{teebc1}) into the
material description. In section \ref{refmat1} we have encountered
the diffeomorphism $(t,X)=(t,f(t,x))$. Applying it to the space-time
co-normal $n_\mu$ we find that it is mapped to the one-form $N_\mu$,
which by definition is connected to $n_\mu$ in the following way:
\begin{equation}
N_\mu dX^\mu = (N_0 + N_A \dot f^A) dt + N_A f^A,_a dx^a = n_\mu
dx^\mu
\end{equation}
Recalling now that the four-velocity was mapped to a multiple of
$\partial_t$, we find that the equation $n_\mu u^\mu=0$ now reads
$N_0=0$, which at once implies the relation
\begin{equation}
n_\mu =f^A,_\mu N_A
\end{equation}
Thus (\ref{teebc1}) becomes
\begin{equation}\label{teebc2}
\tau_{AB}h^{BC}N_C|_{\partial\mathcal{B}}=0
\end{equation}
These are the new boundary conditions. They state that the normal
components of the stress measured along the body have to vanish for
all time.

There is a deep connection between the equations of motion and the
boundary condition. To reveal this connection we note that
\begin{equation}
\frac{\partial L}{\partial F^i,_A}=\gamma^{-\frac{1}{2}}V
\tau_{MN}\frac{\partial h^{MN}}{\partial F^i,_A}
\end{equation}
Using formula (\ref{eiersalat}) for the variation of the strain this
becomes
\begin{equation}
\frac{\partial L}{\partial F^i,_A}=-2
\gamma^{-\frac{1}{2}}V\tau_{MN} f^M,_i h^{NA}
\end{equation}
This means the following: if we replace (\ref{teebc2}) by the
equivalent equation (recall that $f^A,_i$ was an isomorphism)
\begin{equation}\label{teebc}
-2 \gamma^{-\frac{1}{2}}V f^A,_i
\tau_{AB}h^{BC}N_C|_{\partial\mathcal{B}}=0
\end{equation}
we find that we can rewrite the boundary condition in a slightly
different way as
\begin{equation}\label{teebcalt}
\frac{\partial L}{\partial
F^i,_A}N_A|_{\partial\mathcal{B}}=\frac{\partial L}{\partial
F^i,_\mu}N_\mu|_{\partial\mathcal{B}}=0
\end{equation}
since $N_0=0$. Due to this correspondence (\ref{teebc}) (sometimes
the version (\ref{teebcalt})) will be the version of the boundary
condition we are going to use in the following.

So the problem is to find a solution to the initial-boundary value
problem given by the equations of motion (\ref{teeallgglg1})
subjected to the boundary conditions (\ref{teebcalt}).

\begin{bem}
Note that the statements of this section are independent of the
restricted space-time geometry, but apply to the general setup.
\end{bem}

\subsection{Reference deformation}\label{teeref}
It is natural to assume the existence of a "natural state" of the
material, which may be viewed as a trivial realization of the
(abstract) material manifold in space-time.

In the given context we assume that there exists a static reference
deformation, such that the elasticity operator satisfies the strong
point-wise stability condition (\ref{sps1}).

We formulate these requirements first in terms of the space-time
description: There exists a configuration $\tilde f^A$, such that
the four-velocity coincides with the Killing field:
\begin{equation}
\tilde u =\partial_t
\end{equation}
This implies that the reference configuration is time-independent:
\begin{equation}
\partial_t \tilde f^A =0
\end{equation}
From section \ref{radenska1} we know that the reference
configuration by definition has to be stress-free with the
corresponding normalized four-velocity $\tilde u^\mu$ being
geodesic. But the latter point together with the property of being a
static Killing field implies the Born condition, which in turn means
that the reference strain can be viewed as a metric field on the
body. (A simple way of seeing this is by recognizing that $\tilde
h^{AB}$ is time-independent as well as the reference configuration,
which implies that it becomes a field on $\mathcal{B}$, once
translated into the material description.)
\begin{bem}
Note that this nice property of the reference configuration is an
artefact of this special setup. For general space-times as
investigated in the sections dealing  with the unbounded setup,
there will neither exist a Born-rigid motion nor will the strain
define a tensor field on the body.
\end{bem}
\begin{bem}
Note that the requirements on the elasticity operator employed here
are stronger than those needed for the unbounded setup. We will see
that it is a proper way of satisfying the Garding inequality, which
is a necessary component for the Koch theorem.
\end{bem}
Once the properties of the reference state are known in the
space-time description it is quite easy to determine the analogous
statements for the material description:

There exists a reference deformation $\tilde F^i$, which is
time-independent. The reference strain $\tilde h^{AB}$ is a metric
field on the body, and satisfies the following conditions:
\begin{equation}
\epsilon (X, \tilde h) =: \tilde \epsilon (X)>0
\end{equation}
as well as
\begin{equation}\label{teetau=0}
\tau_{AB}(X, \tilde h) =0
\end{equation}
Finally strong point-wise stability for
\begin{equation}
U(X,\tilde h)_{ABCD} =: \tilde U (X)_{ABCD}
\end{equation}
reads (suppressing the $X$-dependence; also compare (\ref{sps1}))
\begin{equation}\label{teesps}
 \tilde U_{ABCD}\mu^{AB}\mu^{CD} \ge C \tilde h_{C(A}\tilde h_{B)D}  \mu^{AB}\mu^{CD}
\end{equation}
for $C>0$.
\begin{bem}
We stress that strong point-wise stability implies the
Legendre-Hadamard condition but not vice versa.
\end{bem}
It is worth noting that our definition of the reference deformation
did not imply any statement on the boundary conditions. However one
finds that the following is true:
\begin{lemma}
The reference deformation also solves the boundary value problem
(\ref{teeallgglg1},\ref{teebc}). In particular this means that
initial data given on the hyper-surface $\{t=0 \}$ of the form
$(\tilde F^A,0)$ generate the reference deformation.
\end{lemma}
The first part of the lemma is a trivial consequence of
(\ref{teetau=0}). The second part however is not that trivial. It is
obvious that the reference deformation is a solution to
(\ref{teeallgglg1},\ref{teebc}) with the claimed initial data. But
in order to exclude that there are other solutions with the same
initial data, which do not coincide with the reference deformation
we have to use the uniqueness part of the Koch theorem. We will show
in the following that the requirements of this theorem are met for
the reference deformation (and hence by continuity for deformation
which are close enough). This will result in the main theorem of
this section, which implies the uniqueness statement needed here.

The principal part given in (\ref{fuertee1},\ref{fuertee2}) takes a
relative simple form for the reference deformation. (Remember that
in these formulas not the principal part $A^{\mu\nu}_{ij}$ was given
but rather $B^{\mu\nu}_{AB}:=A^{\mu\nu}_{ij}F^i,_A F^j,_B$. Another
slight difference comes from the fact that in the definition given
there the factor $V$ coming from the volume form was eliminated,
which is not convenient anymore in this context due to the presence
of the boundary conditions.)

Using the convention that every object evaluated in the reference
deformation is marked by an upper tilde, f.e.
\begin{equation}
A^{\mu\nu}_{ij}|_{F=\tilde F}=\tilde A^{\mu\nu}_{ij}
\end{equation}
we find that
\begin{equation}\label{tee00}
\tilde A^{00}_{ij}=-V \tilde \epsilon \tilde g_{ij}
\end{equation}
as well as
\begin{equation}
\tilde A^{AB}_{ij}=4 V \tilde U_{CEDF}\tilde h^{AC}\tilde h^{DB}
\tilde f^E,_i \tilde f^F,_j
\end{equation}
The remaining terms of the principal part vanish when taken for the
reference deformation:
\begin{equation}
\tilde A^{0A}_{ij}=0
\end{equation}
Apart from other reasons the last equations holds as a result of the
special form of the metric. If a lapse would have been present, the
elasticity operator would have entered $\tilde A^{00}_{ij}$ and
$\tilde A^{0A}_{ij}$ would not necessarily have vanished.

This would have resulted in the problems we know from section
\ref{fuertee3} concerning a proper coordinate choice. This question
of coordinate choice gains extra-finesse since in these different
coordinates the boundary would most probably loose its convenient
coordinate form. All these problems were bypassed by the
restrictions placed on the metric in this section.

We finally observe that by (\ref{tee00}) the coefficient of the
time-derivative $A^{00}_{ij}$ is negative definite for the reference
deformation and thus by continuity for all deformation sufficiently
close.

\subsection{Coerciveness}

A crucial entity for the Koch theorem is the coerciveness of the
spatial principal part. At this stage the equation of state (i.e.
the elasticity operator) comes into play. We will show that in the
given context materials those stored energy satisfies strong
point-wise stability (\ref{teesps}) at the reference state give rise
to this coerciveness condition.

More concrete we have to show that the Garding inequality (\ref{G1})
\begin{equation}\label{teegarding}
\int_\mathcal{B} A_{ij}^{AB}\partial_A\phi^i\partial_B\phi^j d^3
X\ge \epsilon ||\phi||^2_{H^1}- \kappa ||\phi||^2_{L^2}
\end{equation}
holds for all smooth $\phi^i$ where the constants $\epsilon>0$ and
$\kappa \ge 0$. ($H^1= W^{1,2}(\mathcal{B})$ and
$L^2=W^{0,2}(\mathcal{B})$; see e.g. \cite{evans1})
\begin{bem}
Note that the inequality involves no derivatives of the deformation
higher than first order. By this property it becomes a restriction
on possible initial data.
\end{bem}
We will first derive the Garding inequality for the reference
deformation:
\begin{equation}\label{teetrinken}
\int_\mathcal{B} 4 V \tilde U_{CEDF}\tilde h^{AC}\tilde h^{DB}
\tilde f^E,_i \tilde f^F,_j\partial_A\phi^i\partial_B\phi^j\ge
\epsilon ||\phi||^2_{H^1}-\kappa ||\phi||^2_{L^2}
\end{equation}
By continuity (\ref{teegarding}) is then true for all deformation
sufficiently close (the precise form of this statement will be given
later). This includes that the Garding inequality holds for all
initial data sufficiently close to those generated by the reference
deformation.

To prove (\ref{teetrinken}) we have to invoke Korn's inequality (see
f.e. \cite{horgan}) or (to be more accurate) a slight generalization
to Riemannian manifolds due to \cite{chenjost}: Given a vector field
$\psi^A$ on the body, which can be viewed as the Riemannian manifold
$(\mathcal{B},\tilde h_{AB})$. Then define $L(\psi)$ by
\begin{equation}
L(\psi)_{AB}=2 \tilde h_{C(A} \psi^C,_{B)}
\end{equation}
Then the Korn inequality tells us that there exists a positive
constant $C$, such that
\begin{equation}\label{korn}
||L(\psi) ||^2_{L^2}+||\psi||^2_{L^2} \ge C ||\psi ||^2_{H^1}
\end{equation}
Going back to $(\ref{teetrinken})$ we note that by (\ref{teesps})
the integral on the left hand side is taken over the product of two
positive functions (the scalar $V$ and the rest). Since $V$ is also
a continuous function on a compact set, we can estimate it from
below by its minimum, which is positive. Using (\ref{teesps}) again
on the remaining integral we still have to show that
\begin{equation}\label{teemitrum}
\int_{\mathcal{B}} \tilde h_{C(A}\tilde h_{B)D} \tilde
f^B,_i\phi^i,_E\tilde h^{AE} \tilde f^D,_j\phi^j,_F\tilde h^{CF}
\end{equation}
is bounded from below in a suitable way. For this purpose we assume
without restriction of generality that coordinates on $\mathcal{B}$
were chosen in a way such that $\tilde F^i$ becomes the identity
map.
\begin{bem}
It is notable that this specialization directly leads to the
classical setup of non-relativistic elasticity theory in its
classical formulation: that the body is identified with a certain
part of space.
\end{bem}
For this choice of coordinates the deformation gradient coincides
with the Kronecker delta:
\begin{equation}
\tilde F^i,_A =\delta^i_A
\end{equation}
which clearly includes
\begin{equation}
\tilde f^A,_i =\delta^A_i
\end{equation}
Applying this to the integrand in (\ref{teemitrum}) leads us to
\begin{equation}\label{teemitkorn}
\int_{\mathcal{B}} \tilde h_{C(A}\tilde h_{B)D} \tilde
f^B,_i\phi^i,_E\tilde h^{AE} \tilde f^D,_j\phi^j,_F\tilde h^{CF} \ge
C ||L(\tilde f^A,_i \phi^i )||^2_{L^2}
\end{equation}
Note that we would obtain equality, if the norm $|| \cdot ||_{L^2}$
was defined using the reference strain. If one decides to use a flat
metric $\delta_{AB}$ to construct these norms (as is the usual way)
equality does not hold anymore. In any case however the equivalence
of finite-dimensional norms tells us that the above inequality holds
for some positive constant $C$ (which depends on the metric used to
construct the norm on the right hand side). Note that the proper
choice of coordinates was crucial for the argument, since it allowed
us to interchange differentiation with contraction with the inverse
deformation gradient.

Now we can apply the Korn inequality (\ref{korn}) to the right hand
side of (\ref{teemitkorn}). Since $\tilde f^A,_i=\delta^A_i$, we
find that (\ref{teetrinken}) is true, which in turn implies validity
of (\ref{teegarding}) for deformations close enough.

\subsection{Main result}
In this section we will show that all the requirements of the Koch
theorem as given in section \ref{kochsect1} are met in the current
setup. This will ultimately result in a local well-posedness for the
relativistic elastic initial-boundary value problem.

For convenience we recall these requirements (already stated in the
terms used here). First note that the system (\ref{teeallgglg1},
\ref{teebcalt}) has exactly the required connection between boundary
conditions and equations of motion. In addition we have to check
\begin{itemize}
\item the symmetry $A^{\mu\nu}_{ij}= A^{\nu\mu}_{ji}$
\item hyperbolicity (and coerciveness) of the principal part
\end{itemize}
The first point is trivially met, since our equations of motion come
from an action principle which reflects itself in (\ref{teeohnerum})
from which the desired symmetry can readily be read off.

The second point consists of two parts: First we have to show that
$A^{00}_{ij}$ is negative definite. For the reference deformation
this is evident from (\ref{tee00}). From this we concluded the same
property for all deformations "sufficiently close". This statement
will be sharpened in the following:

Assume we are given initial data, i.e. a pair $(F_0, F_1)\in
H^{s+1}(\mathcal{B})\times H^s(\mathcal{B})$ for $s\ge 3$. Then by
Sobolev embedding we find that $(F_0, F_1)\in C^1(\mathcal{B})\times
C^0(\mathcal{B})$ (See f.e. \cite{evans1}). But this means that the
deformation gradient is a continuous function on the initial surface
$\{0\}\times \mathcal{B}$.

Denoting the data induced by the reference deformation (which itself
clearly is assumed to be sufficiently smooth) by $(\tilde F_0, 0)$,
we find that all data close to the reference data in the topology
$H^{s+1}(\mathcal{B})\times H^s(\mathcal{B})$ are close in
$C^1(\mathcal{B})\times C^0(\mathcal{B})$ (which is a point-wise
statement). We conclude that there exists a neighborhood of the
reference data in $H^{s+1}(\mathcal{B})\times H^s(\mathcal{B})$ for
which $A^{00}_{ij}$ is negative.

On the other hand we have to show coerciveness for the spatial
component of the principal part. By the very same argument as we
used for time-components, we conclude that coerciveness holds for
all data close to the reference data in $H^{s+1}(\mathcal{B})\times
H^s(\mathcal{B})$.

We have found that all the requirements of the theorem are met, so
that we obtain the following result:
\begin{theorem}
Given a space-time and a material manifold as described in section
\ref{teeouzo}. Let the stored energy be a smooth function of all its
entries. In addition assume that there exists a smooth static
reference solution as defined in section \ref{teeref}, such that the
elasticity operator satisfies the strong-point-wise stability
condition (\ref{teesps}). Then for all data close to the data
induced by the reference deformation in the topology
$H^{s+1}(\mathcal{B})\times H^s(\mathcal{B})$ that satisfy the
compatibility conditions up to order $s$ there exists a unique $T>0$
and a unique solution $F^i \in C^2([0,T)\times \bar{\mathcal{B}})$
to the initial-boundary value problem (\ref{teeallgglg1},
\ref{teebc}) which depends continuously on the initial data.
\end{theorem}
We end with two remarks concerning the compatibility conditions:
First note that by the lemma \ref{fuerteeforever} from section
\ref{kochsect1} we know that there exists a great number of possible
initial data obeying the compatibility conditions.

Secondly we stress that in the proof of lemma \ref{fuerteeforever}
the compatibility conditions were solved by prescribing certain
normal derivatives of the initial data along the boundary (we refer
to section \ref{kochsect1} for details). In this sense the boundary
conditions (\ref{teebc}) are of Neumann type.

For details concerning the compatibility conditions we refer to
section \ref{kochsect1}.

\chapter{Summary, Conclusions and Outlook}\label{elvisconc1}
What have we learned in the course of the thesis and in what ways
can the results given here hope to be generalized? This will be
answered in the following.

\section{Summary and conclusions}

In this thesis we treated relativistic elasticity as a Lagrangian
field theory. The key objects were configurations (deformations
respectively), which are maps between space-time and an abstract
material manifold. During all of our applications we assumed the
existence of a certain reference state, which played the role of a
natural realization of the material manifold in space(-time). This
assumption constitutes the connection to the classical
non-relativistic setup, where the material in rest is identified
with a certain region in space. Note however that in principle the
(kinetic) theory as presented here can be formulated without the
assumption of any reference state.

There remains another grave difference to the classical
non-relativistic setup: There the material manifold is (spoken in
our language) viewed as a Riemannian manifold equipped with a flat
metric. The reference state is defined to be the state where the
strain coincides with this given flat metric.

For the relativistic formulation we only assume the existence of a
volume form on the material manifold. This assumption turns out to
be sufficient to treat kinematics. If the strain is to define a
metric on the material manifold, there arise certain restrictions on
the space-time geometry (there has to exist a Born-rigid
vector-field).

As in the non-relativistic theory the elastic properties of the
material are basically encoded in the stored energy function, which
is a measure for the elastic energy in dependence of the strain. The
difference between relativistic and non-relativistic theory lies in
the definition of the strain (which can be viewed as a metric for
the deformed state) itself. While in the non-relativistic theory it
is a purely spatial object (compare $k_{AB}$ as defined in
(\ref{k1})) in the given context it is constructed out of the full
space-time metric (compare (\ref{sstrain})) including also
time-derivatives of the configuration.

In the second part of this thesis we derived the equations of motion
governing the time-evolution of the elastic material from an action
principle arising from the variation of a Lagrangian which was
basically given by the total energy density. The resulting equations
were then analyzed both on the linearized and non-linear level. For
the linearized equations on Minkowski space-time we could give an
explicit formula for the solution of the initial value problem
corresponding to an un-bounded homogeneous and isotropic material.
On the non-linear level we discussed conserved quantities and the
connection between the thus defined energy and Killing vector fields
(generating trivial spatial configurations).

Finally we proved well-posedness results for a variety of scenarios
including both bounded and unbounded elastic materials both on a
fixed gravitational background and self-gravitating.

We found that the initial boundary value problem required stronger
restrictions on the material's elastic properties than the other
setups, for which a rather mild assumption (real propagation speeds
for linearized elastic wave solutions) turned out to be sufficient.

In this sense we can say that the theory as presented in this thesis
is well-suited for the description of relativistic elastic
materials.

\section{Outlook and open problems}

Given the results of this thesis what else remains there to do? In
principle there are basically two ways to proceed. One may either
choose to go into detailed analysis for certain kinds of materials
(equations of state) in hope of obtaining some results which could
be compared with observational data (f.e. concerning neutron stars).
This can be seen as a complementary approach to the "ab initio"
attempts of obtaining suitable equations of state (see \cite{hae1}).
It might lead to an understanding of certain effects like the
well-known "glitches" (see \cite{glitches} and references therein).
This task however would require numerical effort rather than
analytical. But even for the numerical treatment well-posedness of
the equations of motion is a necessary ingredient \cite{michaelp1}.

On the other hand one can of course carry on without numerics. Such
analytical work could go in two directions.

First one might try to generalize the result from section
\ref{teeistgut} to include self-gravitation. This task by far
exceeds the one related to the finite body on fixed background,
since one would have to care about
 the propagation of the constraints along the boundary. We have
 learned in section \ref{teeistgut} that it is favorable to formulate
 the boundary value problem in the material description. Section
 \ref{eisauch} however teaches us that the self-gravitating problem
 is conveniently addressed in the space-time setup. One might even
 ask how to formulate the self-gravitating problem in the material
 description in terms of partial differential equations. The
 diffeomorphism between space-time and $\{ $material manifold times
 time$ \}$ introduced in section \ref{refmat1} may be of some help in
 this conquest.

 Secondly one might try to derive global results. At least for small
 initial data (i.e. data close to a natural reference state) it
 should be possible to prove global existence on Minkowski
 space-time. For the non-relativistic theory such results already
 exist for a growing number of scenarios. (See \cite{sid1} for an
 overview.)

 There are at least two important ingredients in the proofs of such
 result. One has to require a certain "null-condition", which prevents
 the self-interaction of the eigen-modes of the elastic material (see \cite{hoer1} for the definition for wave equations and \cite{sid2} for isotropic elastic equations). In
 non-relativistic theory this condition is known to be necessary for
 the long-time existence of solutions (\cite{joh2} (it should be
 possible to obtain analogous non-existence results for the present setup).

 Under the assumption of this null-condition (a generalization of this condition also plays an important role in general relativity, see \cite{cho1}, \cite{lr1}, \cite{lr2}) one then uses generalized energy
 estimates (first introduced by Klainerman in the study of non-linear wave equations \cite{kla1}) which are connected to the symmetry group of isomorphisms
 of space-time. In the non-relativistic theory this symmetry group
 is the Galilei group and hence only 6-dimensional. One has to employ special projection techniques (see \cite{sid1}) to obtain additional
 bounds. In the present relativistic version we have the full 10-dimensional Poincare group
 at our disposal. This might lead to a simplification of the task of
 obtaining the required estimates. One can thus hope that it will be
 rather straightforward to obtain a global existence result for the
 relativistic setup by following the non-relativistic proof.

 The ultimate challenge would be the derivation of statements on the
 asymptotic structure for elastic materials. For unbounded materials
 one might expect
 that the global solution for small initial data approaches the unstressed reference
 solution (on any open neighborhood) as time passes on (perturbation due to non-trivial initial data travel to
 infinity). The analogous problem for finite materials implies
 further problems. Will the material approach a natural state? What
 about translations of the center of mass? How to describe such a
 phenomenon? What is to be expected in the self-gravitating setup?

 Finally (far beyond the scope of any stern approach): what about
 global asymptotic behavior for a self-gravitating finite elastic
 material? What is there to be expected?

This is a whole load of interesting and challenging questions, both
from mathematical and physical point of view. In this sense the
results obtained up to now can be seen as a (necessary) first step
on the way to answering the questions of the type stated above.

\chapter{Appendix}
 This appendix contains some minor results which do not directly fit
 into the main body of this work. It contains an example which is incorporated to become familiar with the objects defined in chapter \ref{chaptergeokin} as well as
 remarks on energy conditions, Newtonian limit of the field equation
 and finally some formulas arising in the context of the 3+1
 decomposition.

\section{An easy example: 1+1 dimensions}
This section contains a simple 1+1 dimensional example to get used
to the definitions introduced in chapter \ref{chaptergeokin}. For
convenience we will specialize to static space-times, in particular
we concentrate on the elastic equations on Schwarzschild background.
Note however that we do not employ any "small-stress" assumptions.

\subsection{Basics}

The unknown is a mapping $f$ from two dimensional space-time $M$
onto a one dimensional manifold $B$ called body. The body is
equipped with a volume form $V dX$, which is a one-form in the given
context.
\begin{eqnarray}
f: \ M &\to& B \\
x^{\alpha} &\mapsto& X=f(x^{\alpha})
\end{eqnarray}
We assume the tangent mapping $df$ to be of maximal rank with
time-like null-space. In the given context this simply means $df\ne
0$. Therefore (as long as $M$ is time-orientable) there exists a
unique future-pointing time-like vector-field $u^\mu$ satisfying the
two conditions
\begin{equation}
\label{u1} u^{\alpha}f_{,\alpha}=0 \ \ \ \ g(u,u)=-1.
\end{equation}
The strain $hdXdX$ is determined by the function $h$ which (in
accordance with (\ref{sstrain})) is given by
\begin{equation}
h = f,_\mu f,_\nu g^{\mu\nu}
\end{equation}
Then the particle density becomes
\begin{equation}
n=V\sqrt{h}
\end{equation}
We also assume the existence of a reference configuration $\tilde
f$. Then we can choose coordinates on the body in such a way that
$\tilde h=1$. For convenience we assume that $V=1$ in these
coordinates, such that the reference strain is a flat metric on the
body and $V$ is the corresponding volume form.

Then a material which is homogeneous and isotropic w.r.t. this
reference configuration is given by a stored energy which depends
only on the invariants of the strain (taken w.r.t. the reference
strain, which is a flat metric). Since in one dimension there is
only one independent invariant we have:
\begin{equation}
\epsilon(n)=\epsilon(\sqrt{h})
\end{equation}
\begin{bem}
In this sense there is no difference between fluids and other
elastic materials in this low dimension. The stored energy does (in
the isotropic case) only depend on the particle number density.
\end{bem}
Consequently we have that the stress is given by
\begin{equation}
\tau=\frac{\partial \epsilon}{\partial h}=\frac{1}{2n}
\frac{\partial \epsilon}{\partial n}
\end{equation}
For the energy-momentum tensor this implies
\begin{equation}
T_{\mu\nu} = \rho u_\mu u_\nu + \frac{\partial \epsilon}{\partial n}
f,_\mu f,_\nu
\end{equation}
From this point on we will specialize to static setups. What we
exactly mean by this we explain in the following:

\subsection{Static setup}

We assume the existence of a Killing field $\xi^\mu$, which in
addition is assumed to be proportional to the four-velocity:
\begin{equation}
u=\phi \xi
\end{equation}
where
\begin{equation}
\phi^2:=(-g(\xi,\xi))^{-1}.
\end{equation}
Under these assumptions we can easily derive the equations of
motion. We have seen in  section \ref{steom11111} that they are
equivalent to divergence-less-ness of the energy-momentum tensor.
Recalling the continuity equation (\ref{cont1}) this condition is
equivalent to
\begin{equation}\label{1+1eom1}
0=n u^\mu (\epsilon u_\nu);_\mu + \left( \frac{\partial
\epsilon}{\partial n} f,^\mu f,_\nu \right);_\mu
\end{equation}
Since $\xi^\mu$ is a Killing field, we have
$\xi^{\mu}\phi,_{\mu}=0$. As a consequence
$u^{\mu};_{\nu}=\phi,_{\nu}\xi^{\mu}+\phi\xi^{\mu};_{\nu}$ yields
$u^{\mu};_{\mu}=0$. We also find that $u^\mu u_\nu ;_\mu
=\frac{1}{2}\phi^2(\phi^{-2}),_\nu$. From this we find that
(\ref{1+1eom1}) contracted with the four-velocity leads to
\begin{equation}
0=-n u^\mu \epsilon ,_\mu + \frac{\partial \epsilon}{\partial n}
f,^\mu f;_{\nu\mu} u^\nu
\end{equation}
But since
\begin{equation}
u^\nu \epsilon,_\nu =\frac{1}{n}\frac{\partial \epsilon}{\partial
n}f,^\mu f;_{\nu\mu} u^\nu
\end{equation}
we find that $T^\mu_\nu;_\mu u^\nu=0$ is satisfied identically by
virtue of the continuity equation as was to be expected from section
\ref{sectem1}.

The remaining component is the only non-trivial one. It is obtained
by contraction of (\ref{1+1eom1}) with $f,^\nu$.
\begin{equation}\label{1+1eom}
0=- n\epsilon u^\mu u^\nu f;_{\mu\nu} + \left(\frac{\partial
\epsilon}{\partial n} f,^\mu f,_\nu \right);_\mu f,^\nu
\end{equation}
The bracket term includes both first and second variations of the
stored energy. Besides these non-vanishing stress-terms the above
equation looks a lot like the equations of motion as given in
section \ref{steom11111}. They constitute a second order equation
for the configuration $f$. In fact it isn't even a PDE but can
rather be cast in the form of an ODE, since we assumed the material
velocity to be proportional to a Killing field. To see this more
clearly we again specialize our example by fixing a particular
background metric, namely the Schwarzschild metric.

\subsection{Elasticity on Schwarzschild background}

We now use the Schwarzschild metric
\begin{equation}
ds^2=-\left( 1-\frac{2m}{r} \right)dt^2+ \left( 1-\frac{2m}{r}
\right)^{-1}dr^2.
\end{equation}
Therefore we use the following Killing field
\begin{eqnarray}
\xi &=&\partial_{t} \\
\phi &=&\left(1-\frac{2m}{r}\right)^{-1/2}.
\end{eqnarray}
The first consequence is that the configuration is time-independent,
i.e.
\begin{equation}
f,_t=\xi^\mu f,_\mu = \phi^{-1} u^\mu f,_\mu = 0
\end{equation}
From this we find that
\begin{equation}
u^\mu u^\nu f;_{\mu\nu}=-\frac{1}{2}(\phi^{-2}),_r f,_r
\end{equation}
Thus (\ref{1+1eom}) becomes a single ordinary differential equation
in $r$.

We find that the second term in (\ref{1+1eom}) becomes
\begin{equation}
\left( \frac{\partial \epsilon}{\partial n}  \phi^{-2} f,_r^2
\right),_r f,_r \phi^{-2} +\frac{1}{2}\frac{\partial
\epsilon}{\partial n}  f,_r^3 \phi^{-2}(\phi^{-2}),_r
\end{equation}
Altogether this means that (\ref{1+1eom}) is given by (using the
identity $n^2=h= f,_r^2\phi^{-2}$)
\begin{equation}
0=\frac{1}{2}n (\phi^{-2}),_r  f,_r   \left(  \epsilon + n
\frac{\partial \epsilon}{\partial n}  \right) + \left(
\frac{\partial \epsilon}{\partial n}  n^2  \right),_r n \phi^{-1}
\end{equation}
Multiplying by $\phi n^{-1}$ leads to
\begin{equation}
0=(\ln \phi ^{-1}),_r n \left(  \epsilon + n \frac{\partial
\epsilon}{\partial n}  \right)+\left( \frac{\partial
\epsilon}{\partial n}  n^2  \right),_r
\end{equation}
Assuming that $\left(  \epsilon + n \frac{\partial
\epsilon}{\partial n}  \right) \ne0 $, we conclude that
(\ref{1+1eom}) is equivalent to
\begin{equation}
0=(\ln \phi ^{-1}),_r + \left( \ln\left(  \epsilon + n
\frac{\partial \epsilon}{\partial n}  \right) \right),_r
\end{equation}
Thus we can solve by integration. We obtain
\begin{equation}\label{raki}
 \epsilon + n
\frac{\partial \epsilon}{\partial n}= C \phi
\end{equation}
where the constant $C$ has to be determined via the boundary
conditions entering the integration. This is an algebraic equation
for $n$ and hence for $f,_r$. If we can solve algebraically for the
configuration gradient, we can obtain the configuration itself via
integration.

For convenience we define
\begin{equation}
A(n):=e+ne,_{n}=\rho,_{n}
\end{equation}
Using the definition for the stress
\begin{equation}
p:=n\frac{\partial \rho}{\partial n}-\rho
\end{equation}
we can translate $A$ into
\begin{equation}
A(n)=\frac{p+\rho}{n}
\end{equation}
Recall that we used the assumption $A\ne 0$. In fact we will also
assume that $A$ is positive:
\begin{equation}
A>0
\end{equation}
and monotone
\begin{equation}
\frac{\partial A}{\partial n}>0
\end{equation}
This implies that the energy density $\rho$ has to be convex.

 We now turn to the analysis of (\ref{raki}) for certain special
setups.

First we concentrate on the case of a hanging body (supported at the
upper end).

\begin{figure}[h]
\center \psfrag{a}{$(n_u ,r_{u})$} \psfrag{b}{$(n_l ,r_{l})$}
\includegraphics[totalheight=4cm]{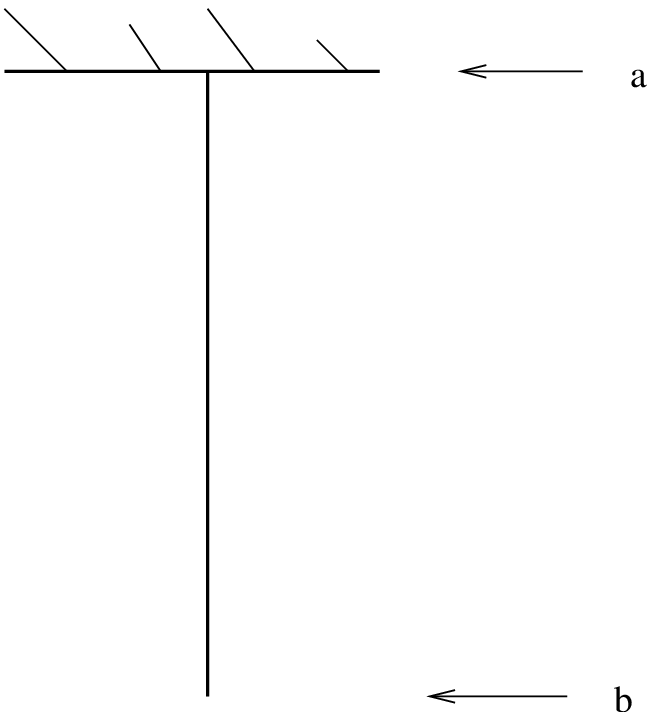}
\end{figure}

It is natural to assume $r_{u}$ and $n_{l}$ to be given. The first
is the suspension point. The second is given through $p(n_{l})=0$,
since we require the pressure to vanish at the lower end. Note that
we can solve $p(n_l)=0$ for $n_l$, since by
\begin{equation}
\frac{\partial p}{\partial n}= n \frac{\partial A}{\partial n}
\end{equation}
the pressure $p$ is monotone in $n$. For the described scenario
(\ref{raki}) becomes
\begin{equation}\label{dieunglg}
A(n_u)= A(n_l)\phi(r_u) \phi^{-1}(r_l)=:C_1 \left( 1-\frac{2m}{r_l}
\right)^{\frac{1}{2}}
\end{equation}
where $C_1>0$. This equation only gives us information on the
endpoints of the material.

To obtain a formula which gives the particle density for general $r$
(away from the endpoints), we assume for the moment that $r_l$ is
given (this assumption has to be removed later on). Then
(\ref{raki}) becomes
\begin{equation}\label{gin}
A(n)=A(n_l)\phi^{-1}(r_l) \phi(r)
\end{equation}
Assume for a moment that $m=0$ (i.e. vanishing gravity). Then
$n=n_l$ which means that the particle density is constant along the
stick and thus that there is no pressure present when gravity is
turned off.
\begin{bem}
Recall that $n_l$ was uniquely determined by the condition
$p(n_l)=0$.
\end{bem}


Since we are interested in a finite body we assume that its
intrinsic length $L$ defined by
\begin{equation}
L=\int_B V =\int_B dX
\end{equation}
is finite. This implies (the configuration is viewed as mapping
between body and the space orthogonal to the Killing flow)
\begin{equation}\label{ilaenge}
L=\int_{f^{-1}(B)} f^* V = \int_{r_{l}}^{r_{u}} f_{,r}
dr'=\int_{r_{l}}^{r_{u}}\frac{n}{\left(
1-\frac{2m}{r'}\right)^{1/2}} dr'
\end{equation}
We can use this equation to get rid of the $r_l$ entering formula
(\ref{gin}). We will see that we can solve (\ref{ilaenge}) for $r_l$
in terms of $r_u$, $n_l$, $L$ and $m$.

For this we have to show that
\begin{equation}\label{1+1wiedermal1}
0\ne \frac{\partial L}{\partial r_l}=\int_{r_{l}}^{r_{u}} \phi(r)
\frac{\partial n}{\partial r_l} dr -  n(r_l)\phi(r_l)
\end{equation}
This condition is necessary and will give rise to a restriction on
$A$ and hence on the stored energy $\epsilon$.

To see this note that the second term is negative while for the
first term (\ref{gin}) gives
\begin{equation}
\frac{\partial A(n)}{\partial r_l}=  A(n_l) \phi(r) \frac{\partial
\phi^{-1}(r_l)}{\partial r_l}= A(n) \phi(r_l )^2 \frac{m}{r_l^2}
\end{equation}
which is positive. By the monotonicity of $A$ (recall
$\frac{\partial A}{\partial n}>0$) this includes the analogous
statement on the variation of $n$. Thus from our original (mild)
assumptions on $A$ we cannot conclude a sign for the right hand side
of (\ref{1+1wiedermal1}. We thus assume the validity of
(\ref{1+1wiedermal1}) in the following. It is reasonable to assume
that $\frac{\partial L}{\partial r_l}$ to be negative, since this
implies that a longer material is also longer in space. In a more
explicit form this condition reads
\begin{equation}\label{1+1hirter}
0>\frac{\partial L}{\partial r_l} =\int_{r_l}^{r_u} \phi
\frac{A}{A,_n}\phi(r_l)^2 \frac{m}{r_l^2} dr- n_l \phi(r_l)
\end{equation}
Without analyzing this condition any further at the moment we shall
assume it as given, since it obviously plays the role of a minimal
requirement for a reasonable model.

To assign a coordinate-independent meaning to a certain length in
space(-time), we define an intrinsic length using the induced volume
element of $t=const$ surfaces of space-time:
\begin{equation}
\label{mlaenge}
l_{m}:=\int_{r_{l}}^{r_{u}}\frac{dr}{\left(1-\frac{2m}{r}\right)^{1/2}}.
\end{equation}
Since both $r_u$ and $r_l$ can be obtained from the data, we can
assign a spatial length to the material.

One would expect that the elastic material is stretched when gravity
becomes stronger. To actually see this we compute $l_m,_m$. For this
computation recall that $r_l$ depends on $m$ via (\ref{ilaenge}).
Consequently
\begin{equation}\label{1+1lm,m}
l_m,_m= \int_{r_l}^{r_u}\frac{dr}{r
\left(1-\frac{2m}{r}\right)^{3/2}} -
\frac{r_l,_m}{\left(1-\frac{2m}{r_l}\right)^{1/2}}
\end{equation}
The first term is positive, which is in accordance with what one
would expect. In the second term we have to get rid of the $m$
derivative of $r_{l}$, since we know a priori nothing about its
sign. This is done by differentiating (\ref{ilaenge}) with respect
to $m$ and using the resulting equation to determine the sign of
$r_l,_m$.

Since $L$ is by construction independent from $m$, we have
\begin{equation}\label{1+1teekocher}
0= \frac{\partial L}{\partial r_l} r_l,_m + \int_{r_l}^{r_u}\phi
\left( \frac{n}{r}\phi^2 +\frac{A}{A,_n} \left[   \frac{\phi^2}{r}
-\frac{\phi(r_l)^2}{r_l} \right]  \right) dr
\end{equation}
To obtain the desired result, namely that $r_l,_m <0$ we have to
show that the integral has is negative.

Since doing these estimates for general $m$ would go beyond the
scope of a simple example we shall be satisfied with a simpler case,
namely by showing the above claims for small $m$. For this first
note that (\ref{1+1hirter}) is indeed satisfied:
\begin{equation}
\frac{\partial L}{\partial r_l}|_{m=0} = -n_l <0
\end{equation}
By continuity the above inequality and the following results also
hold for small $m$.

But we can do better. In fact we can derive an explicit formula for
$r_l$. We can use that the solution for the elastic equation
(\ref{gin}) for $m=0$ implies that the particle density has to be
constant to evaluate (\ref{ilaenge}) at $m=0$:
\begin{equation}
r_l |_{m=0}=r_u - \frac{L}{n_l}
\end{equation}
Thus in order to obtain an explicit expression for (\ref{1+1lm,m})
for $m=0$, we only have to compute $r_l,_M|_{m=0}$ This is done by
evaluating (\ref{1+1teekocher}) at $m=0$.
\begin{equation}
r_l,_m|_{m=0}=\int_{r_l}^{r_u} \left( \frac{1}{r} +
\frac{A}{nA,_n}|_{n=n_l}\left[ \frac{1}{r}-\frac{1}{r_l} \right]
\right)dr
\end{equation}
Thus according to (\ref{1+1lm,m}) we have
\begin{equation}
l_m,_m |_{m=0}=\frac{A}{nA,_n}|_{n=n_l} \int_{r_l}^{r_u}\left(
\frac{1}{r_l}-\frac{1}{r} \right)dr
\end{equation}
Since the integrand is positive we conclude that there  an actual
growth in the length of the material in space at least in first
order.

Using the above results we can expand the length of the elastic
material into a formal Taylor series in  the Schwarzschild mass $m$.
Using only the given data, which are $(n_l, r_u)$ as well as $A$ and
$L$ we obtain
\begin{equation}
l_m=\int_{r_u -\frac{L}{n_l}}^{r_u}\frac{dr}{r} +m
\frac{A}{nA,_n}|_{n=n_l}\int_{r_u -\frac{L}{n_l}}^{r_u}\left(
\frac{1}{r_u-\frac{L}{n_l}}-\frac{1}{r} \right)dr + \mbox{h.o.t.}
\end{equation}


The next example is simpler in nature, since we need not use $l_m$.
Assume now to hang the body into the gravitational field. Then place
the same body such that its lower end stands, where the hanging body
ended.

\begin{figure}[h]
\center \psfrag{n1,r1}{$(n_{u},r_{u})$}
\psfrag{n1,r}{$(n_{0},R_{u})$} \psfrag{n,r2}{$(n_{0},r_{l})$}
\psfrag{n2,r2}{$(n_{l},R_{l}=r_{l})$}
\includegraphics[totalheight=4cm]{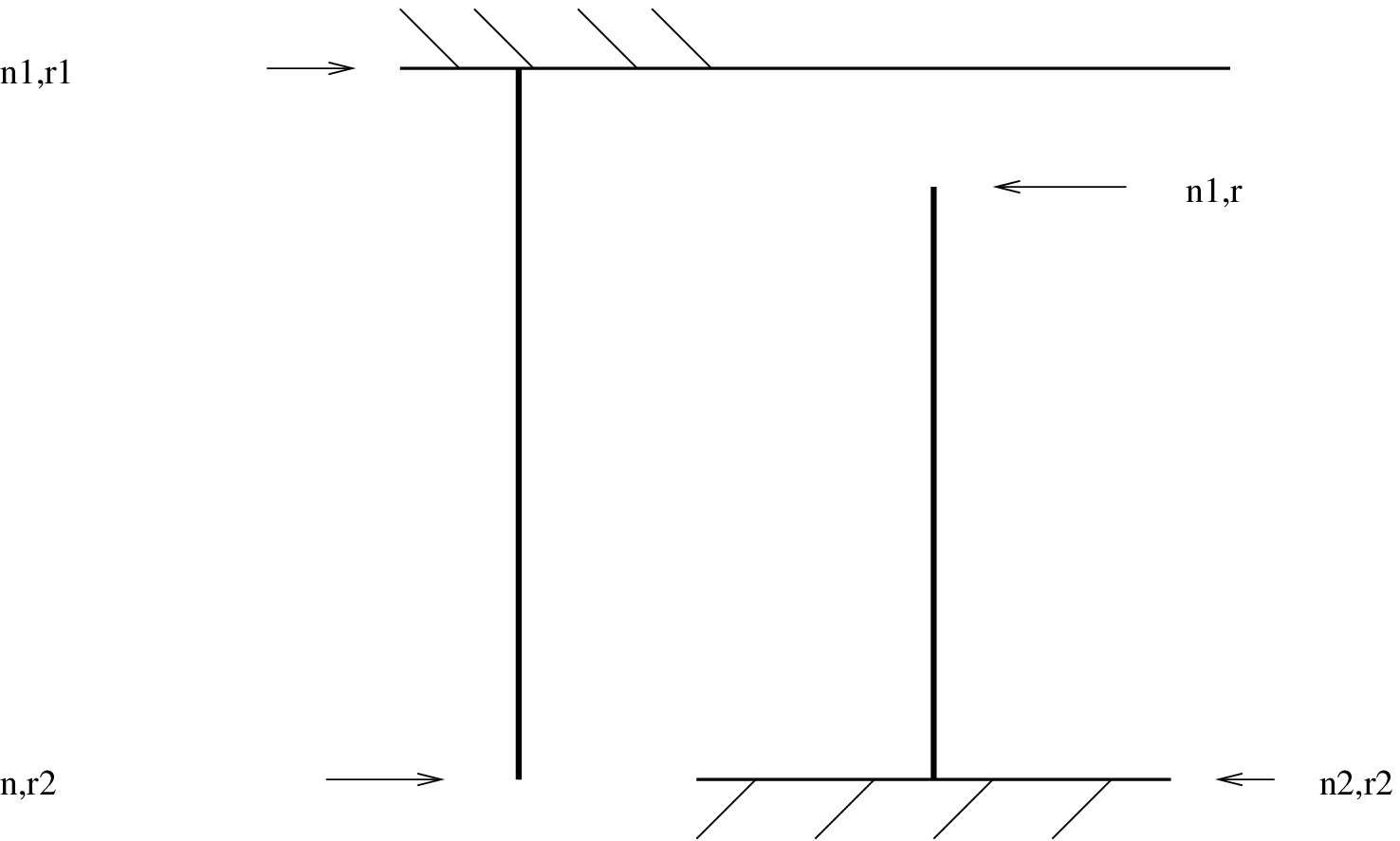}
\end{figure}

We may now compare the two solutions corresponding to the picture
(Compare (\ref{dieunglg})), where $n_h(r)$ denotes the particle
density along the hanging body (compare (\ref{gin})), while $n_s(r)$
corresponds to the hanging body (note that in this setup now $n_0$
defines the pressure-free state):
\begin{eqnarray}
\label{e1}
A(n_{h}) \phi(r_l) =A(n_{0})\phi \\
\label{e2} A(n_{s}) \phi(R_u)= A(n_0) \phi
\end{eqnarray}
These equations give $n_{h}$ and $n_{s}$ as functions of $r$. To
prove that the standing material does not reach the original
suspension point of the hanging material given by $r_u$, we proceed
by contradiction. Since both scenarios $n_s$ and $n_h$ correspond to
the same material, they both lead to the same intrinsic length $L$.
Assume now that $R_U\ge r_u$. Since the integrand is positive this
includes
\begin{equation}
\int_{r_{l}}^{r_{u}}n_{h} \phi dr= \int_{r_l}^{R_u}n_s \phi dr \ge
\int_{r_{l}}^{r_{u}}n_s \phi dr
\end{equation}
This means that there have to exist at least some points $\tilde r$
where $n_h(\tilde r) \ge n_s(\tilde r)$. By the monotonicity of $A$
this includes $A(n_h(\tilde r))\ge A(n_s (\tilde r))$.

Now we make use of the equations (\ref{e1}) and (\ref{e2}). Taken
together they imply
\begin{equation}
A(n_{h}) \phi(r_l)=A(n_{s}) \phi(R_u)
\end{equation}
Combining this with the last observation we obtain
\begin{equation}
\phi(R_u) \ge\phi(r_l)
\end{equation}
This is equivalent to
\begin{equation}
R_u \le r_l
\end{equation}
which is a contradiction to the setup. Thus we have proved the claim
that $r_u >R_u$ and have thus shown that the hanging material is the
longer one.

\begin{bem}
Note that one can formulate the above problem in a coordinate
invariant way by comparing the spatial length $l_m$. But it is easy
to see that this problem can be reduced to the one handled here.
\end{bem}

\begin{bem}
Note that in these examples we had no grave restrictions on the
equation of state nor on the state itself. In contrary to the main
body of this work the results derived here hold for materials
subjected to high stress.
\end{bem}

\section{A short remark on energy conditions}

This section will be concerned with the energy conditions and
whether they hold for elastic materials. This will be a rather short
discussion since the energy-momentum tensor given by (see equation
(\ref{stemt111})):
\begin{equation}
T_{\mu\nu}= \rho u_\mu u_\nu +2n\tau_{AB}f^A,_\mu f^B,_\nu
\end{equation}
includes the stress. Our conditions did only involve the elasticity
operator. But the following is trivially true: for states
sufficiently close to a reference state, the energy-momentum tensor
is arbitrarily close to that of a pressure-free perfect fluid.

In this sense all the energy conditions hold. For a stress-free
state we have
\begin{itemize}
\item the weak energy condition:
\begin{equation}
T_{\mu\nu}V^\mu V^\nu =\rho (u_\mu V^\mu)^2 \ge 0
\end{equation}
for all time-like $V^\mu$.
\item the dominant energy condition, since in addition to the weak
energy condition
\begin{equation}
-T_{\mu}^{\nu}V^\mu=-\rho (u_\mu V^\mu) u^\nu
\end{equation}
is future-pointing and time-like for time-like $V^\mu$ since $V^\mu
u_\mu<0$.
\item the strong energy condition: for all $V^\mu$ with $V_\mu V^\mu
=-1$ we have
\begin{equation}\nonumber
T_{\mu\nu}V^\mu V^\nu =\rho (u\cdot V)^2 \ge \rho
(-u^2)(-V^2)=\rho=-T_{\mu\nu}g^{\mu\nu}>-\frac{1}{2}T_{\mu\nu}^{\mu\nu}
\end{equation}

\end{itemize}

\begin{bem}
By the above observations at least some of the states appearing in
our analysis in the last chapter obey the energy conditions.
\end{bem}
It may be interesting to see to which degree the states for which
local existence can be obtained satisfy energy conditions (and which
of them in particular). Both statements use smallness assumptions on
the stress. Formulating those in a proper way may (or may not) lead
to interesting agreement.

\section{Non-relativistic limits}

Here we show that the non-relativistic limit of the elastic field
equations coincides with the known non-relativistic field equations.
Since the non-relativistic limit for the space-time description can
be found in \cite{bs1}, we will concentrate on the material
description. The non-relativistic field equations are given by (see
f.e. \cite{gur1})
\begin{equation}
-\ddot F^i + \frac{d}{dX^A}\left( \frac{\partial \tilde
\epsilon}{\partial F^i,_A} \right) =0
\end{equation}
which can be obtained from an action principle of the form (see
again \cite{gur1}
\begin{equation}
\int_{I\times \mathcal{B} }\left(   - \frac{1}{2}\dot F^2 + \tilde
\epsilon (k) \right)  dt d^3X
\end{equation}
We will do the non-relativistic limit only for the Lagrangian, since
the corresponding limit for the equations of motion is then a
trivial consequence if the involved quantities are regular.

The action for the material description according to
(\ref{maction1}) is given by
\begin{equation}
S[F]=\frac{1}{c}\int_{I\times \mathcal{B} } N
\gamma^{-\frac{1}{2}}\epsilon V dt d^3X
\end{equation}
Note the modification: we inserted an extra $\frac{1}{c}$ to keep
the limit regular. Otherwise the total Lagrangian would diverge with
order $c$.

In a first step we set $V=1$. The metric is chosen to be
Minkowskian, i.e.
\begin{equation}
\eta_{\mu\nu}dx^\mu dx^\nu = -c^2 dt^2 +\delta_{ij}dx^i dx^j
\end{equation}
Note that this choice already implies that the foliation, w.r.t. to
which the material description was introduced (recall that such a
foliation was necessary) is flat.

By the above restrictions $N=c$, while the shift vanishes. We find
that
\begin{equation}
\gamma^{-1}=1-W^2=1-\frac{1}{c^2}\dot F^i \delta_{ij}\dot F^j
\end{equation}
Thus developing into a formal Taylor series in terms of
$d:=\frac{1}{c}$ we find that
\begin{equation}
\gamma^{-\frac{1}{2}}= 1- d^2\frac{1}{2} \dot F^2 + \mbox{h.o.t.}
\end{equation}
It remains to evaluate the stored energy. For this purpose we have
to say a word on the strain, which is given by
\begin{equation}
h^{AB}=k^{AB}-w^A w^B
\end{equation}
where $k^{AB}=f^A,_a \delta^{ab}f^B,_b$ is the non-relativistic
version of the strain, while $w^A=d f^{A},_a  \dot F^a$ (the $d$
comes from the fact that we use $t$ now instead of $ct$ as time
variable).

Note that by the expression for $w^B$ the strain is of order
\begin{equation}
h^{AB}=k^{AB} + O(d^2)
\end{equation}
The stored energy consists of a rest-energy portion and a variable
part $\tilde \epsilon$
\begin{equation}
\epsilon(h)=c^2 + \tilde \epsilon(h)
\end{equation}
Using the above observation on the strain, we find that
\begin{equation}
\epsilon(h)=c^2 + \tilde \epsilon(k) + O(d^2)
\end{equation}
Thus the Lagrangian becomes
\begin{equation}
\mathcal{L}=  (c^2 + \tilde \epsilon(k)) (1- d^2 \frac{1}{2}\dot
F^2) + O(d^2)= (c^2  - \dot F^2 + \tilde \epsilon (k)) + O(d^2)
\end{equation}
There are essentially three parts: the first part is constant and
thus does not contribute to the field equations, since it vanishes
under variations. The last term vanishes in the non-relativistic
limit $d\to 0$ (which is of course equivalent to $c\to \infty$).
Thus the action one obtains from the non-relativistic limit is
\begin{equation}
S_{nonrel}[F]=\int_{I\times \mathcal{B} }\left(   - \dot F^2 +
\tilde \epsilon (k) \right)  dt d^3X
\end{equation}
which obviously coincides with the action given above.

\begin{bem}
Comparing this non-relativistic Lagrangian with the relativistic
one, we find the following: if one linearizes the above Lagrangian,
the resulting field equations would exactly coincide with the ones
derived from the linearized material Lagrangian (\ref{ooooo111}).
There is thus no way to distinguish between relativistic and
non-relativistic elasticity on the linearized level.
\end{bem}

\section{3+1 decompositions}

At some stage of this work (section \ref{refmat1}) we needed the
form of the Christoffel symbols in a 3+1 decomposition. For matters
of completeness these shall hence be given at this point.

Let the metric be given in the ADM form
\begin{equation}
ds^2=-N^2 dt^2 +g_{ij} (dx^i + Y^i dt) (dx^j + Y^j dt)
\end{equation}
Then the inverse metric reads
\begin{equation}
g^{\mu\nu}\partial_\mu \partial_\nu=-N^{-2}(\partial_t
-Y^i\partial_i )^2 +g^{ij}\partial_i\partial_j
\end{equation}
Then one can compute the Christoffel symbols in terms of lapse,
shift and induced metric. The result is
\begin{eqnarray}
2 N^2 \Gamma^{0}_{00}= -\frac{\partial}{\partial t}(-N^2+Y^2) + Y^{i}\left( -(-N^2+Y^2),_{i}+2\dot Y_{i}\right) \\
2 N^2 \Gamma^{0}_{0j}=-(-N^2+Y^2),_{j}+Y^{l}\left(2Y_{[l},_{j]}+\dot g_{lj}\right) \\
\label{15}  2N^2\Gamma^{0}_{ij}=\dot g_{ij}-2Y_{(i},_{j)} -Y^{l}g_{ij},_{l}+2Y^{l}g_{l(i},_{j)}\\
2 N^2 \Gamma^{i}_{00}= Y^{i}\frac{\partial}{\partial t} (-N^2+Y^2)+\left( N^2 g^{il} - Y^{i}Y^{l}\right)( (N^2-Y^2),_{l}+ 2 \dot Y_{l} ) \\
2N^2 \Gamma^{i}_{0j}=-Y^{i}(N^2-Y^2),_{j}+\left( N^2 g^{il}- Y^{i}Y^{l}\right) (2 Y_{[l},_{j]}+\dot g_{lj} )\\
 \label{3}  2N^2 \Gamma^{i}_{jk}=-Y^{i}\dot g_{jk} + 2 Y^{i}Y_{(j},_{k)}+\left( N^2 g^{il}- Y^{i}Y^{l}\right)(-g_{jk},_{l}+2g_{l(k},_{j)})
\end{eqnarray}

\bibliography{elastodynamics}





\end{document}